\documentclass{cup-hpl}

\usepackage{graphicx}% Include figure files
\usepackage{float}
\usepackage{adjustbox}
\usepackage{dcolumn}% Align table columns on decimal point
\usepackage{bm}% bold math
\usepackage[utf8]{inputenc}
\usepackage[T1]{fontenc}
\usepackage{mathptmx}
\DeclareGraphicsExtensions{.pdf,.png,.jpg,.gif}
\graphicspath{{Images/}}
\usepackage[hidelinks]{hyperref}
\usepackage{caption,subcaption}
\begin{document}
	
	\shorttitle{Complex interferometry of magnetized plasma}                                   
	\shortauthor{Iu.~Kochetkov et al.}
	\title{Complex interferometry of magnetized plasma: accuracy and limitations}
	\author[1]{Iu.~Kochetkov \corresp{Iu. Kochetkov, NRNU MEPhI, 115409, Kashirskoe shosse 31, Moscow, Russia. \email{iu.kochetkov@gmail.com}}}
	\author[2]{T.~Pisarczyk}
	\author[3]{M.~Kalal}
	\author[2]{T.~Chodukowski}
	\author[2]{A. Zaras-Szydlowska}
	\author[2]{Z. Rusiniak}
	\author[4,3]{R.~Dudzak}
	\author[3,4]{J.~Dostal}
	\author[3,4]{M.~Krupka}				
	\author[1,5,6]{Ph.~Korneev}
	\address[1]{National Research Nuclear University MEPhI, 31 Kashirskoe shosse, 115409 Moscow, Russian Federation}
	%\address[osaka]{Osaka university,1-1 Yamadaoka, Suita, Osaka Prefecture 565-0871, Japan}
	%\address[bordo]{University of Bordeaux, CNRS, CEA, CELIA, 33405 Talence, France}
	\address[2]{Institute of Plasma Physics and Laser Microfusion, Hery Street 23, 01-497 Warsaw, Poland}
	\address[3]{Institute of Plasma Physics of the Czech Academy of Sciences, Za Slovankou 1782/3, 182 00 Prague, Czech Republic}
	\address[4]{Institute of Physics of the Czech Academy of Sciences, Na Slovance 2, 182 21 Prague, Czech Republic }
	\address[5]{P.N. Lebedev Physical Institute of RAS, 53 Leninskiy Prospekt, 119991 Moscow, Russian Federation}
	\address[6]{Joint Institute for High Temperatures RAS, 13-2 Izhorskaya st., Moscow, 125412 Russian Federation}

	\begin{abstract}
		
		Expanding laser plasmas, produced by high energy laser radiation, possess both high thermal and magnetic field energy density. Characterization of such plasma is challenging but needed for understanding of its physical behaviour. Among all standard experimental techniques for plasma diagnostics, classical interferometry is one of the most convenient, informative and accurate.
		Attempts to extract more information from each single laser shot have led to development of complex interferometry, which under certain processing allows to reconstruct both plasma electron density and magnetic field strength distributions using just one data object. However, the increase in complexity and universality of the diagnostics requires a more detailed analysis of the extracted values and their accuracy. 
		This work focuses on axisymmetric interaction geometry. We present general analysis, starting from the basic principles and trace main error sources, finally ending up with plasma density and magnetic field distributions with the derived error bars. Use of synthetic profiles makes the analysis universal, though we present an example with real experimental data in the Appendix. 
	\end{abstract}

	\maketitle
	
	\section{Introduction}
	
	In laser plasma experiments, optical diagnostics take their particular place. With beam splitting, a leakage of a main beam, sometimes modified by nonlinear transformations, is often used in  interferometric and polarimetric measurements. Combination of different beam sources also may be implemented, although their synchronization is a sophisticated task \cite{Dostal-17}.
	
	The interference phenomenon is used in classical interferometry to obtain information about the phase shift between a reference wave and a probe electromagnetic wave, propagating through an optically thin object, e.g. laser-produced plasmas \cite{Plasma_Interferometry_2017}. This information is represented by interfringe pattern distortion and can be extracted by appropriate phase detection (retrieving phase map) and unwrapping (calculating a single continuous field from measured one, that is divided into separate regions with phase values $\in\left(-\pi ,\pi\right]$) techniques.
	Another very powerful diagnostic tool is polarimetry - measurement and interpretation of polarization state of the probe beam. 
	In the most general case the polarization state can be fully characterized by means of four Stokes parameters \cite{Hough}, which can be reduced to three in assumption of beam depolarization absence. They can be calculated using three experimentally measured independent parameters – e.g. two orthogonal amplitudes and their phase difference. However, under certain stronger assumptions, even one parameter may be enough to characterize the change of polarization state. This change may be induced by magneto-optical effects, and to connect measured data with magnetic field parameters the additional information about the medium in which the polarization plane rotates is required.	 
	In particular, the polarimetry as an independent diagnostics can be used for measurement of quasistationary vacuum magnetic fields by placing the birefringent crystal with known parameters in the field area \cite{Santos-NJP2015, Pisarczyk2019}.
	
	The two aforesaid methods can be used either separately or together, when they complement each other and allow simultaneous determination of plasma electron density and plasma magnetic fields \cite{Pisarczyk90,Pisarczyk2015}. The spontaneous magnetic fields (SMFs) generation is one of the most interesting phenomena, occurring in laser plasma \cite{Korobkin1966, STAMPER1971, Basov, TadeuszPPCF}. Those fields participate in plasma evolution influencing both hydrodynamics and kinetics of the plasma \cite{TadeuszPPCF,kinetic}. Understanding and control of SMFs is a significant part of laser-plasma studies dealing with particle acceleration \cite{Gong}, laboratory astrophysics \cite{RevModPhys.78.755}, inertial confinement fusion \cite{strozzi_2015, Hohenberger} and many other applications.   
	
	The standard approach for SMF measurements implies usage of two separate diagnostics  - polarimetry and interferometry, and subsequent combination of experimental data, which may cause certain accuracy and interpretation problems due to difference in optical paths and magnifications. The complex interferometry, which combines classic interferometry and polarimetry in a one diagnostic tool, allows to record and subsequently reconstruct several sets of independent data from a single data object named complex interferogram \cite{Pisarczyk2015,Kalal2003,Kalal:87, AgaArt,AgaBelfast}. Despite its complexity the complex interferometry method possess certain advantages due to fewer sources of errors, leading to increased accuracy, and the setup relative compactness.
	
\begin{figure*} 
		\begin{center}
			\includegraphics[width=0.93\linewidth]{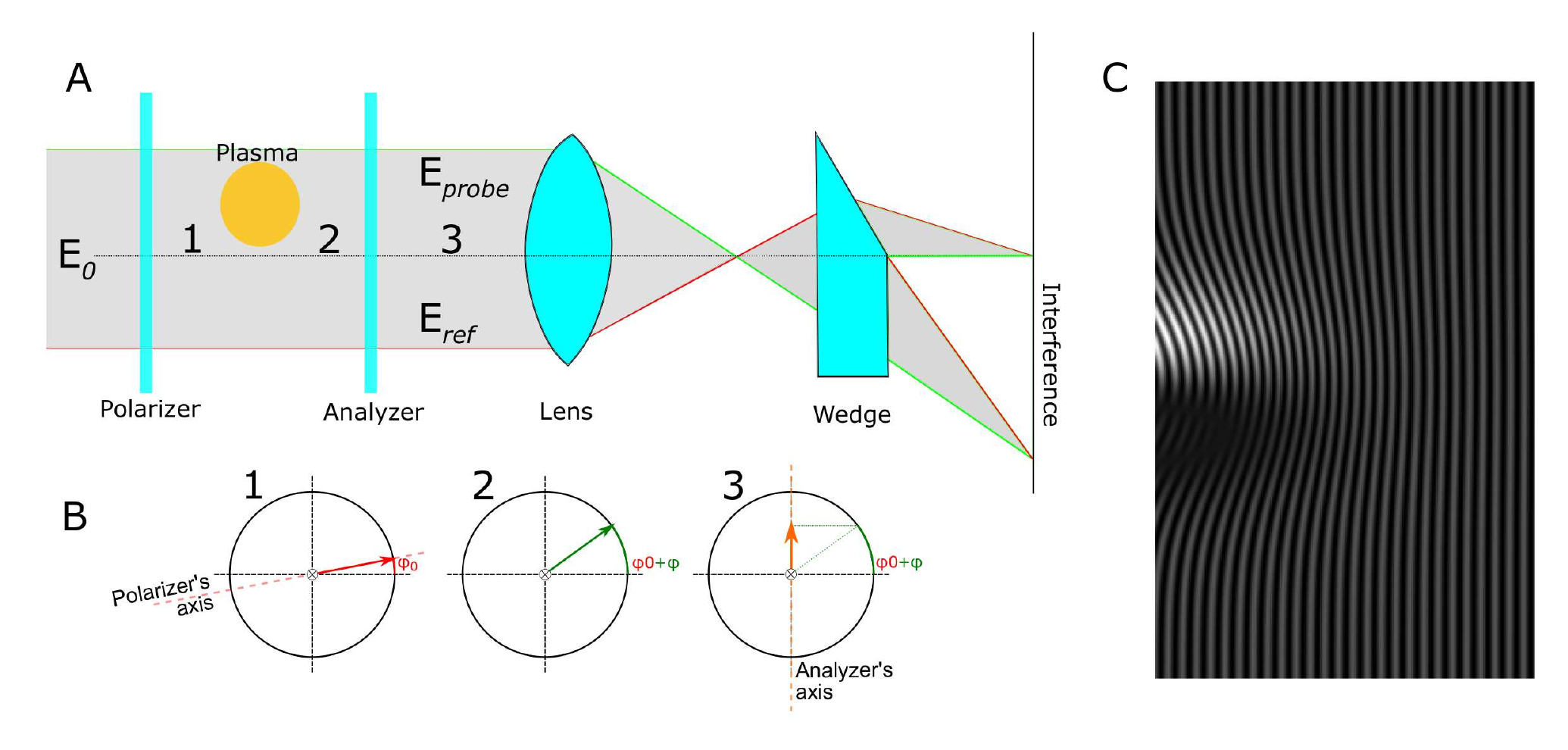}
			\caption{Basic optical scheme of single-frame polaro-interferometer module operating in the complex interferometry regime (A), probe beam polarization plane rotation (B), typical complex interferogram (C).}
			\label{img:scheme}
		\end{center}
	\end{figure*}

	The complex interferometry concept dates back to first SMF studies, and provides excellent spatial resolution \cite{AgaArt,AgaBelfast,Kalal1998}. However, due to setup and data processing complexity it was not widely used until recently, when a femtosecond complex interferometry system had been tested and implemented at the PALS facility in a single-frame regime \cite{Pisarczyk2015}, and later was upgraded to 3-frame system \cite{Pisarczyk2019}. This setup (see simplified optical scheme of one channel at Fig. \ref{img:scheme}) is based on two polarizers, with nearly ninety-degrees crossed axes (small uncrossing $\varphi_0$ is chosen for optimal registration conditions as will be shown later) for a polarimetric part and a specially designed wedge for an interferometric one (lateral shear interferometry).  

	In case of normal incidence of an axisymmetrical laser beam on a plane target, obtained plasma profile and distributions of physical quantities therein are also axisymmetric up to some extend.  In this quasisymmetric situation, assuming only small deviations from axisymmetrical plasma profiles, a tomographic setup is not required to fully characterize plasma parameters.   
	Ideally, using the axial symmetry in the plasma it is possible to reconstruct entire $B(r)$ distribution from a path-integrated projected data. However, in a realistic situation, even a small uncertainty and non-ideality of the plasma expansion may considerably increase the error of the measurements, especially around the plasma axis. In this work, the most important error sources are considered in the context of the complex interferometry of quasi-axisymmetrical plasma flows. Possible error handling algorithms are proposed and discussed.  
		
	\section{Mathematical basis of complex interferometry in a quasistationary process approximation}\label{sec:math}
	The full mathematical formulation of complex interferometry can be found in \cite{Kalal:87, Kalal1998}, and here it is reformulated in a simplified form to highlight the critical points of the processing algorithms. For simplicity, we will hereafter consider only quasistationary processes, thus avoiding time-averaging uncertainty. In case of a reasonably long probe beam, providing data averaged over a small time interval of the whole plasma expansion process, the analysis presented in this work remains valid. Note that implementation of a short auxiliary laser beam with a long main driver \cite{Dostal-17}, presented in Appendix, corresponds namely to this situation.
	%\subsection{Complex interferometry basics}
	
	First of all consider the intensity, measured by the detector. Taking $E_{probe}$ and $E_{ref}$ as the amplitude of electric field that passed through the plasma region and the undisturbed one (see Fig. \ref{img:scheme}A), $\omega_0$ and $\nu_0$ as the spatial frequencies in horizontal (y) and vertical (z) directions, which depend on probe and reference beam convergence angle and wavelength, $\delta(x,y)$ as the phase shift between probe and reference beams, and omitting temporal evolution of these parameters, registered intensity is:
	\begin{equation}\label{eq:I}
	\begin{aligned}
	I(y,z)=E_{probe}^2 + E_{ref}^2 +\\
	+2E_{ref}E_{probe}\cos[2\pi(\omega_0 y+\nu_0 z) +\delta(y,z)].
	\end{aligned}
	\end{equation}
	For classical interferometry and for polarimetry it is preferable to obtain two interferograms - a 'reference' one, before plasma is formed, and a 'shot' one - interferogram taken when diagnosing the plasma. As the complex interferometry is a combination of these two diagnostics methods, a reference complex interferogram is essential to have an acceptable accuracy. In real experiment, laser pulse total energies may vary from shot to shot, so to take it into account we introduce a ratio $p=\hat{E}/E$, where $E$ and $\hat E$ are electric field amplitudes during the reference and shot stages respectively. Hereafter we will denote parameters for the shot with the hat.
	
	Before passing through a magnetized plasma region, electric field amplitudes $E_{probe0}=E_{ref0}=E_{0}$. Then $E_{probe0}$ is rotated by some angle $\varphi(y)$ due to Faraday effect (see Fig. \ref{img:scheme}B), which is an important quantity for further analysis.
	Both beams pass through the analyzer set up with small uncrossing $\varphi_0$, that results in change of their amplitudes by Malus's law:
	
	\begin{equation}\label{eq:Eref}
	\begin{aligned}
	E_{ref}(y,z)= E_{0}(y,z)\cdot \cos(\varphi_0), \\
	E_{probe}(y,z) = E_{0}(y,z)\cdot \cos(\varphi_0), \\
	\hat{E}_{ref}(y,z)=pE_{0}(y,z)\cdot \cos(\varphi_0),\\
	\hat{E}_{probe}(y,z) = pE_{0}(y,z)\cdot \cos[\varphi_0 + \varphi(y,z)].
	\end{aligned}
	\end{equation}

	In order to determine SMF distribution, as will be shown below, phase shift $\delta(y,z)$ and rotation of polarization plane $\varphi(y,z)$ should be determined. While phase shift can be determined by standard fringes detection and phase unwrapping techniques, the $\varphi(y,z)$ can be extracted from a background functions $b(y,z)$ and $\hat{b}(y,z)$, defined as follows
	
	\begin{equation}\label{eq:b}
	\begin{aligned}
	b(y,z)=E_{probe}^2 + E_{ref}^2 = 2E_{0}^2(y,z)\cdot \cos^2(\varphi_0),\\
	\hat{b}(y,z)=\hat{E}_{probe}^2 + \hat{E}_{ref}^2.
	\end{aligned}
	\end{equation}
	The background is defined by the central lobe of Fourier spectrum of the interferogram and can be extracted by it's low-pass filtering and inverse Fourier transform.
	
	From (\ref{eq:Eref}) it follows that 
	\begin{equation}\label{eq:arccos}
	\begin{aligned}
	\varphi(y,z) = \arccos	\left( \frac{\hat{E}_{probe}}{\hat{E}_{ref}}\cdot \cos(\varphi_0)\right) - \varphi_0 = \\
	=\arccos(\gamma\cdot \cos(\varphi_0)) - \varphi_0,
	\end{aligned}
	\end{equation}
	where the amplitudes ratio $\gamma$ we can be calculated from known parameters
	
	\begin{equation}\label{eq:gamma}
	\begin{aligned}
	\gamma(y,z)=\frac{\hat{E}_{probe}}{\hat{E}_{ref}}=\sqrt{\frac{\hat{b}(y,z)-\hat{E}_{ref}^2}{\hat{E}_{ref}^2}}=\sqrt{\frac{2\hat{b}(y,z)}{p^2 b(y,z)}-1} .
	\end{aligned}
	\end{equation}
	
	%\subsection{Magnetic field calculation}
	
	The most general equations, connecting experimentally obtained parameters, such as  the shift of interference fringes $\delta(y,z)$ and rotation of polarization plane $\varphi(y,z)$ with plasma parameters, can be found in literature, see e.g. \cite{Pisarczyk90}. For a fixed coordinate on an axis $z$ perpendicular to the target surface, in CGS system units they read
	
	\begin{equation}\label{eq:phase}
	\begin{aligned}
	\delta(y) = \frac{1}{\lambda}\int_{0}^{L}\Delta N(x)dx= \frac{\lambda e^2}{2\pi c^2 m_e}\int_{0}^{L}n_e(x)dx\\
	=4.49\cdot 10^{-14}\lambda\int_{0}^{L}n_e(x)dx,
	\end{aligned}
	\end{equation}
	
	\begin{equation}\label{eq:rotation angle}
	\begin{aligned}
	\varphi(y) =\frac{\pi}{\lambda}\int_{0}^{L}(n_2-n_1)dx = \frac{e^{3}\lambda^{2}}{2\pi m_e^{2}c^{4}}\int_{0}^{L}n_{e}B_{||}dl\\
	=2.62\cdot 10^{-17}\lambda^{2}\int_{0}^{L}n_{e}B_{||}dx,
	\end{aligned}
	\end{equation}
	where $x$ is the coordinate along the beam propagation direction, $L$ is the geometrical path length inside the plasma, $\lambda$ is a wavelength, $c$ - speed of light, $m_e$ - the electron mass, $e$ the electron charge, $n_e$ - electron density, $B_{||}$ - magnetic field projection on probing direction in G.
	
	Equation (\ref{eq:phase}) provides the linear (integrated) density $\int\limits_0^L n_e(x) dx$. 
	The integrated magnetic field may be found as:
	\begin{equation}\label{eq:field integrated}
	B_{\int||}(y)=\frac{\int_{0}^{L}n_{e}B_{||}dx}{\int_{0}^{L}n_e(x)dx}=\frac{1.7\cdot 10^3}{\lambda}\cdot \frac{\varphi(y)}{\delta(y)}.
	\end{equation}
	
	In case of axially symmetric distributions it is possitble to reconstruct not only the integrated field, but the entire distribution of $\vec{B}(r)$. Rewriting (\ref{eq:phase}) and (\ref{eq:rotation angle}) in the cylindrical coordinates, we have:
	\begin{equation}\label{eq:phase_cyl}
	\delta(y) = 2\cdot 4.49\cdot 10^{-14}\lambda  \int_{y}^{R} \frac{n_e(r)rdr}{\sqrt{r^2-y^2}},
	\end{equation}
	
	\begin{equation}\label{eq:rot_cyl}
	\varphi(y) = 2\cdot2.62\cdot 10^{-17}\lambda^2  \int_{y}^{R} \frac{B(r)n_e(r)ydr}{\sqrt{r^2-y^2}},
	\end{equation}
	where $\vec{B}(r)$ is perpendicular to $\vec{r}$, $R$ is the maximal calculation radius.
	Choice of this radius in not a trivial question. On the one hand, in order for the Abelized distributions to have a better accuracy, R should be maximized. On the other hand, due to possible errors in data the radius should be minimized to exclude them. Therefore, the optimal R value should be chosen analysing the Abelized function distribution.  
	
	Using a canonical form of Abel transformation and denoting $\mathcal{A}$ as an operator of abelization:
	
	\begin{equation}\label{eq:abel_canon}
	F(y) = \mathcal{A}(f(r))=2\int_{y}^{R} \frac{f(r)r}{\sqrt{r^2-y^2}}dr,
	\end{equation}
	
	equation (\ref{eq:phase_cyl}) results in
	\begin{equation}\label{eq:f_phase}
	f_{\delta}(r) =4.49\cdot 10^{-14}\lambda n_e(r),
	\end{equation}
	
	\begin{equation}\label{eq:F_phase}
	F_{\delta}(y)=\delta(y);
	\end{equation}
	
	and equation (\ref{eq:rot_cyl}) gives
	\begin{equation}\label{eq:f_rot}
	f_{\varphi}(r) =2.62\cdot 10^{-17}\lambda^2 \left( \frac{B(r)n_e(r)}{r} \right) ,
	\end{equation}
	
	\begin{equation}\label{eq:F_rot}
	F_{\varphi}(y)=\varphi(y)/y .
	\end{equation}
	
	To calculate function $f$ from $F$, inverse Abel transformation should be performed \cite{Bracewell}
	\begin{equation}\label{eq:abel_inv}
	f(r) =\mathcal{A}^{-1}(F(y))= -\frac{1}{\pi}\int_{r}^{R} \frac{dF}{dy} \frac{dy}{\sqrt{y^2-r^2}} + \frac{F(R)}{\pi \sqrt{R^2-r^2}}. 
	\end{equation}
	
	Abel inversion solution methods can be divided in two groups, interpolation and approximation. The direct solution is effective only for ideally smooth distributions due to presence of sensitive to noise derivative in the integral (see \cite{Pisarczyk90, Kasper1978, Hansen:85}). In this work, several methods were tested to substantiate the results presented below for the synthetic data: Mach-Shardin (also known as Pierson) method as interpolation and Gegenbauer and Fourier as approximation methods.
	
	Finally, by experimental measurements of parameters $\delta(y)$ and $\varphi(y)$, parameters $F_{\delta}(y)$ and  $F_{\varphi}(y)$ are calculated, and the inverse Abel transformation then allows to calculate  $f_{\delta}(r)$ and $f_{\varphi}(r)$. The latter are directly connected to plasma parameters and can be used to define distribution of the magnetic field $B(r)$ as
	
	\begin{equation}\label{eq:B(r)}
	B(r)=1.7\cdot 10^3\cdot\frac{rf_{\varphi}(r)}{\lambda f_{\delta}(r)}.
	\end{equation}
	
	In real experiment there are multiple factors, influencing the measured intensity, not included in this mathematical model and therefore impair the diagnostic accuracy. Among them the most significant are: (i) a probe beam depolarization in plasma, (ii) an influence of plasma self-luminosity, and (iii) a shot-to-shot instability of energy distribution in a cross-section of the probing beam. Hence the intensity, registered on a detector, can be described (see \cite{AgaArt}) by an equation:
	\begin{equation}\label{eq:I_real}
	I_{real}=I_{0}(K+k+\text{sin}^2(\varphi_0+\varphi))+I_{pl},
	\end{equation}
	where $I_0=E_{0}^2$ - initial probe beam intensity, $K$ is a polarization coefficient of the probing beam, $k$ is  the contrast of polarizers and $I_{pl}$ is an intensity of plasma self-luminosity. Comparing this intensity with the parasitic signal of background present even in with absence of SMF and described as :
	\begin{equation}\label{eq:I_real}
	I_{bg}=I_{0}(K+k+\text{sin}^2(\varphi_0))+I_{pl},
	\end{equation}
	and maximising the ratio of useful signal to background signal $(I_{real}/I_{bg})$ gives an optimal angle of polarizers uncrossing $\varphi_0$:
	
	\begin{equation}\label{eq:Phi_optimal}
	\varphi_0\approx arcsin\sqrt{\frac{1+\epsilon(K+k)}{\epsilon[1-2(K+k)]-2}} ,
	\end{equation}
	where $\epsilon=I_0/I_{pl}$ is a ratio between the probing beam intensity and the the intensity of plasma self-luminosity.
	
	\section{Data analysis} \label{sec:analysis}
	
\begin{figure*}
		\begin{center}
			\includegraphics[width=0.33\linewidth]{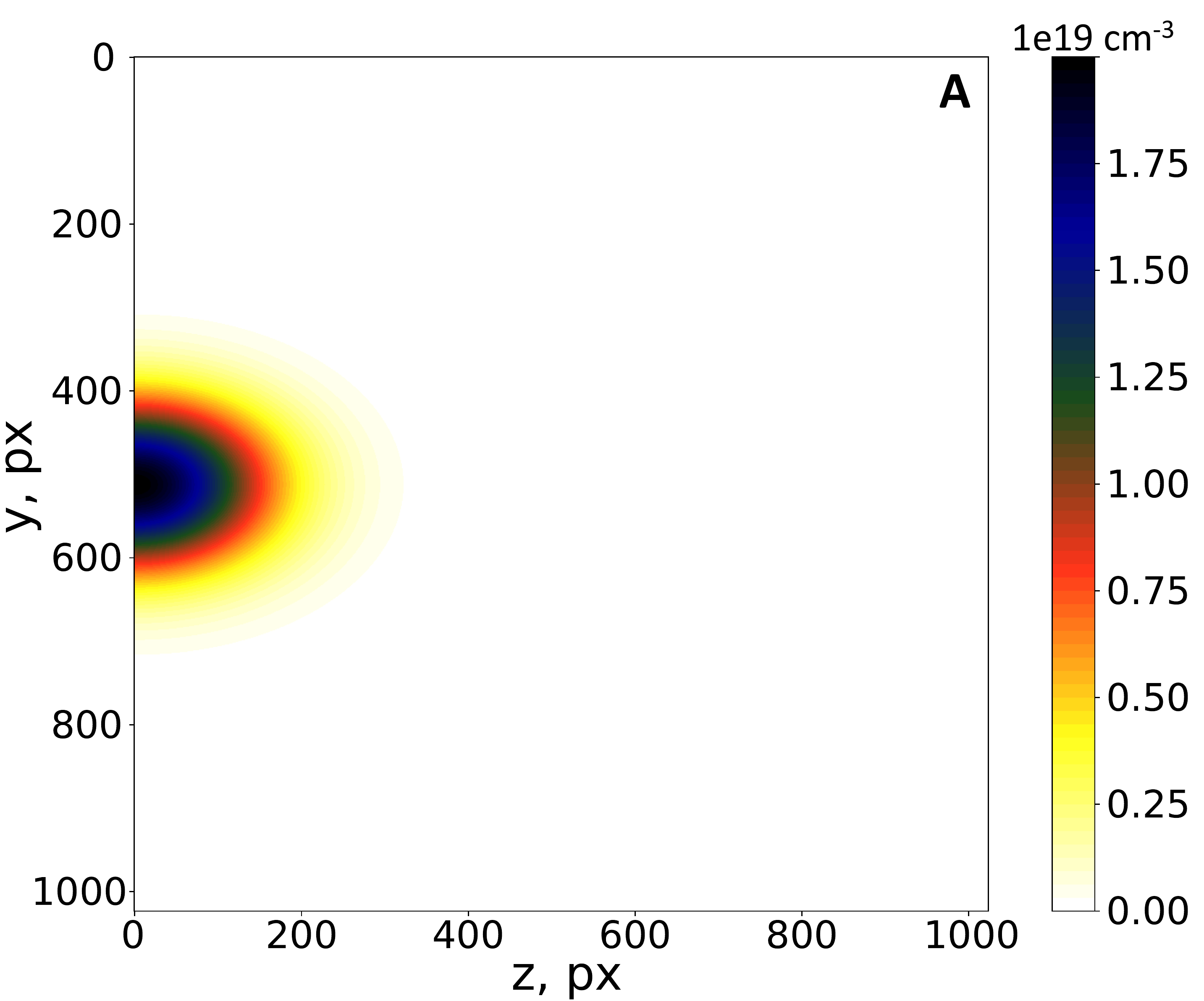}
			\hfill 
			\includegraphics[width=0.33\linewidth]{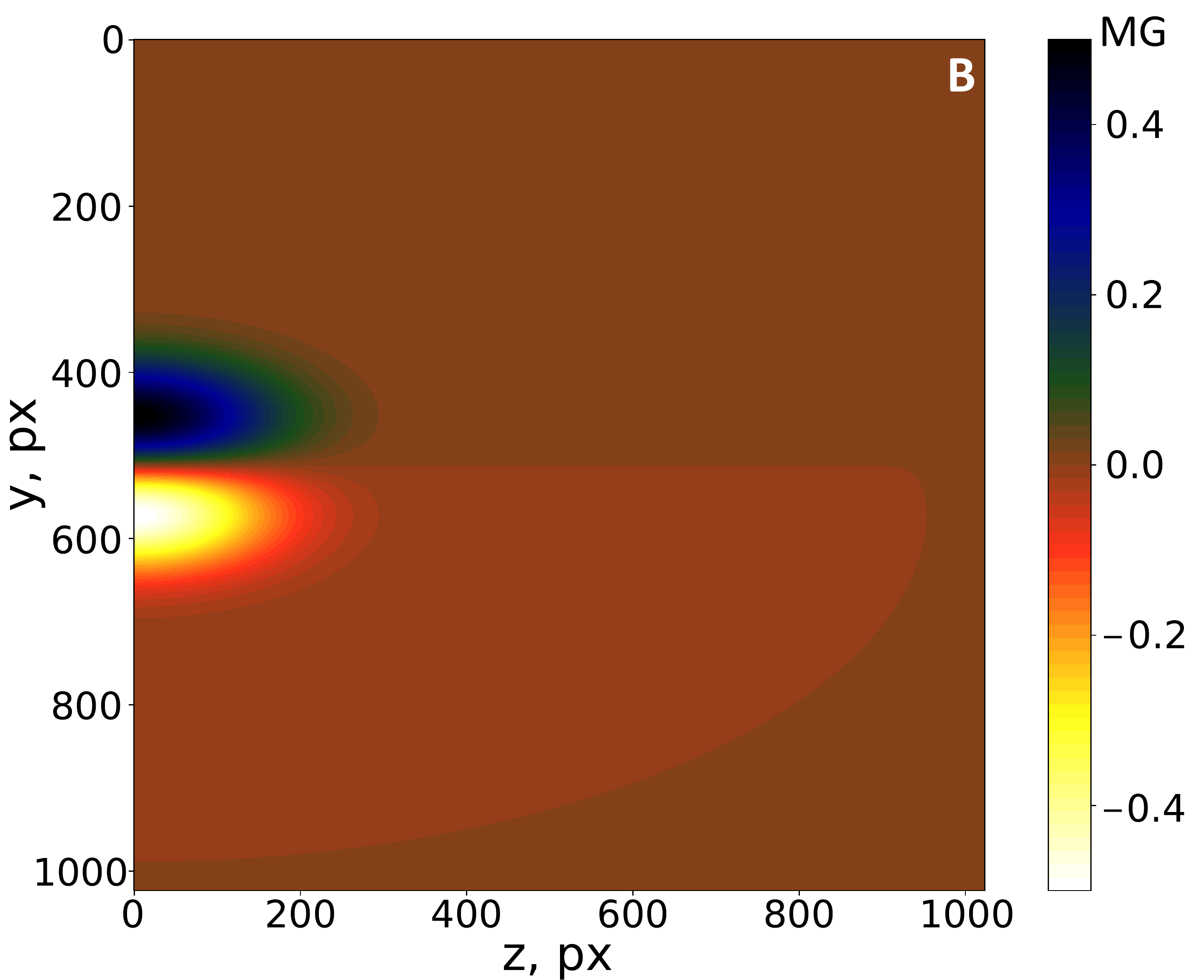}
			\hfill 
			\includegraphics[width=0.29\linewidth]{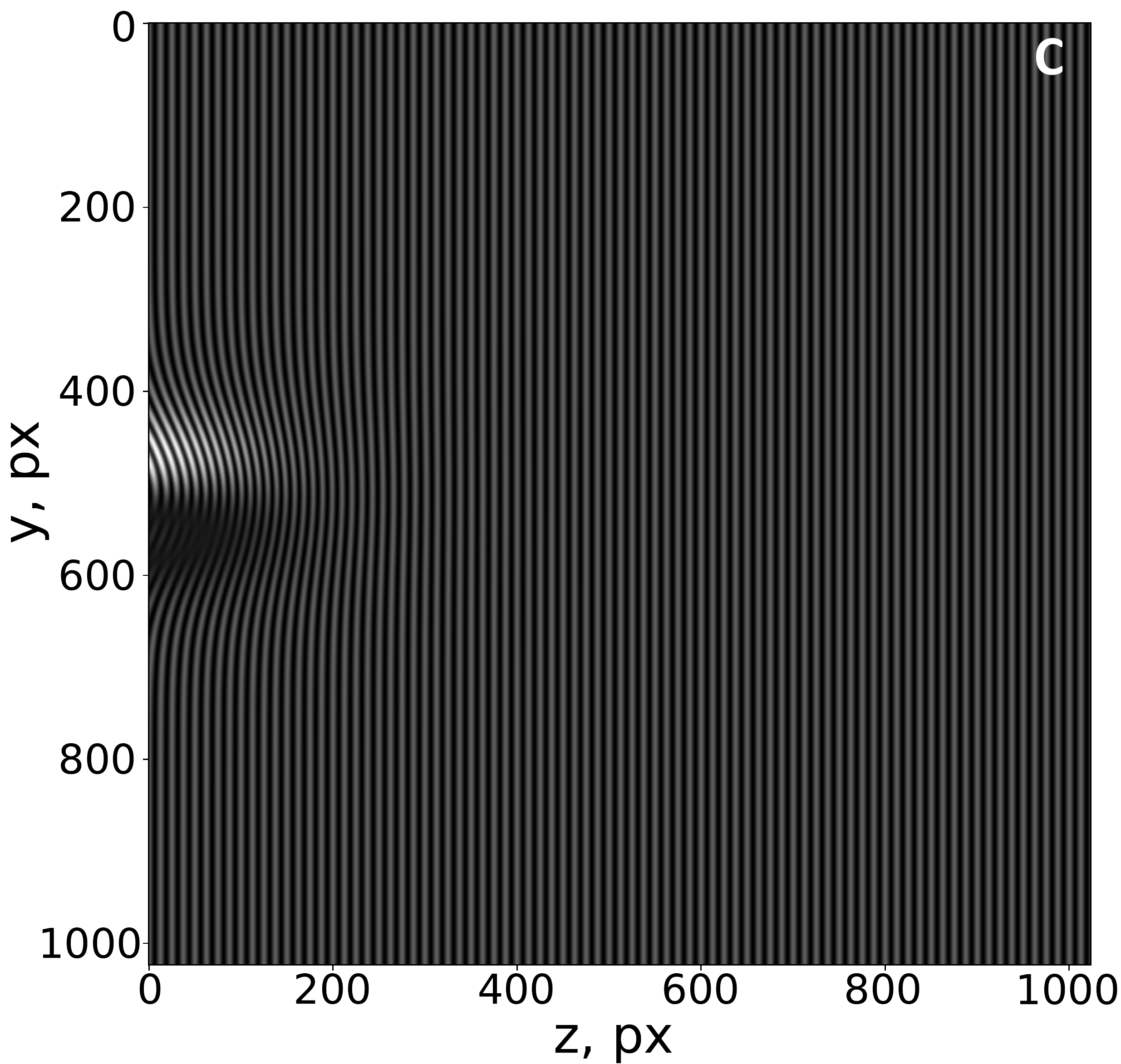}
			\caption{Example of density (A), magnetic field (B) distributions, and synthetic complex interferogram (C) used for analysis.}
			\label{img:Simulated_Interferograms}
		\end{center}
	\end{figure*}

\begin{figure}[H]
		\begin{center}
			\includegraphics[width=0.99\linewidth]{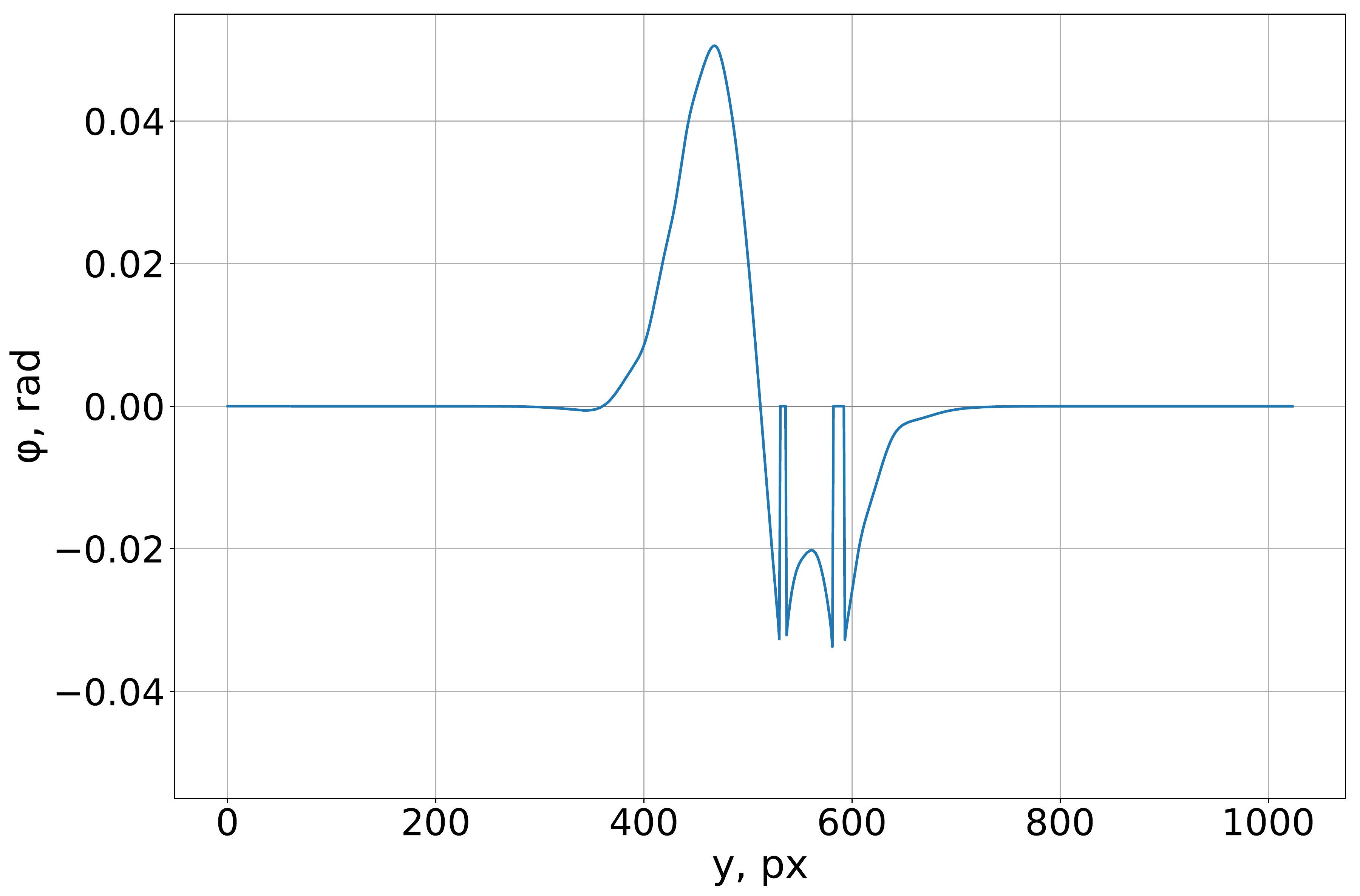}
			\caption{Reconstruction of angle value from simulated data at $z$=30 px.}
			\label{img:dips}
		\end{center}
	\end{figure}

	Further analysis is based on the processing forth and back synthetic polaro-interferograms. For that a set of synthetic complex interferograms were prepared using equation (\ref{eq:I}). A plasma density and magnetic field distributions were defined analytically as a combination of gaussian profiles. Each of the 2D Gaussian distributions is characterized by it's central point, placed at the left border of interferogram, imitating target surface, and two parameters representing the function decrease rate at perpendicular directions. In a simpler case it is possible to analytically calculate the resulted phase shift, in other cases forward Abel transformation or direct numerical integration can be carried out - since the function is analytic and therefore smooth, this doesn't introduce a noticeable error.
	The same is applied to the polarization plane rotation angle, calculated using density and field distributions.
	In the presented analysis the maximum density value is $2\cdot10^{19}$ cm$^{-3}$ and maximum field value is 0.5 MG (see Fig. \ref{img:Simulated_Interferograms}), which for the chosen distributions result in maximum polarization plane rotation of 0.05 rad. This value is bigger than optimal initial polarizes uncrossing of $\varphi_0=0.032$ rad, calculated using equation (\ref{eq:Phi_optimal}) and experimental setup parameters found at \cite{Pisarczyk2015}: $K$=0, $k$=5$\cdot 10^{-6}$, $\epsilon\approx 10^3$. Such synthetic density and field data were chosen as in the real experiment the polarization plane rotation angle maximum value is impossible to predict and can exceed the optimal angle $\varphi_0$. Under such conditions the reliable reconstruction of polarization plane rotation angle distribution can be carried out only using a half of this antisymmetric distribution, where the rotation direction coincide with initial polarizers uncrossing direction. The angle in an opposite half will have dips when the shot background value $\hat{b}(y,z)$ becomes much smaller than the reference one $b(y,z)$ due to negative values under the root function at equation (\ref{eq:gamma}), and will have opposite derivative sign when it's real value exceeds $\varphi_0$ (see Fig. \ref{img:dips}).

	Note that for further graphical interpretation the cross-sections of studied distributions are presented. The cross-sections are taken at fixed coordinate on "z" axis. This coordinate is 100 px throughout this paper, unless specifically stated otherwise.
	
	\subsection{Experimental uncertainties}
	
	The main uncertainties considered in this work as sources of possible errors, originate from both interferometric and polarimetric parts of the diagnostics. Although various physical processes are responsible for the each of the sources discussed below, it is not usually necessary to define where these uncertainties come from, which may be impossible in realistic situation. For this reason we introduce the two classes of the uncertainties: $\mathbf{(G)}$ -- \textit{geometrical} and $\mathbf{(V)}$ -- \textit{value} errors. 
	
	\textit{Geometrical} errors include deviation of both plasma and polarization plane rotation angle axes from their (identical) ideal positions, their distortion and distributions asymmetry. There are many physical effects 
	contributing to such errors: non-ideality of irradiation, inaccuracy of expected angles in the optical scheme of an experimental setup, some of the plasma instabilities, etc. The considered error sources, which are applied to the synthetic interferograms, are $\mathbf{G_{angle}}$ -- deviation of the symmetry axis from the normal to the target plane; $\mathbf{G_{shift}}$ -- a shift between ideally coincidental axes. Note that the symmetry axis in real experiment may be distorted, however in this paper we will consider only a straight axis case. The presented analysis remains valid for a distorted axis case as all the calculations are carried out individually for each cross-section of considered distributions.
	
	\textit{Value} errors originate from plasma self-illumination, absorption, second-order on the magnetic field value  effects, such as the Cotton-Moutton one, and from shot to shot instability in  probe laser beam energy distribution. According to our estimations, the Cotton-Mouton effect in case of an axially symmetrical distribution is strongest on the symmetry axis, and reaches about 1$\%$ of maximum brightness modulation caused by Faraday effect for the real experimental situation \cite{TadeuszPPCF}. Laser energy distribution instability, on the other hand, can cause a value error of more than 20$\%$. The considered error sources, which are applied to the synthetic interferograms, are $\mathbf{V_{int}}$ -- change in the intensity i.e. darkening or brightening of the whole image; $\mathbf{V_{noise}}$ -- noise applied to the whole image. The noise was added to synthetic image Fourier spectrum and was represented by an image-sized array with normally distributed values with mean value of zero and variance defined from experimental images, multiplied by 2D Gaussian function with bell curve maxima at synthetic image's zero frequency. 
	%(see Fig. \ref{img:noise_add}). 
	Two cases were tested: addition of the same noise distribution to both reference and shot interferograms and addition of individually generated for each image noise distributions.

	%	\begin{figure}[H]
	%		\begin{center}
	%			\includegraphics[width=0.3\linewidth]{spectrum_slice1.pdf}
	%			\hfill 
	%			\includegraphics[width=0.3\linewidth]{spectrum_sim_slice1.pdf}
	%			\hfill 
	%			\includegraphics[width=0.3\linewidth]{brigtness_vary1.pdf}
	%			\caption{Cross-section (10 pixels from zero frequency point) of spectrums of real and simulated interferogram (A), cross-section (10 px from zero frequency point) of spectrums of real interferogram and simulated noise (B) and vertical (along the bright fringe) slice (z=400 px) of real and simulated noisy interferogram (C).}
	%			\label{img:noise_add}
	%		\end{center}
	%	\end{figure}
	
	\subsection{Plasma electron density calculation}
	
	The phase distribution $\delta(y,z)$ can be extracted from complex interferograms using various methods \cite{Pisarczyk90, Kasper1978, TadKasp2001}. The processing can be divided into two stages: phase detection and phase unwrapping. Phase values retrieved lie within the interval (-$\pi$, +$\pi$]. Hence, the phase map consists of bands, has discontinuities at their boundaries and is called "wrapped". Process of elimination of phase discontinuities (wraps) recovering a correct phase map is called phase unwrapping. 
	Ideally, the unwrapping problem is trivial for phase maps of a good quality where the signal is free of noise and the phase changes are not very abrupt. However in the presence of noise, speckles, regions with abrupt phase changes, unwrapping becomes more challenging \cite{Ghiglia1998TwoDimensionalPU}. Since last few decades many algorithms has been proposed for solving phase unwrapping problems \cite{Phase2017}.
	
	In this study several different methods of phase detection and unwrapping were used. The Fourier \cite{Takeda82} and wavelet \cite{Zhong:05} phase detection can give very good and accurate results, however struggle in a high-density region of plasma due to a low fringes contrast. Method of a direct tracing of fringes, when they are outlined by a scientist, is free from this drawback, but takes incomparably more time.
	First two methods are implemented in several freely available interferometry processing software - IDEA \cite{IDEA} and Neutrino \cite{Neutrino}. The later one is more flexible as it has many parameters that can be varied and several different unwrapping techniques built in - Quality guided \cite{Quality}, Goldstein and some simpler ones \cite{Ghiglia1998TwoDimensionalPU}. 
	
	In real experiment plasma can not be \textit{ideally} symmetrical due to many factors - instabilities of plasma configuration, unperfect laser focal spot energy distribution, surface inclination etc. However to use axial symmetry approximation assumed for SMF distribution calculation the axis has to be chosen. In this paper we choose the axis as a straight line, however, as all of the calculations are carried out individually for plasma cross-sections at each point on the "z" axis, all the conclusions are valid for non-straight axis case.
	
	There may be different approaches to select the axis, which can lead to slightly different results. The approach we used was based on specifying and minimizing the objective function:
	\begin{equation}\label{eq:symm func}
	f(y)=\sum_{z_{min}}^{z_{max}}\sum_{i=1}^{R_z} |n_e(y-i,z)-n_e(y+i,z)|,
	\end{equation}
	where $n_e(y,z)$ is a considered electron density distribution, $R_z$ - radius of plasma in the cross-section $z$. It should be noted that due to the lack of perfect symmetry the calculated axis of the plasma $y_{plasma}$ is just our approximation, that contributes to the resulting error of SMF calculation.

	\subsection{Polarization plane rotation angle calculation}
	
\begin{figure}[H]
		\begin{center}
			\includegraphics[width=0.49\linewidth]{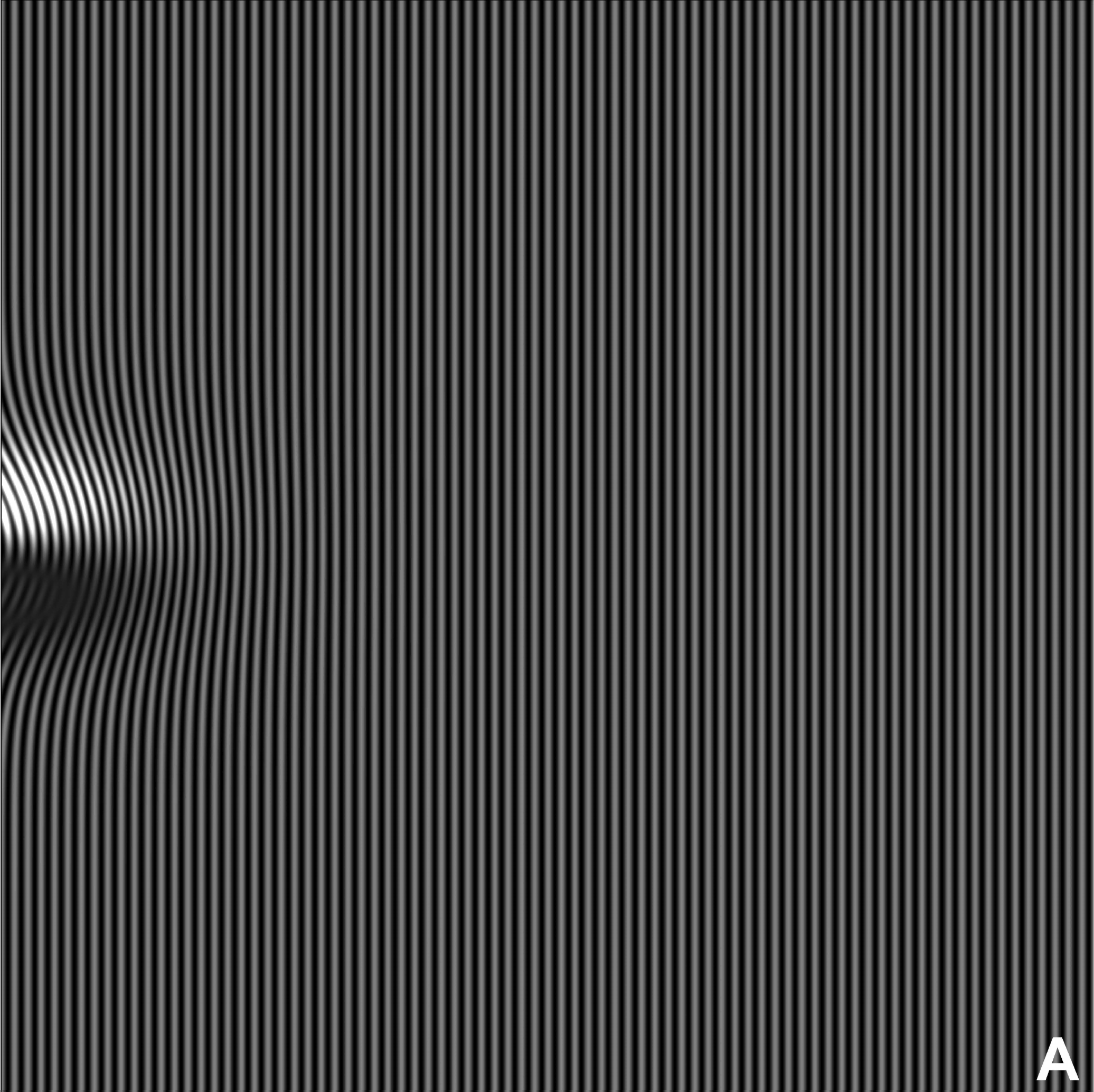}
			\hfil 
			\includegraphics[width=0.49\linewidth]{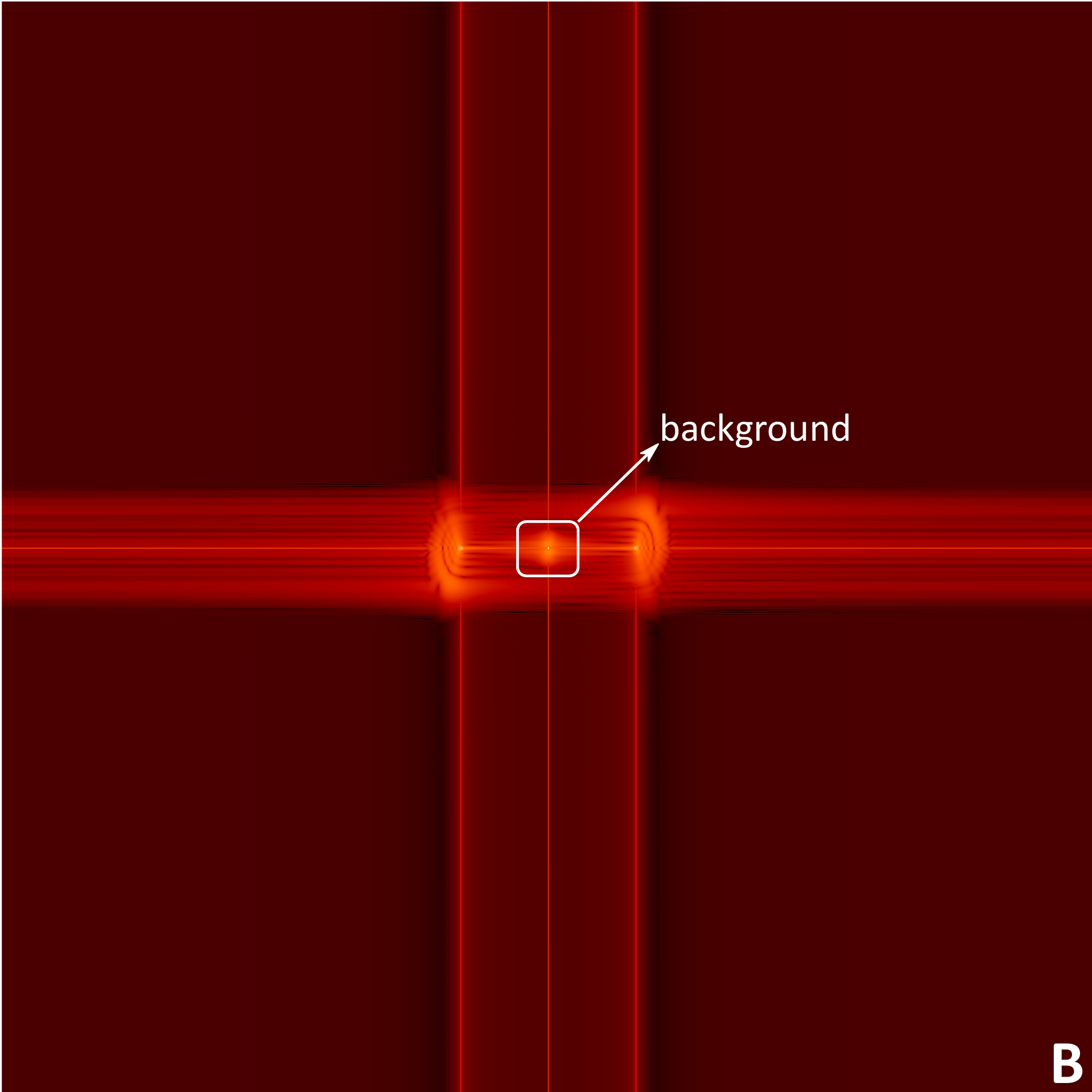}
			\caption{Reference complex complex interferogram (A) and it's spectrum (B).}
			\label{img:int and spec}
		\end{center}
	\end{figure}

	As described in presented mathematical methodology, the polarization plane rotation angle $\varphi(y,z)$ is encrypted in background function and can be calculated by equation (\ref{eq:arccos}) and equation (\ref{eq:gamma}). To extract the background value, Fourier spectrum should be calculated, it's central lobe should be selected, removing the side lobes, and then the reverse Fourier transform should be carried out (see Fig. \ref{img:int and spec}). As was mentioned before, laser intensity may be different in reference and probe shots, which can be due to the output energy instability, or done on purpose, e.g. at the single shot and long recuperation time laser facilities the reference shot energy is lower. To account for this, the coefficient $p$, used in equation (\ref{eq:gamma}), may be calculated using the experimental energy measurements. However, this approach will not work if some filters were put in a beam's way for a probe shot to avoid camera saturation or if the camera parameters were changed. A more convenient and accurate way is to calculate the ratio between probe and reference interferogram's mean brightness in plasma-free region, which gives the needed parameter $p$.
	
\begin{figure*}
		\begin{center}
			\includegraphics[width=0.32\linewidth]{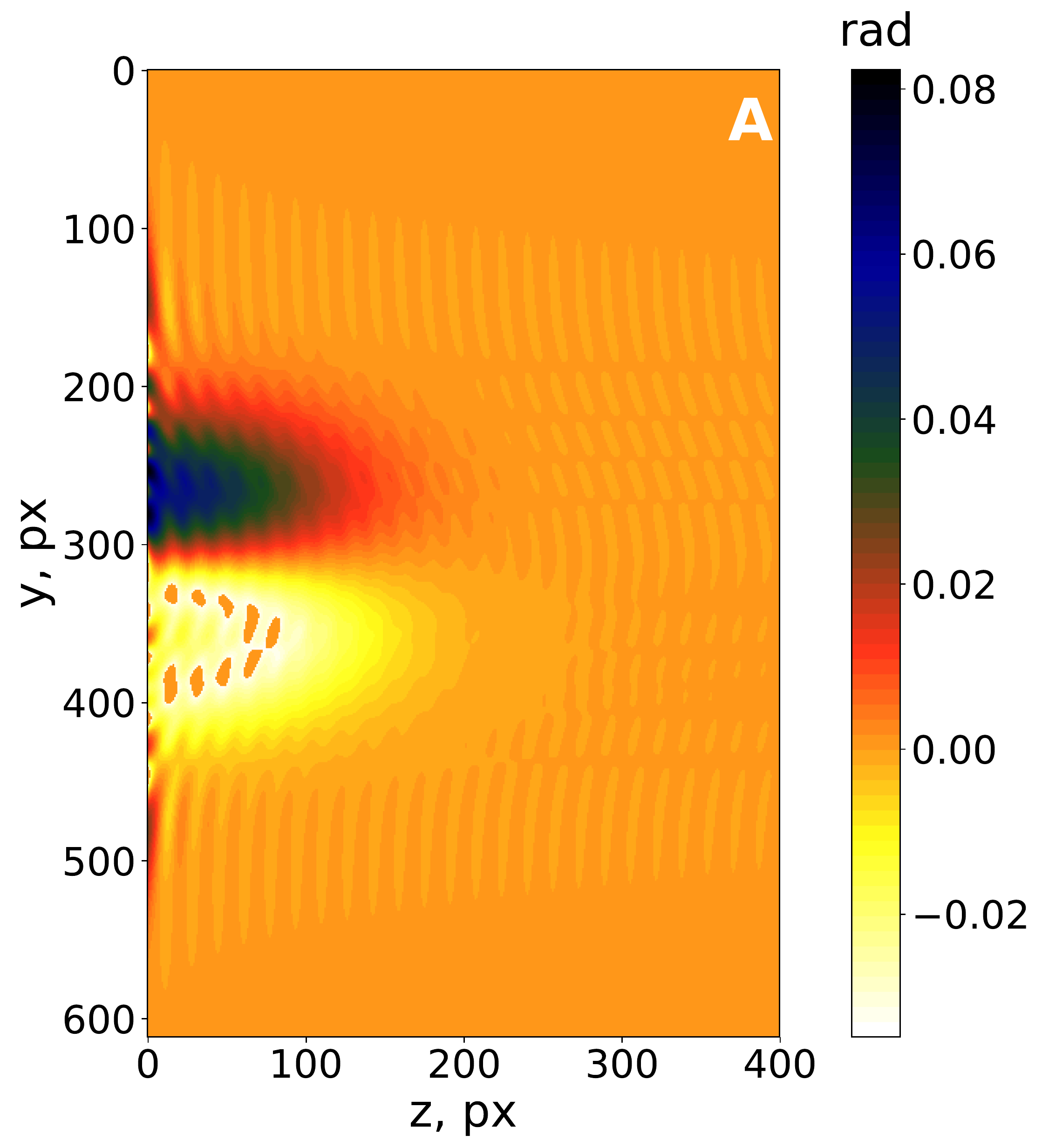}
			\hfill 
			\includegraphics[width=0.32\linewidth]{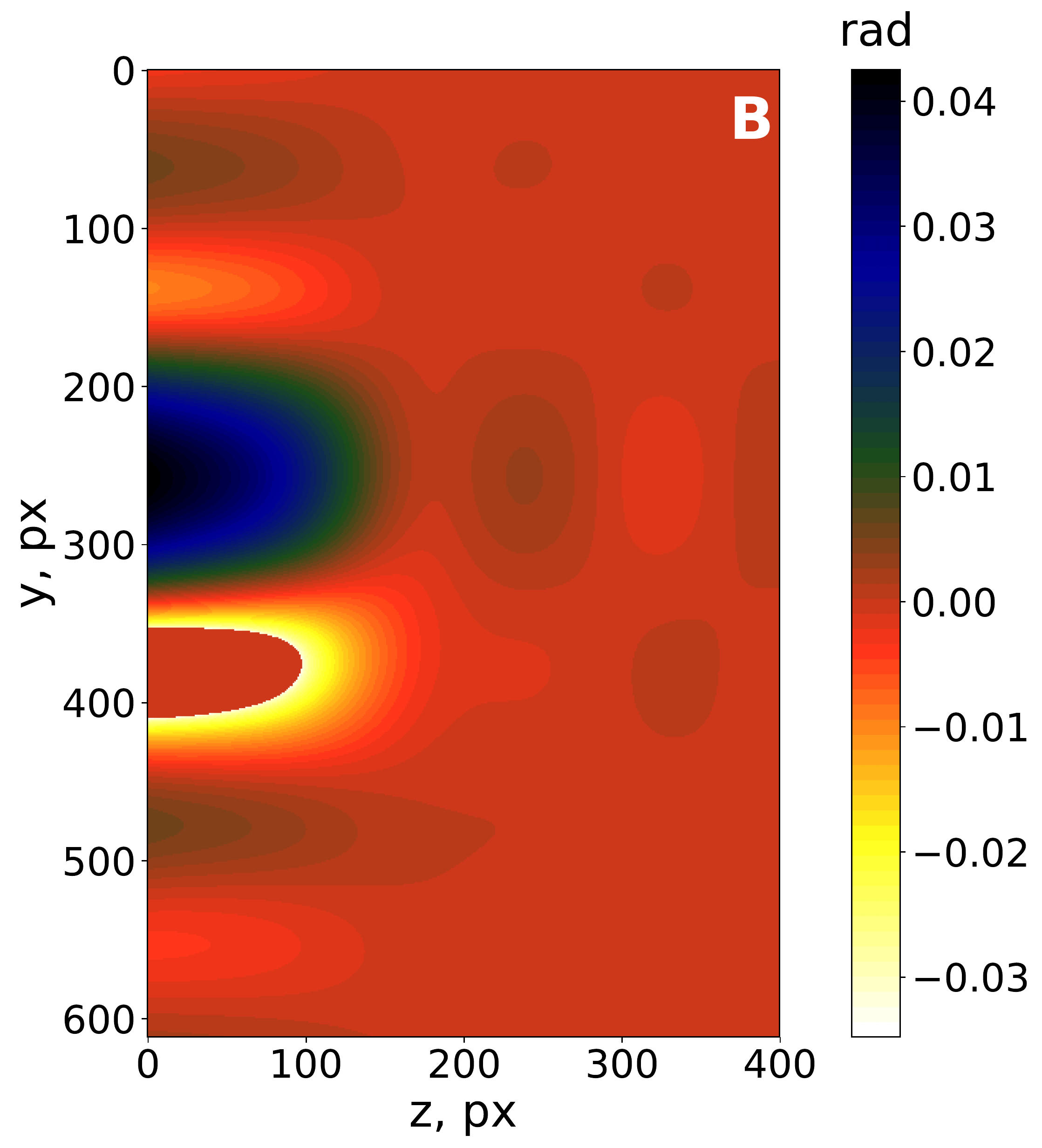}
			\hfill 
			\includegraphics[width=0.32\linewidth]{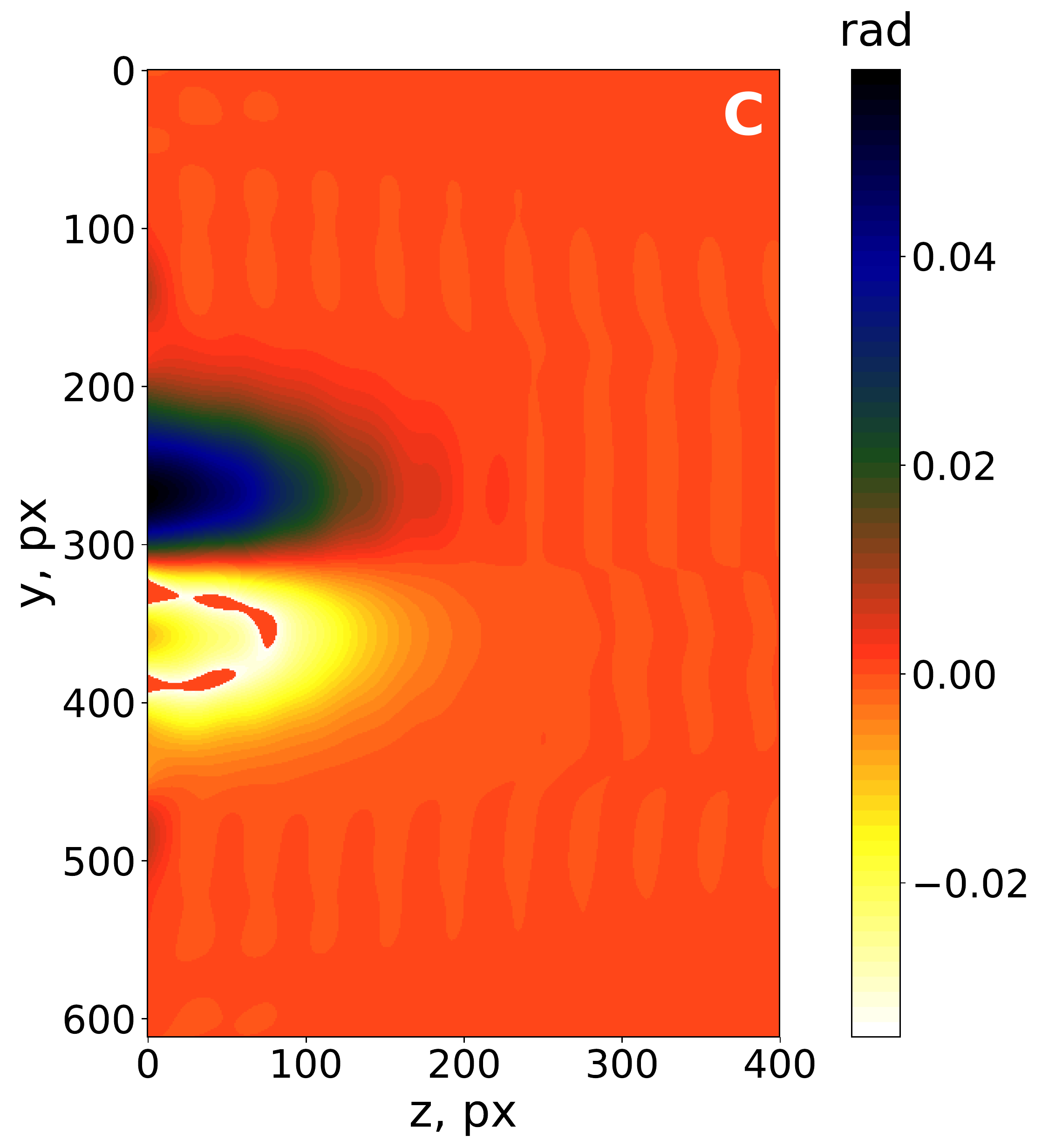}		
			\caption{Extracted angle using filtering with 100x100 pixels window (A), using filtering with 10x10 pixels window (B), using filtering with 30x40 pixels window (C).}
			\label{img:filter}
		\end{center}
	\end{figure*}

\begin{figure}[H]
		\begin{center}
			\includegraphics[width=0.99\linewidth]{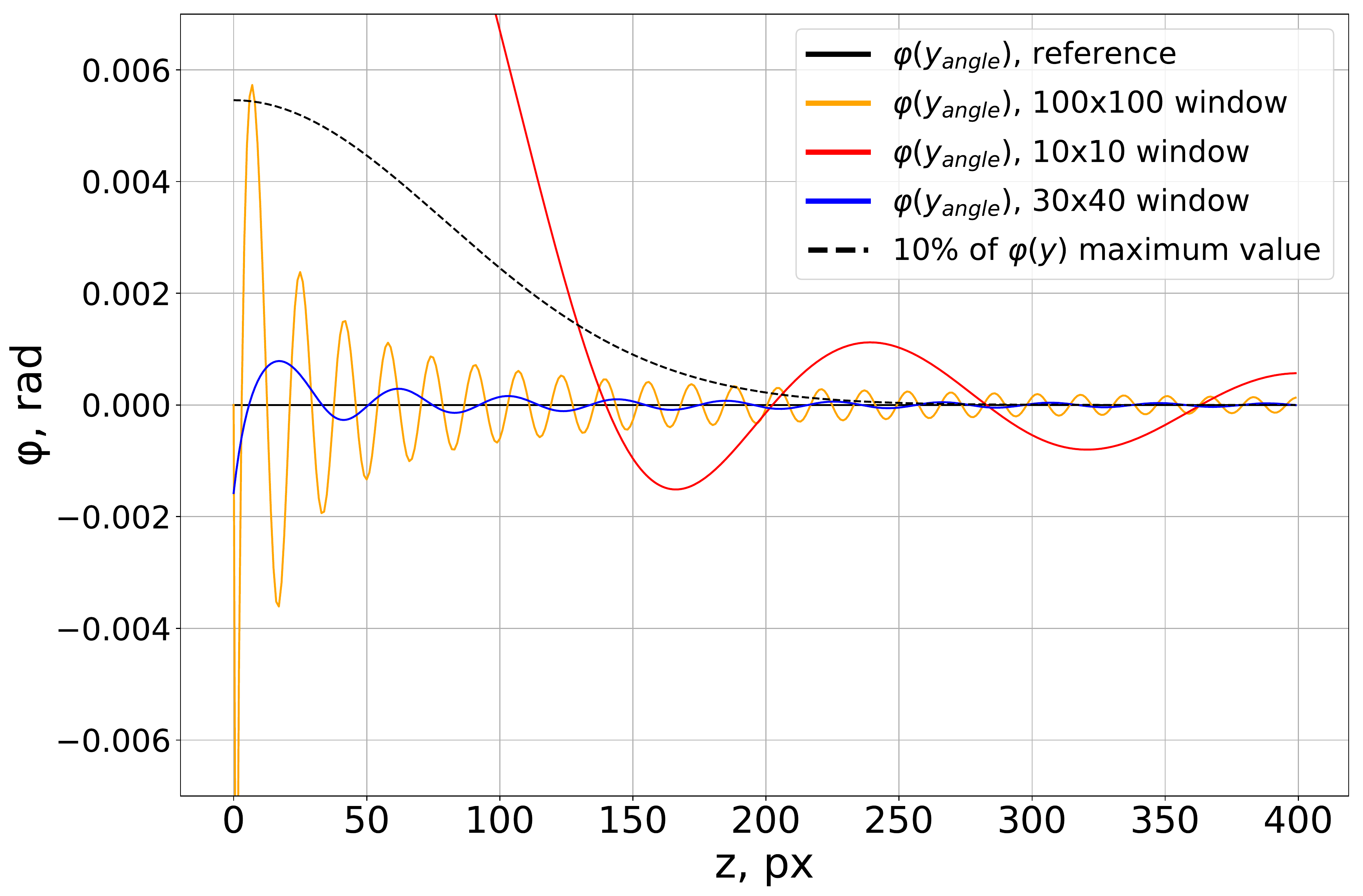}
			\caption{Calculated angle value at the anti-symmetry axis for simulated 1024x1024 interferogram (solid lines), maximum angle value at the corresponding cross-section (dashed line).}
			\label{img:oscill}
		\end{center}
	\end{figure}

	One problem, that can arise during processing is that in case of the rotation angle reaching values comparable to the set uncrossing between analyzer axis and the non-rotated wave (which is supposed to be small enough for good sensitivity, but big enough for unequivocal rotation measurement) the extracted coefficient $\gamma$ may be calculated correctly only in the brighter area.
	This is a significant diagnostic system limitation when studying non-symmetric distributions, however in case of systems with axial symmetry this problem is inconsequential. 
	An important point is a filtering window size for the accurate background extraction. Too big window size will lead to the presence of a residual fringe pattern while too small will distort the background distribution leading to loss of details (see example for synthetic data filtered using different windows at Fig. \ref{img:filter}). The optimal window size should be selected individually for every diagnostic system, as it depends primarily on the diagnostic system resolution, magnification and fringe frequency. This can be done by creating synthetic data using  experimental parameters and minimizing the deviation of the extracted background from the reference one by adjusting the window size.  
	During plasma density analysis the plasma symmetry axis $y_{plasma}$ was defined. This axis is our best assumption of $y_{real}$ (unknown in real experiment), and as we are working in assumption of axially symmetrical distributions it is supposed to be also the axis of the angle of polarization plane rotation. But it was found out that the angle distribution, calculated using Fourier filtering for background extraction, contains parasitic oscillations with the period depending on the window size. Studying the extracted angle value at the reference axis $y_{real}$, chosen when creating synthetic interferograms, we can see these oscillations. Their amplitude is gradually decreasing with the decrease of maximum angle v alue in the corresponding cross-section $z$ (see example for synthetic interferograms filtered using different windows in Fig. \ref{img:oscill}). The deviation of the measured angle value at the axis from ideal zero can exceed tens of percent of maximum angle value for incorrect filtering window size. In case of the correct one, it is about 2 percent, while the measured angle zero point position deviation from a reference one is about 1 px. This purely mathematical effect is small compared to various physical effects, but also can lead to complications of the SMF calculation. 
	
	In the experiment there are a lot of essential effects for laser plasma optical diagnostic  that influence measurements: plasma self-luminescence, Cotton-Mouton effect, unstable laser intensity distribution, laser beam depolarization etc.
	Their contribution to the experimental data (see Appendix) leads to highly unstable angle value at the chosen plasma symmetry axis $y_{plasma}$ and oscillations with a much higher amplitude than caused by Fourier filtering. Measured angle zero point can be shifted by up to 10 pixels from the plasma axis. We will hereafter call the $(\varphi=0)(z)$ curve as $y_{angle}$. The angle value at plasma-free region also doesn't always drop down to zero, which can help us to estimate the errors level. 
	Unfortunately this noise can't be completely removed but special measures may be taken while conducting the experiment to reduce it (e.g. measurement of additional information - intensity distributions in the reference and probe beams \cite{Kalal_2016}).

	\subsection{Magnetic field reconstruction}\label{sec:reconstruction}
	
	\begin{figure*}
		\begin{center}
		\includegraphics[width=0.45\linewidth]{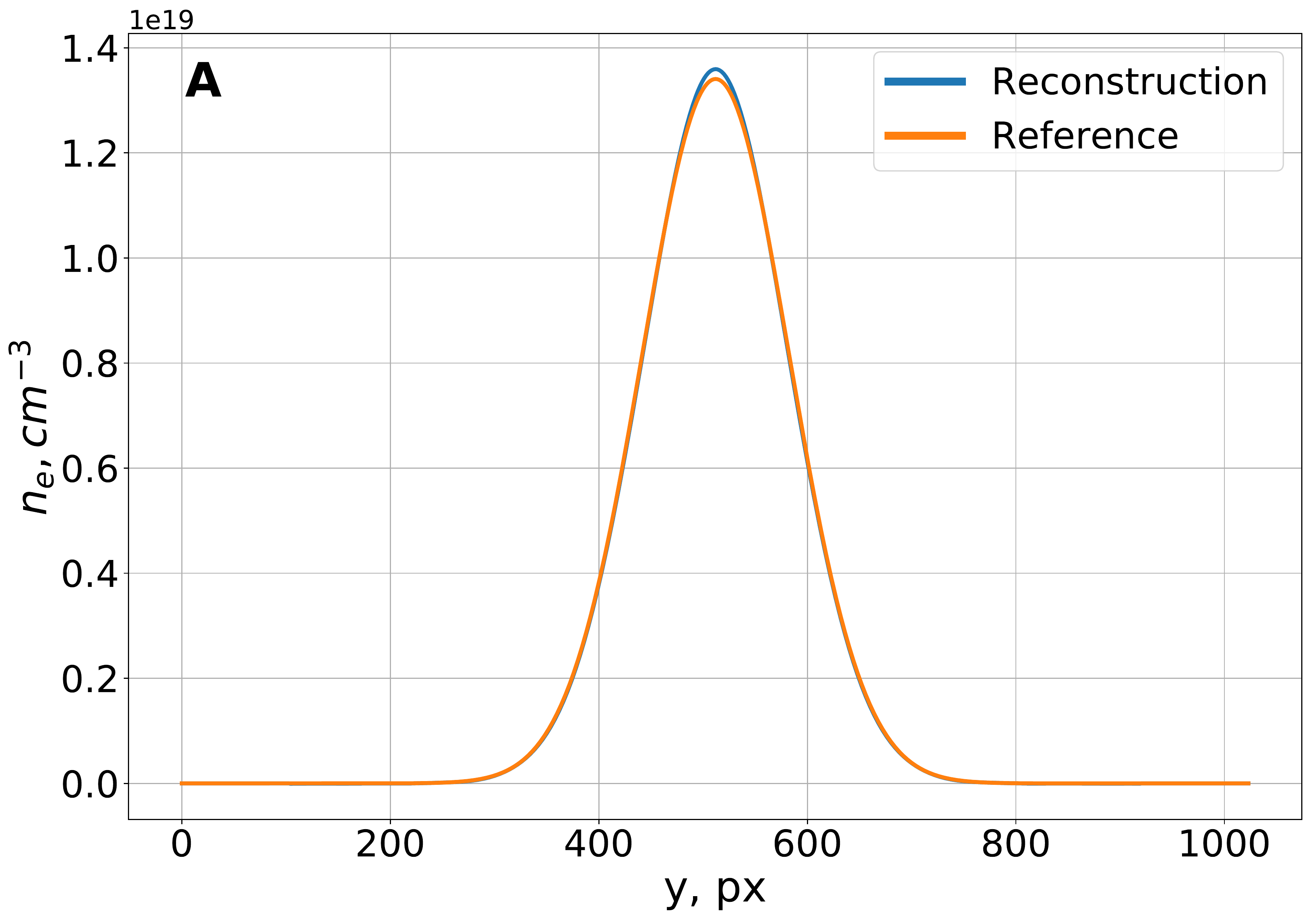}\hfil
		\includegraphics[width=0.45\linewidth]{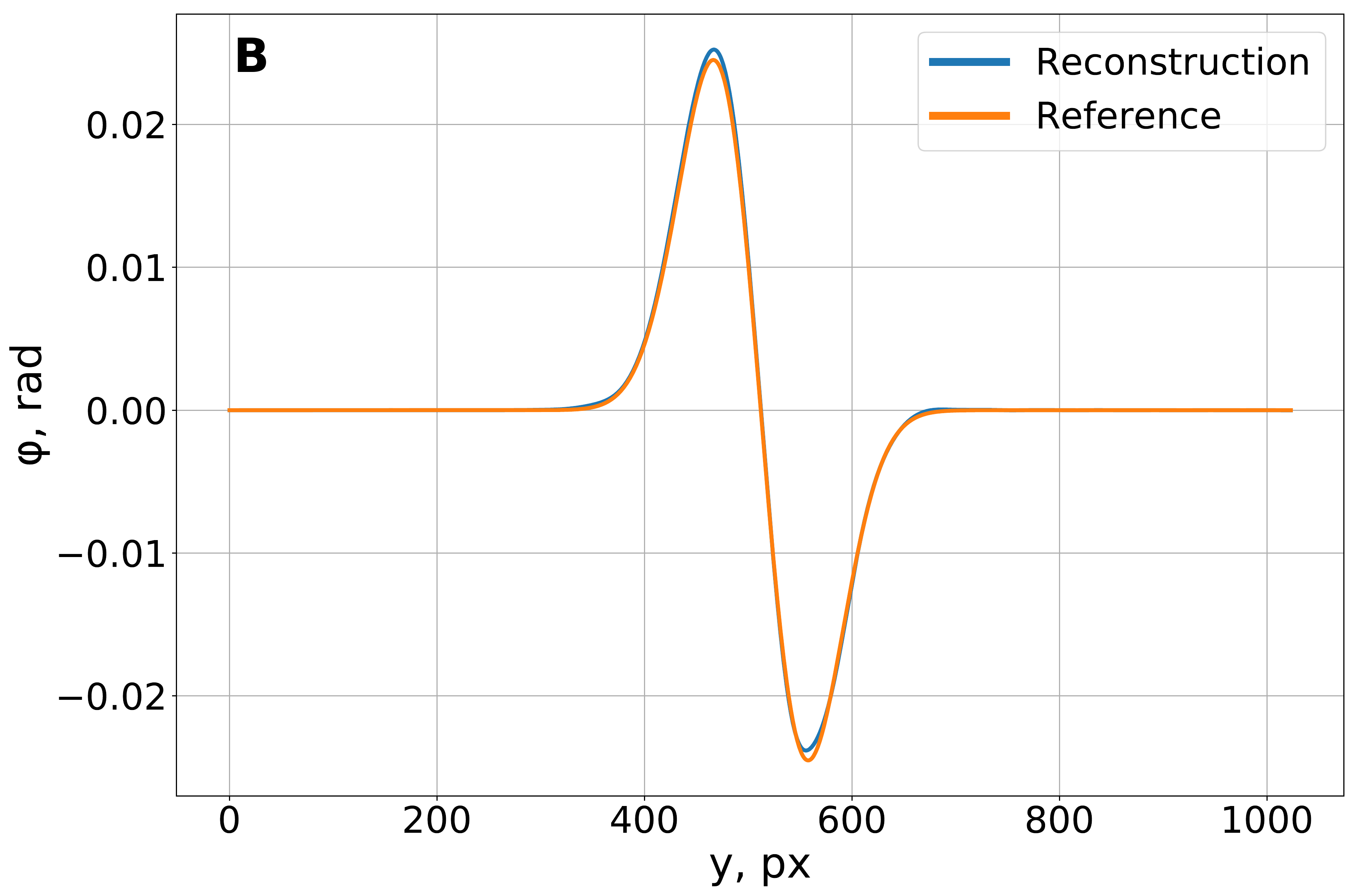}\\[2mm]
		\includegraphics[width=0.45\linewidth]{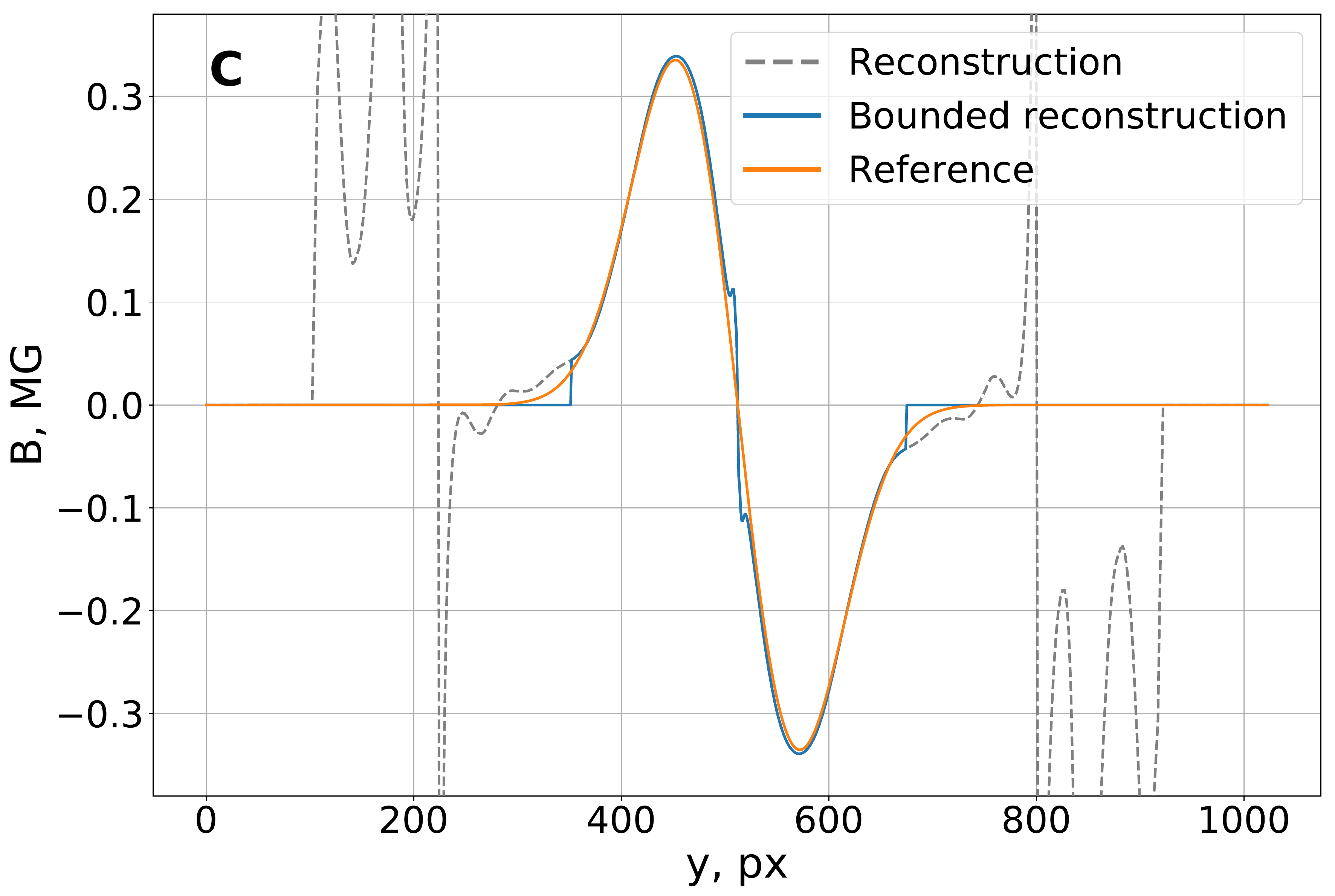} \hfil
		\includegraphics[width=0.45\linewidth]{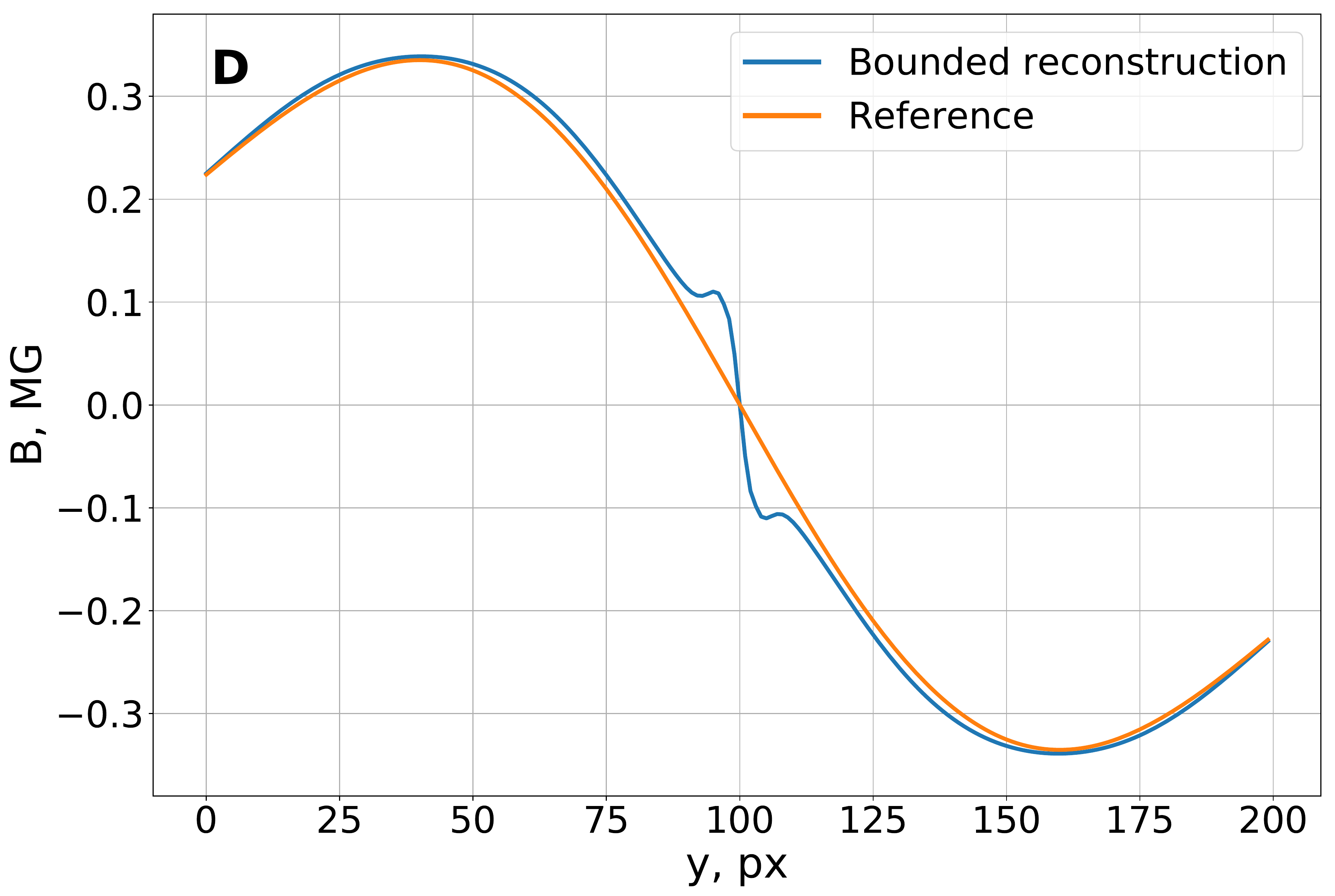}

		\caption{Cross-sections retrieved from simulated complex interferogram without artificial uncertainties  -- density (A), polarization plane rotation angle for 30x40 filtering window (B), calculated SMF (C), calculated SMF near the axis area (D).}
		\label{img:no noise}
		\end{center}
	\end{figure*}	
	For SMF calculation the retrieved $\varphi(y,z)$ and $\delta(y,z)$ distributions should be symmetrized, corresponding $F_{\varphi}$ and $F_{\delta}$ functions and their inverse Abel transformation - calculated and then substituted into equation (\ref{eq:B(r)}). The major problem arises at the symmetrization step due the previously noted difference between plasma symmetry axis $y_{plasma}$ and the angle anti-symmetry curve $y_{angle}$, which should coincide in order for us to use described mathematical approach. To understand the possible error we made a number of calculations with simulated data: with and without added error sources $\mathbf{V_{int}}$, $\mathbf{V_{noise}}$, $\mathbf{G_{angle}}$ and $\mathbf{G_{shift}}$. For the beginning we used the simplest case of simulated data, where plasma density is given only by one Gaussian profile, and polarization plane rotation angle - by a combination of two Gaussians (see Fig. \ref{img:Simulated_Interferograms}).
	\begin{figure*}
		\begin{center}
			\includegraphics[width=0.28\linewidth]{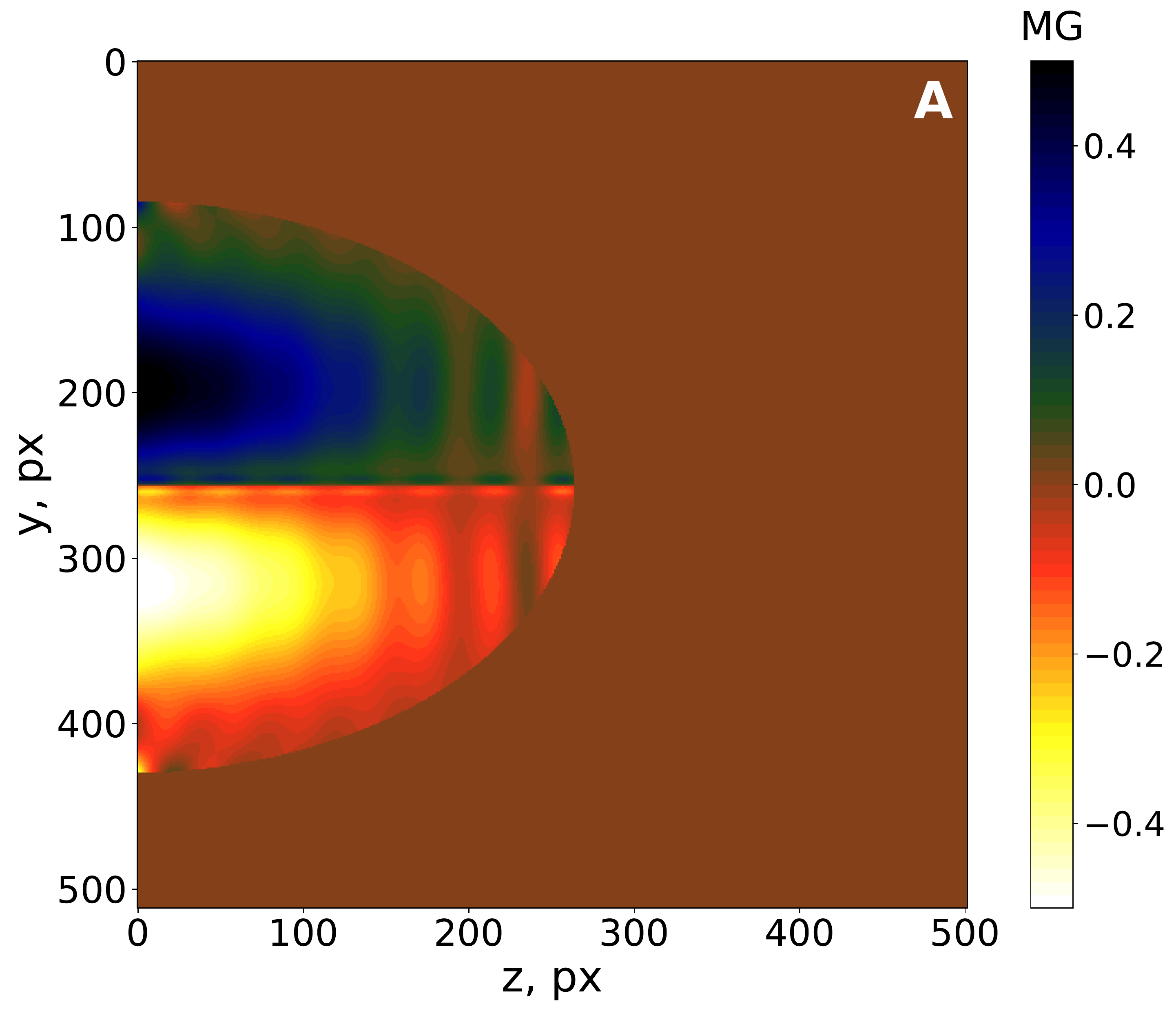}
			\includegraphics[width=0.34\linewidth]{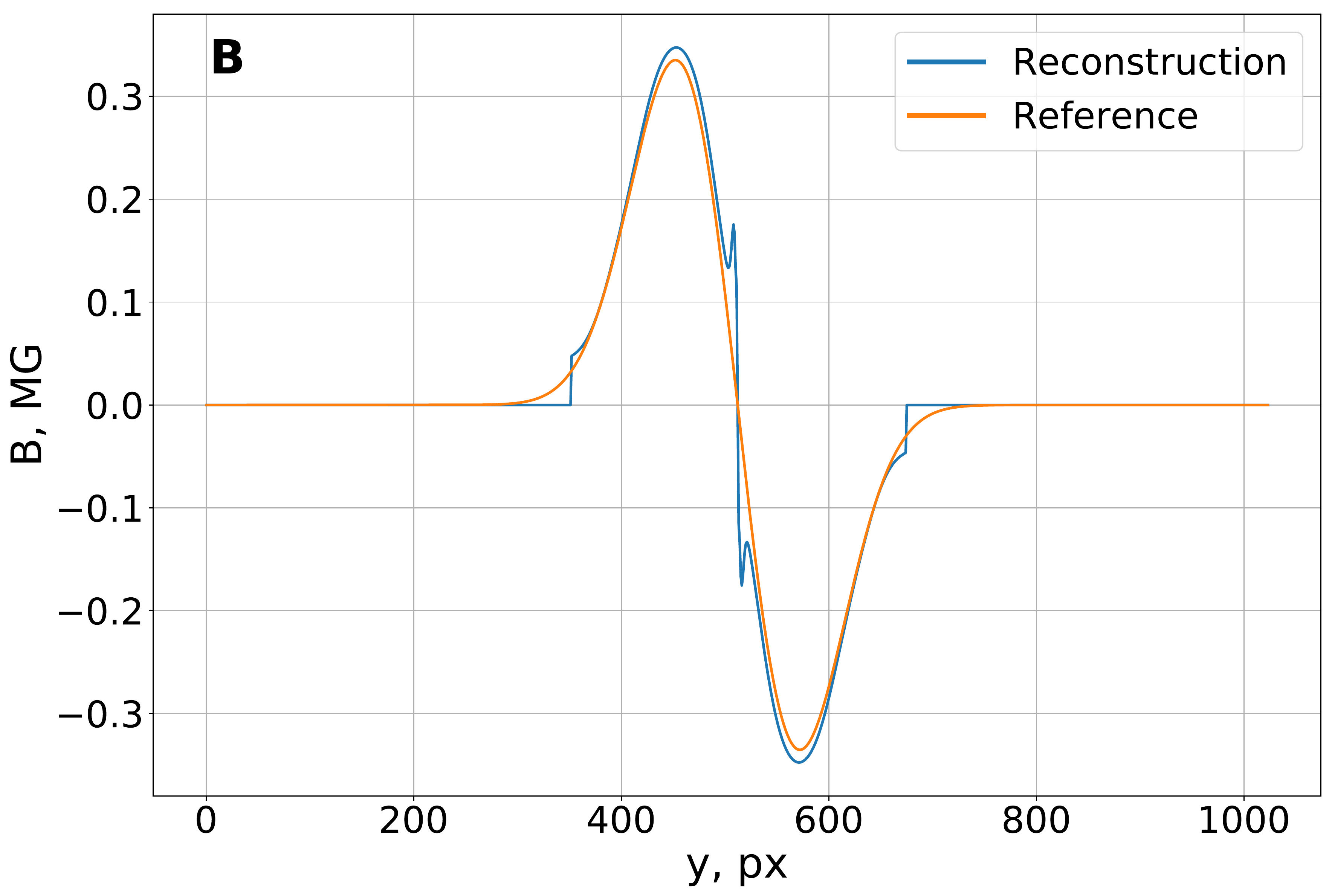}
			\includegraphics[width=0.34\linewidth]{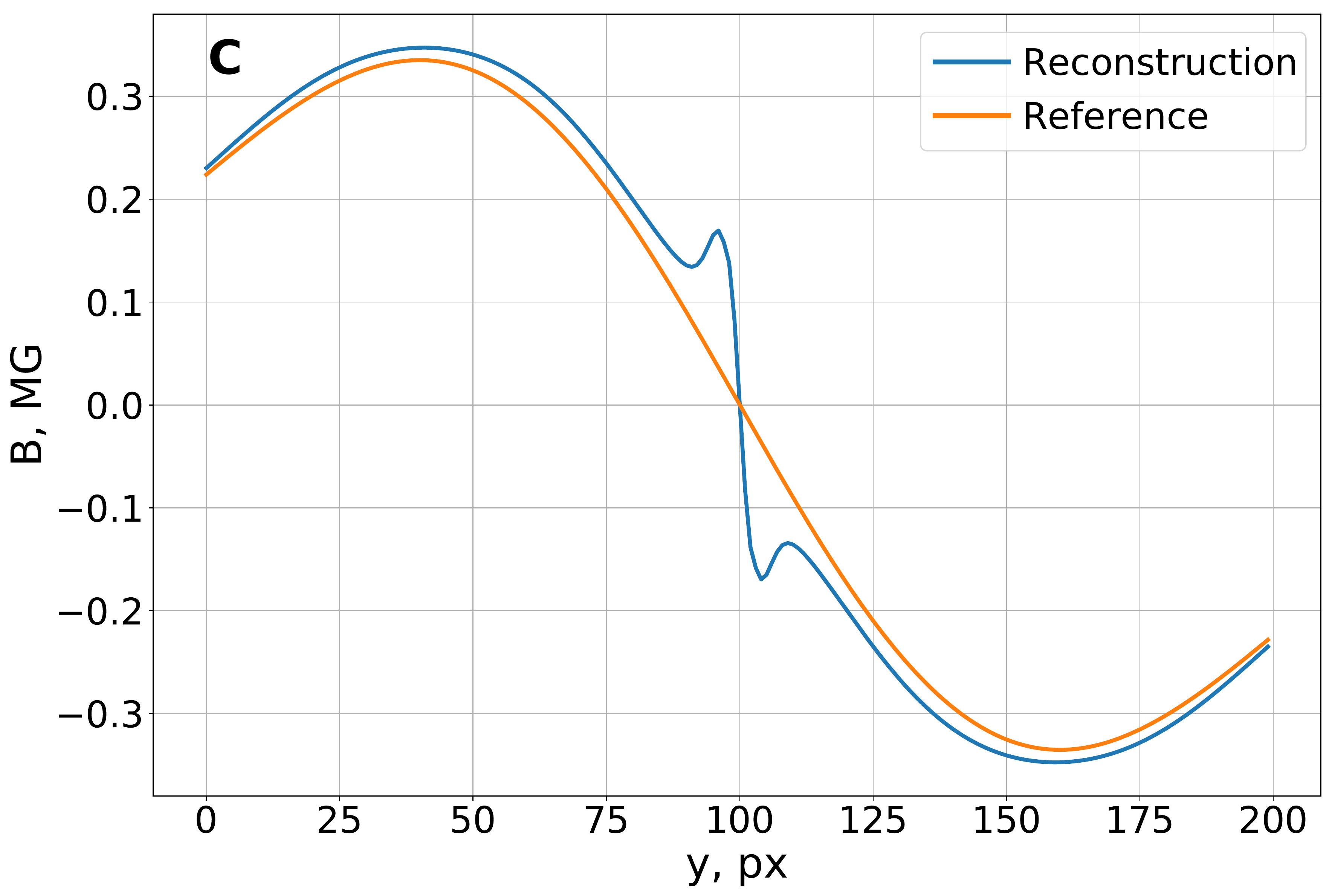}
			\caption{SMF reconstruction in case of 2\% $\mathbf{V_{int}}$ error. Calculated field distribution (A), one slice of the field distribution(B) at $z$=100 px, the same zoomed in (C).}
			\label{img:2 percent big error}
		\end{center}
	\end{figure*}

\begin{figure*}
		\begin{center}
			\includegraphics[width=0.45\linewidth]{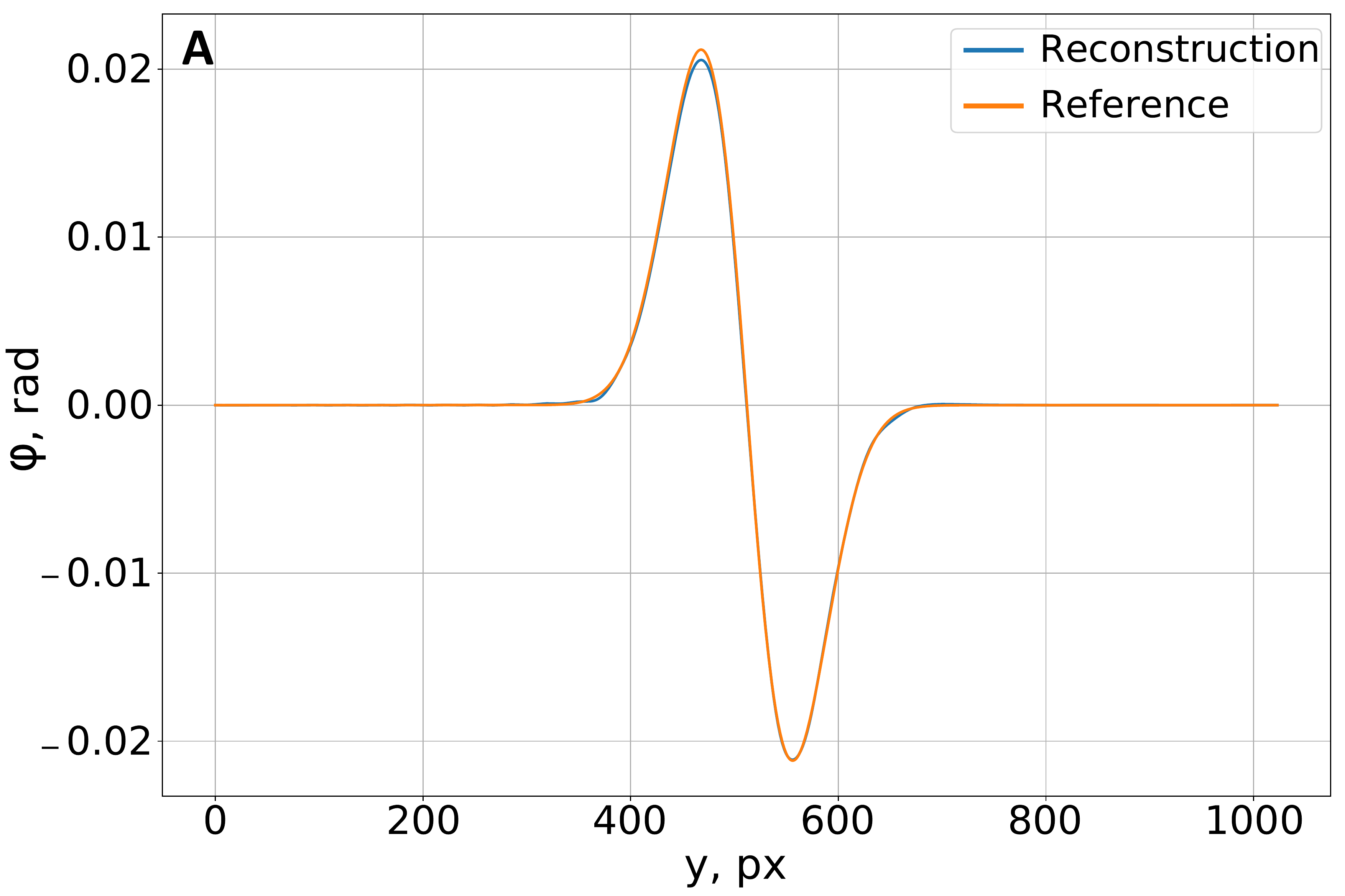}\hfil
			\includegraphics[width=0.45\linewidth]{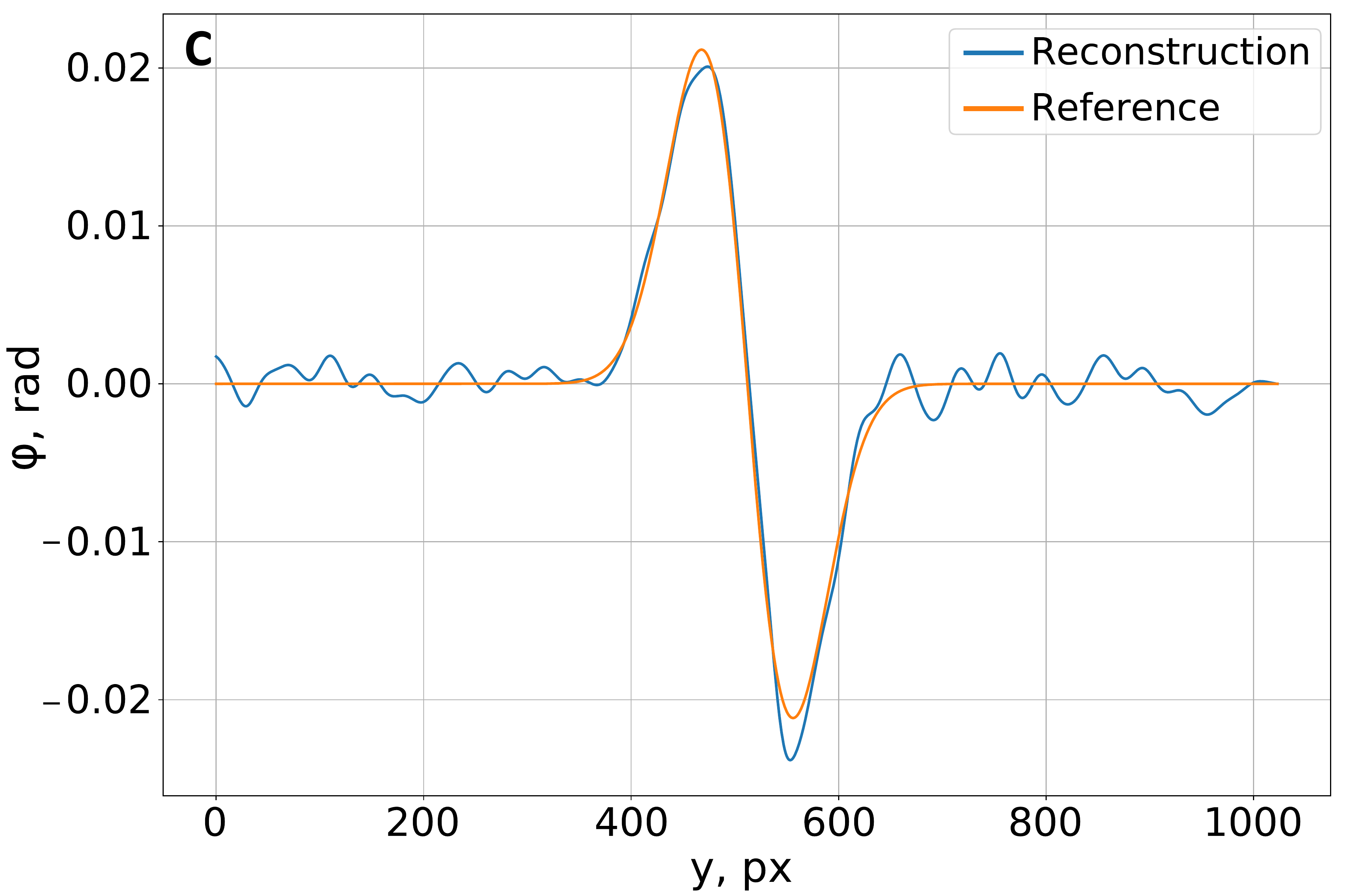}
			\\
			\includegraphics[width=0.45\linewidth]{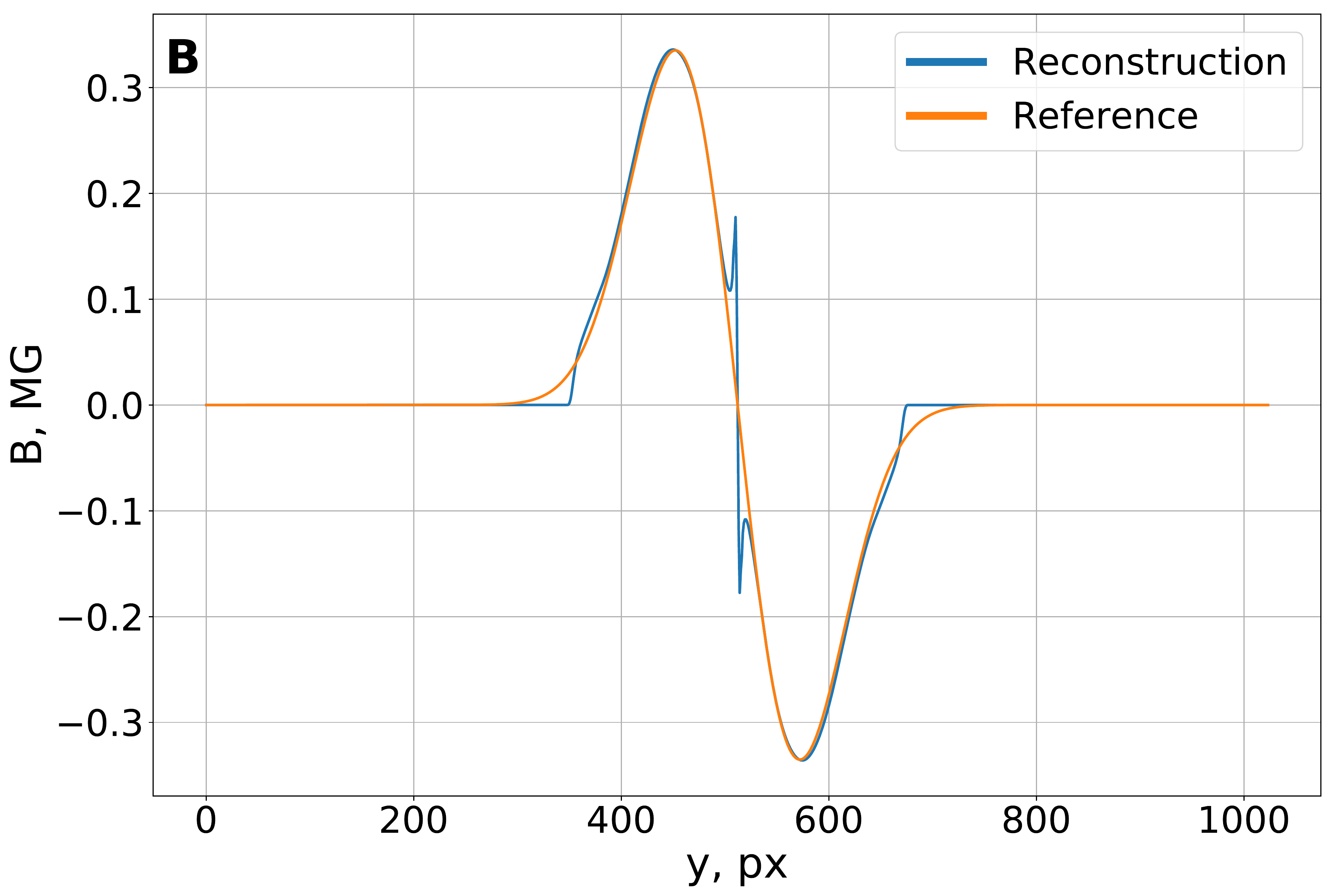} \hfil 
			\includegraphics[width=0.45\linewidth]{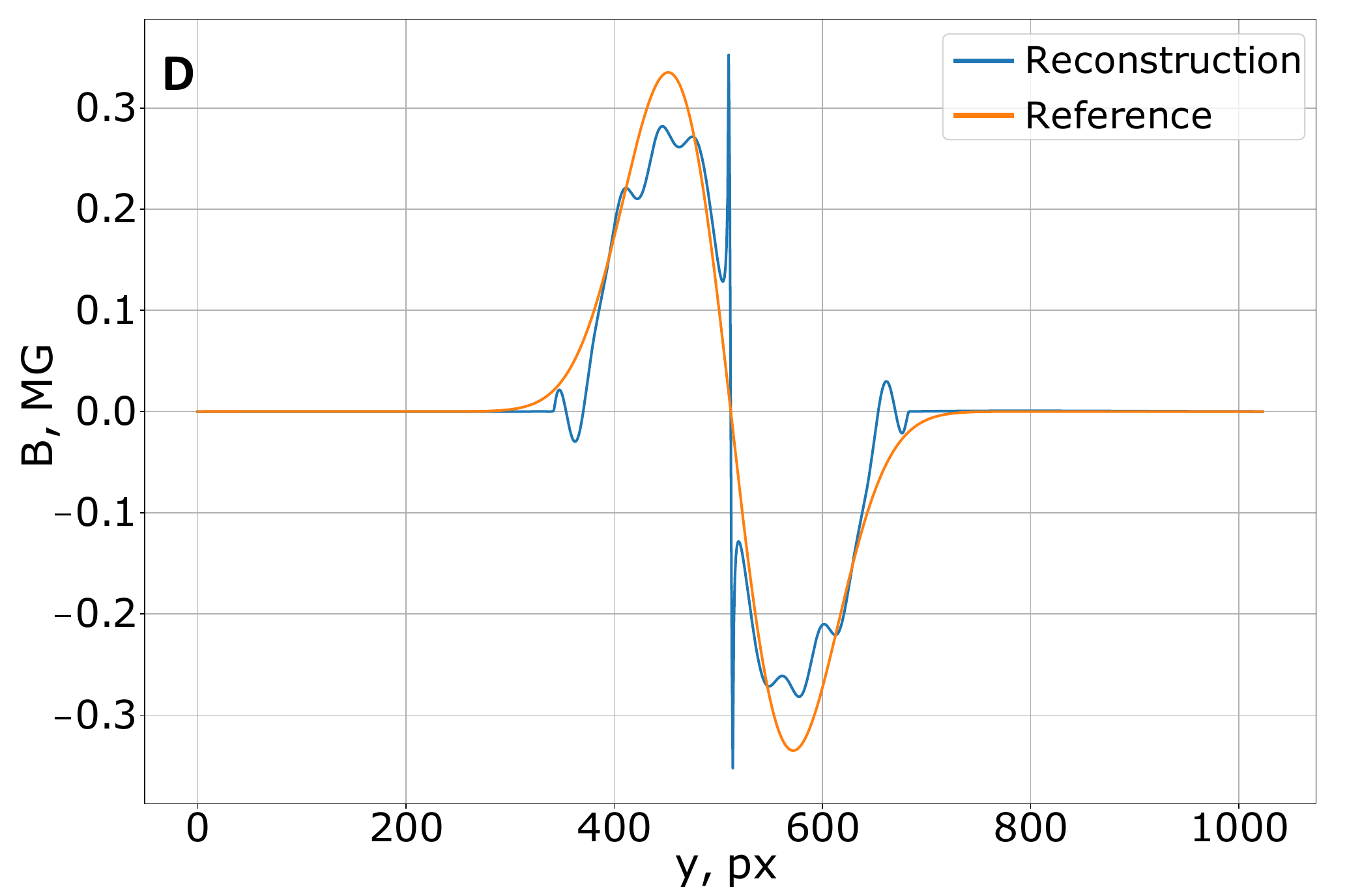}
			\caption{Extracted angle (A) and calculated field (B) in case of identical noise added to interferograms, extracted angle (C) and calculated field (D) in case of different noise added to interferograms.}
			\label{img:noise}
		\end{center}
	\end{figure*}

\begin{figure*}
	\begin{center}
		\includegraphics[width=0.9\linewidth]{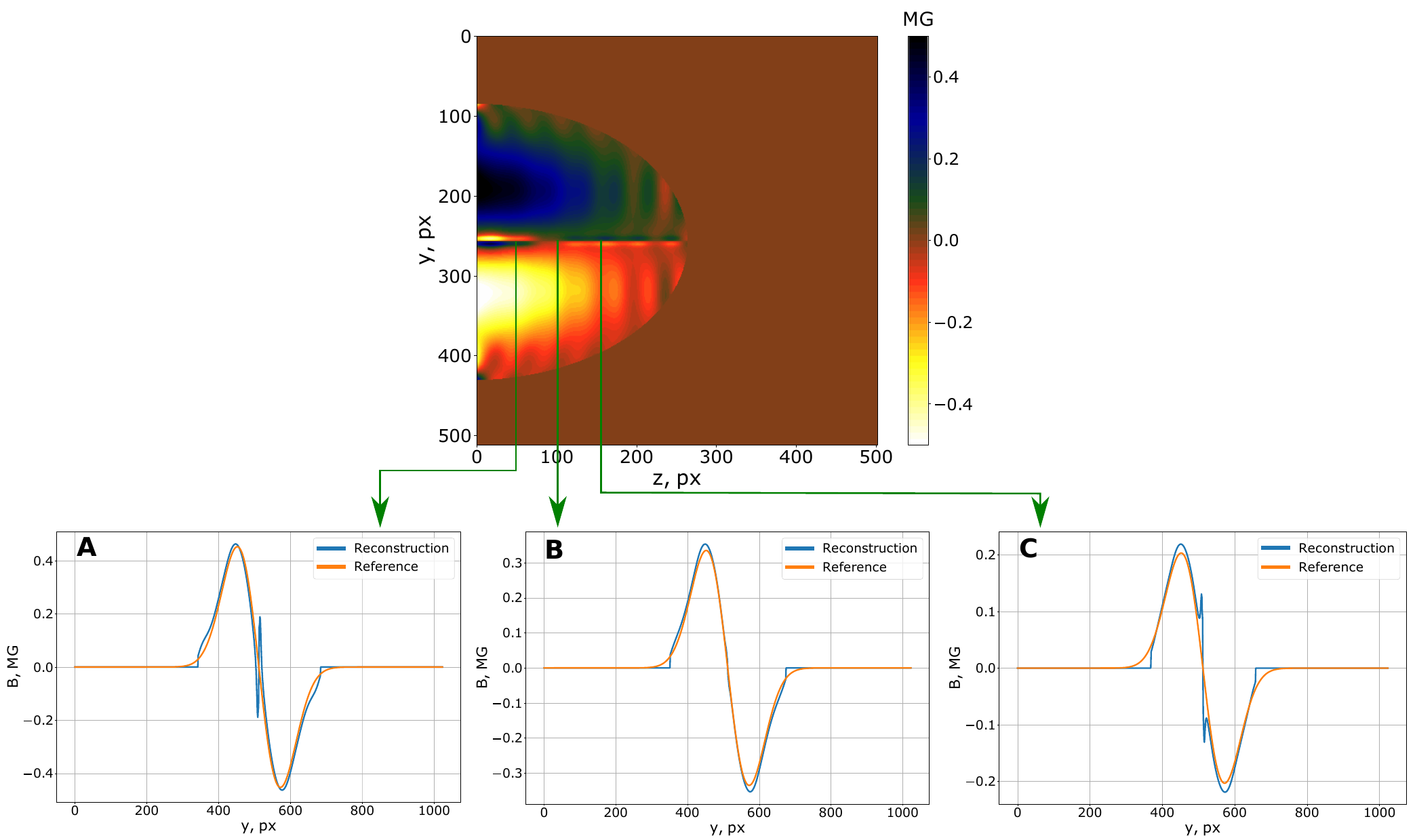}	
		\caption{SMF reconstruction in case of $\mathbf{G_{angle}}$ error. Field distribution(top), slices of the field distribution for $z$=50 px (A), $z$=100 px (B) and for $z$=150 px (C).}
		\label{img:rotat}
	\end{center}
\end{figure*}

	In case of simulated data without any artificial uncertainties, density and rotation angle reconstructions seem to have a very good accuracy (Fig. \ref{img:no noise}A,B), therefore radius R was maximized and placed near the edge of interferogram. Despite that, the field reconstruction yields two erroneous areas - far from the symmetry axis and near it (see Fig. \ref{img:no noise}C). The first one is caused by imperfect angle and density reconstructions, which have small absolute error but considerable relative error in the investigated area. This error can not be fixed, therefore the field distribution has to be cut, i.e. the additional boundary condition has to be set. Throughout this paper, this boundary was chosen at $n_e=10^{18} cm^{-3}$ level and referred to as $R_{mask}$.	A small bump near the axis (Fig. \ref{img:no noise}D) is cased by the previously described angle axis $y_{angle}$ oscillation around $y_{plasma}$ caused by Fourier filtering. The filtering window of 30x40 pixels was used, and as can be seen from Fig. \ref{img:oscill} the angle deviation there at the considered coordinate $z$=100 px is small, however it still leads to noticeable deviations in calculated field value, proving the necessity of further analysis.
	Firstly, to study the $\mathbf{G_{angle}}$, $\mathbf{V_{int}}$, $\mathbf{V_{noise}}$ errors influence on field reconstruction we created three sets of synthetic data: (i) using rotated by 2$^o$ distributions, (ii) adding 2$\%$ of background to a shot interferogram and (iii) adding a noise to the spectrum of synthetic images. For the last point two cases were analyzed - the same noise distribution added to both shot and reference interferograms and addition of individually generated noise to each interferogram. The analysis showed that in every case except of identical noise addition calculated SMF distribution has big error at the axis (see Fig. \ref{img:2 percent big error}). The error in case of identical noise addition is close to that of the noise-free case. This was expected as the reference image is taken particularly to remove errors, that repeat from shot to shot (due to backgrounds \textit{ratio} present in formula \ref{eq:gamma}), and can not neutralize any other errors (see Fig. \ref{img:noise}A,B). The error in case of  $\mathbf{G_{shift}}$, $\mathbf{V_{int}}$ and different noise distributions addition $\mathbf{V_{noise}}$ seems to be identical with a more "wavy" structure for $\mathbf{V_{noise}}$ case (see Fig. \ref{img:noise}C,D), and tends to increase with the increase of the axes offset, background value, noise intensity. 	In the presence of $\mathbf{G_{angle}}$ the error behavior changes: there is only one cross-section with correct SMF reconstruction, and the bump in the SMF reconstruction near the axis has an opposite sign at the opposite sides of this cross-section (Fig. \ref{img:rotat}).
		

\begin{figure*}
	\begin{center}
		\begin{subfigure}[b]{0.99\linewidth}
			\centering
			\includegraphics[width=0.44\linewidth]{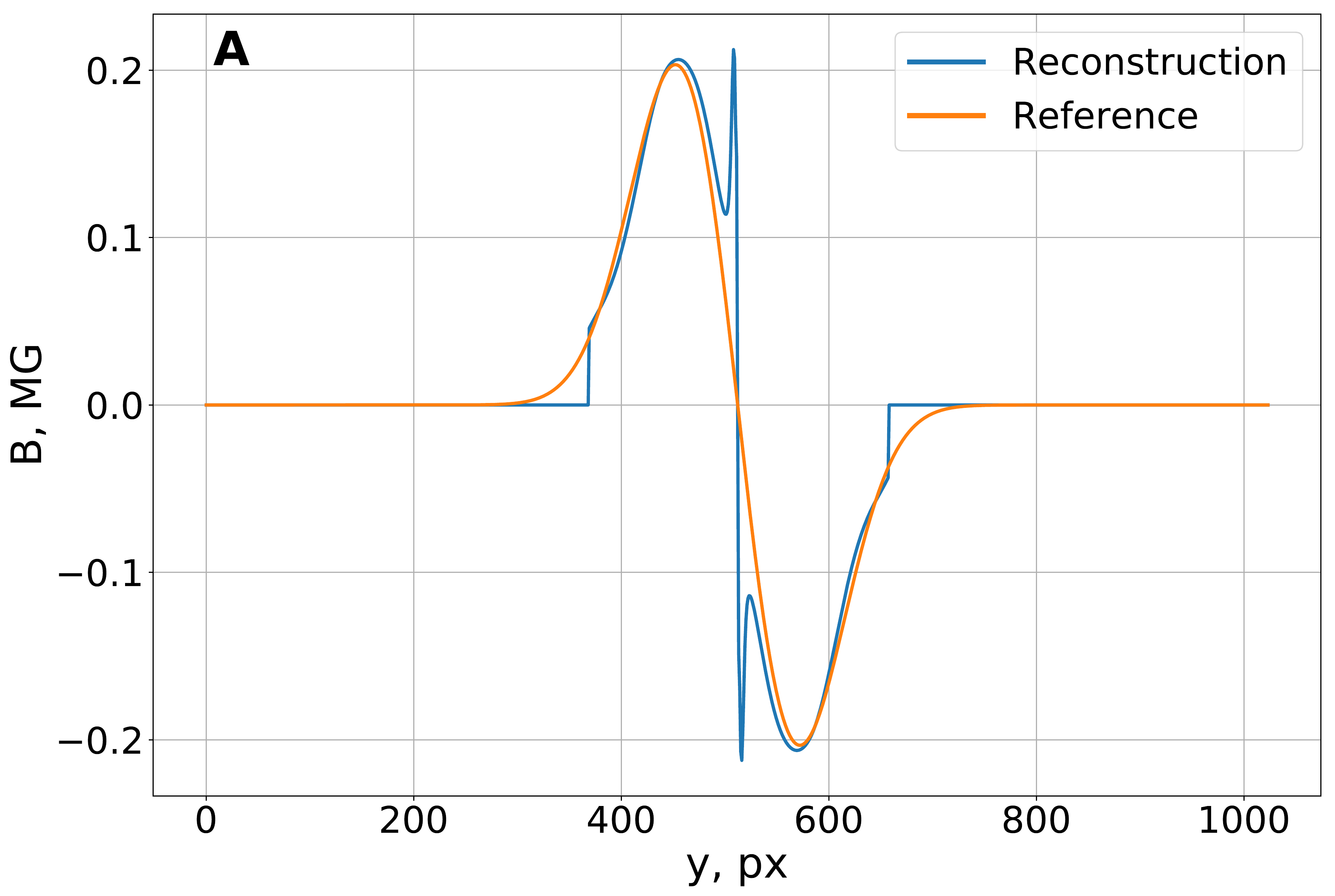} \hfil
			\includegraphics[width=0.44\linewidth]{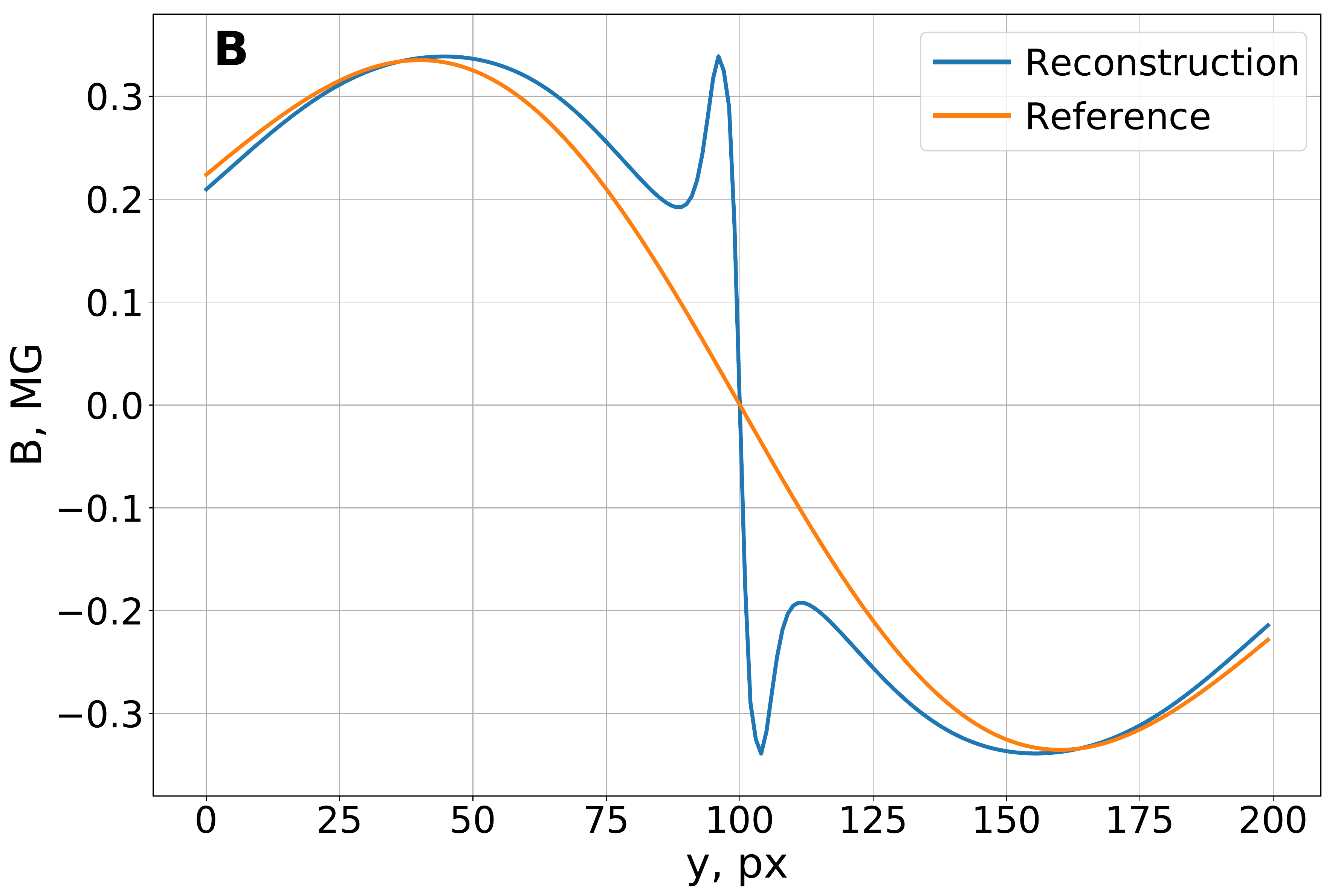}
		\end{subfigure}\\[1mm]
		\begin{subfigure}[b]{0.99\linewidth}
			\centering
			\includegraphics[width=0.44\linewidth]{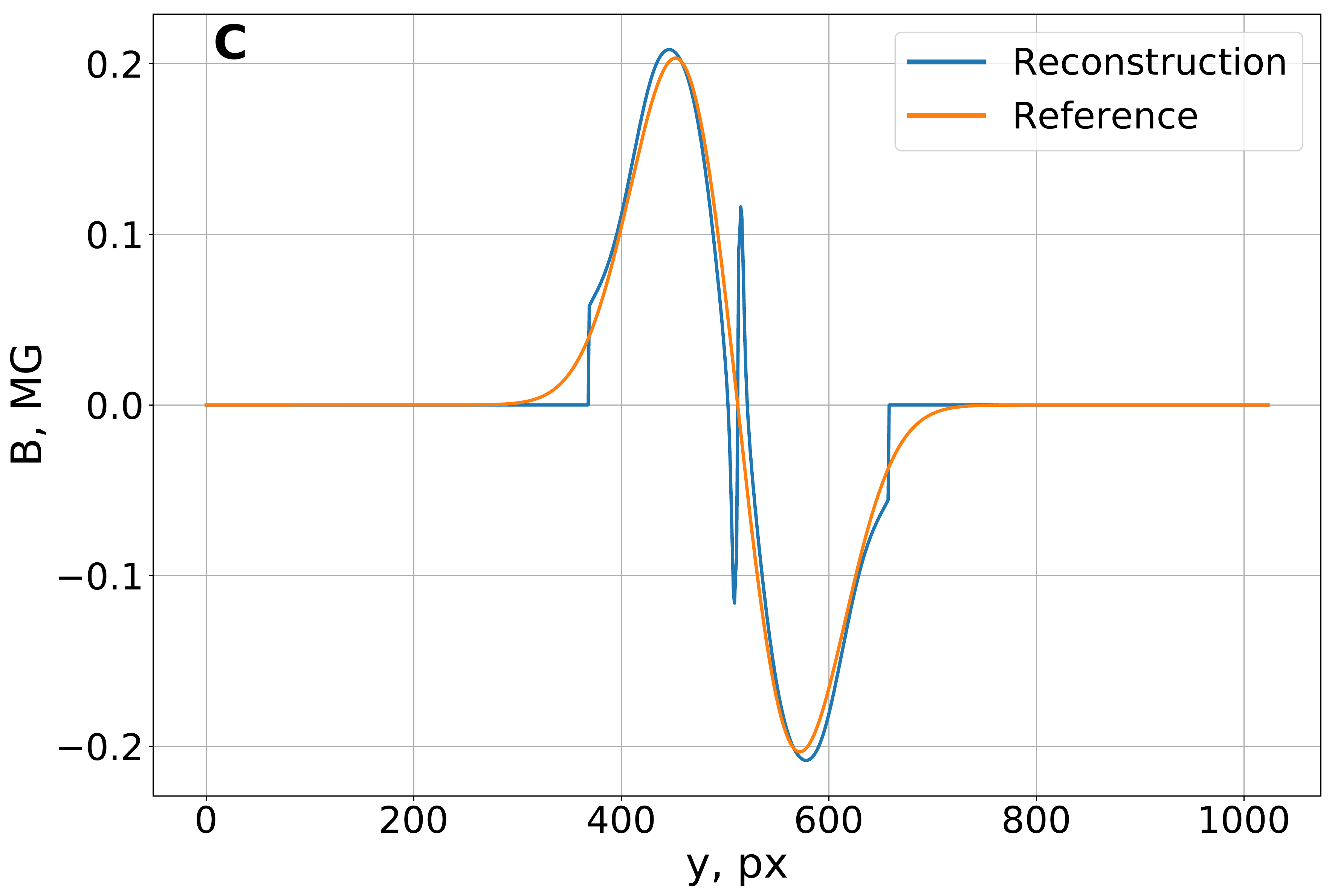}\hfil
			\includegraphics[width=0.44\linewidth]{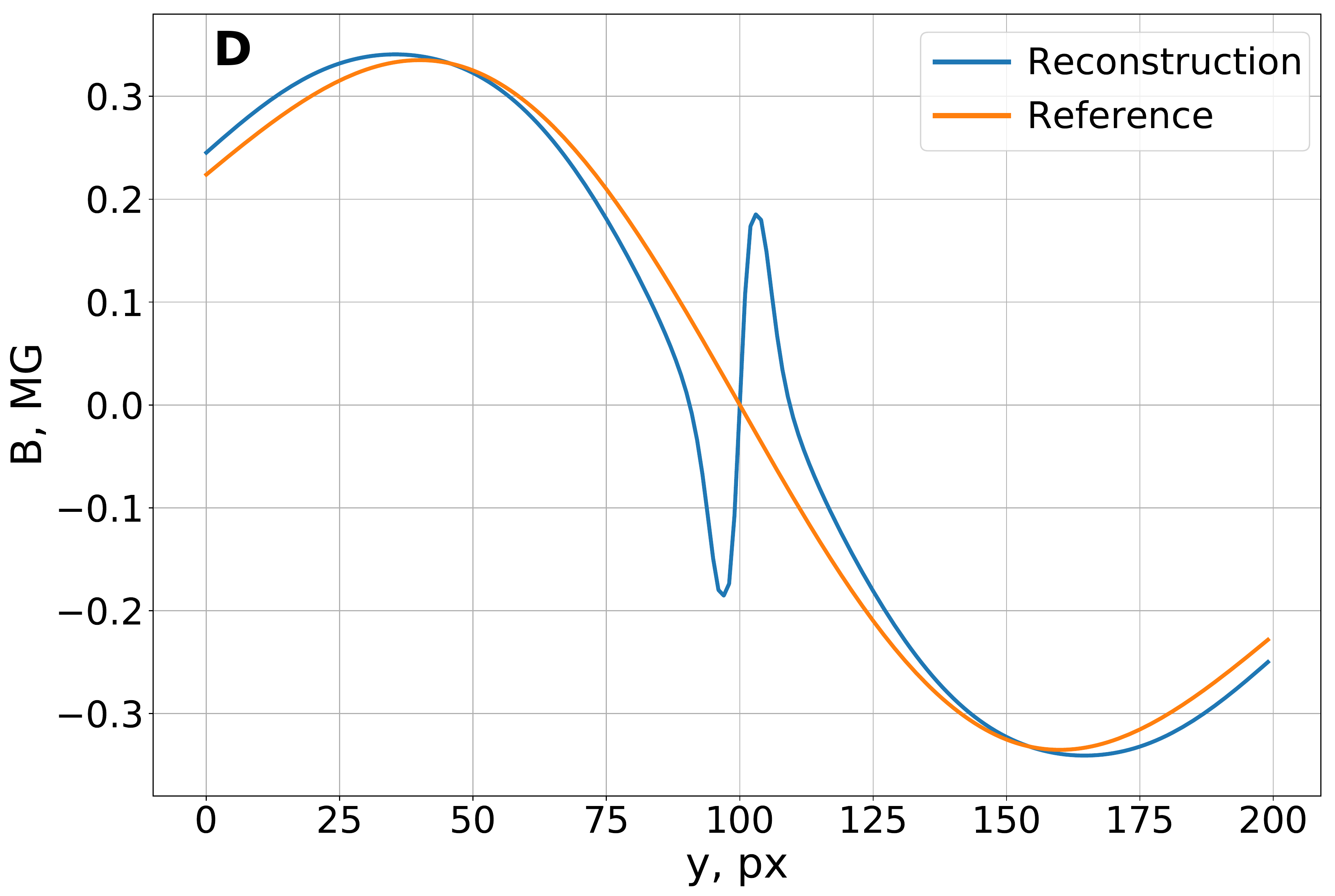}
		\end{subfigure}
		\caption{SMF reconstruction in case of 2 pixels axis offset (A, B) and in case of -2 pixels offset (C,D). }
		\label{img:2px error}
	\end{center}
\end{figure*}
		
	Analysis of these results shows, that all of the considered factors lead to one consequence - to the increase of $|y_{plasma}-y_{angle}|$ difference, while the theory behind the calculations considers the plasma and angle distributions to have a common axis $y_{real}$. Logically, axes inclination in respect to each other will lead to the same result with cross-section at the axes intersection point only yielding a good reconstruction, which was confirmed in the analysis. To check this assumption we also made calculations in the absence of artificial uncertainties, but defining the $y_{plasma}$  with an error of merely 2 pixels (corresponding to 7 $\mu$m in both our simulations and experiments) and received the same results, see Fig. \ref{img:2px error}.

\begin{figure}[H]
	\begin{center}
		\includegraphics[width=0.85\linewidth]{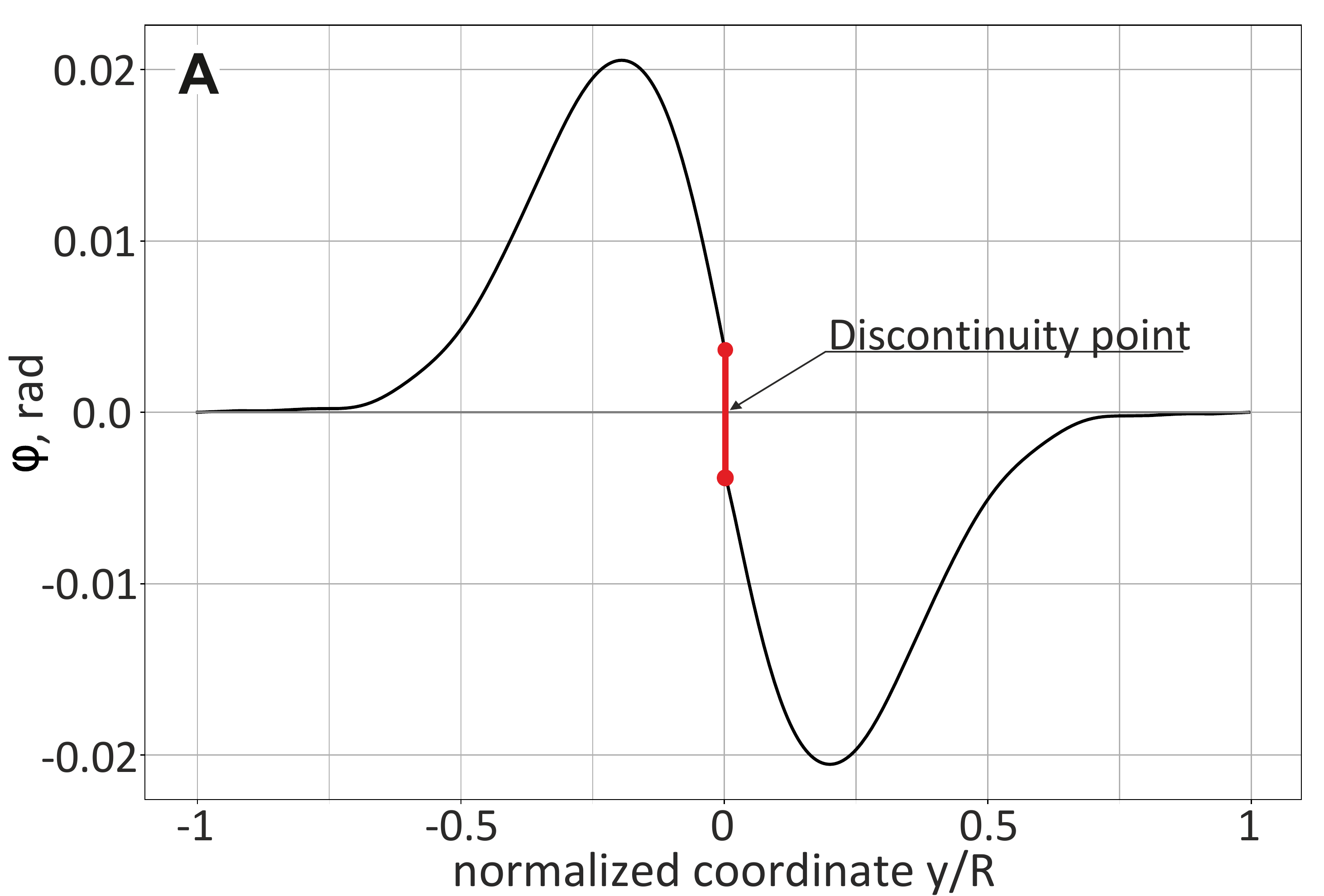}
		\\[2mm]
		\includegraphics[width=0.85\linewidth]{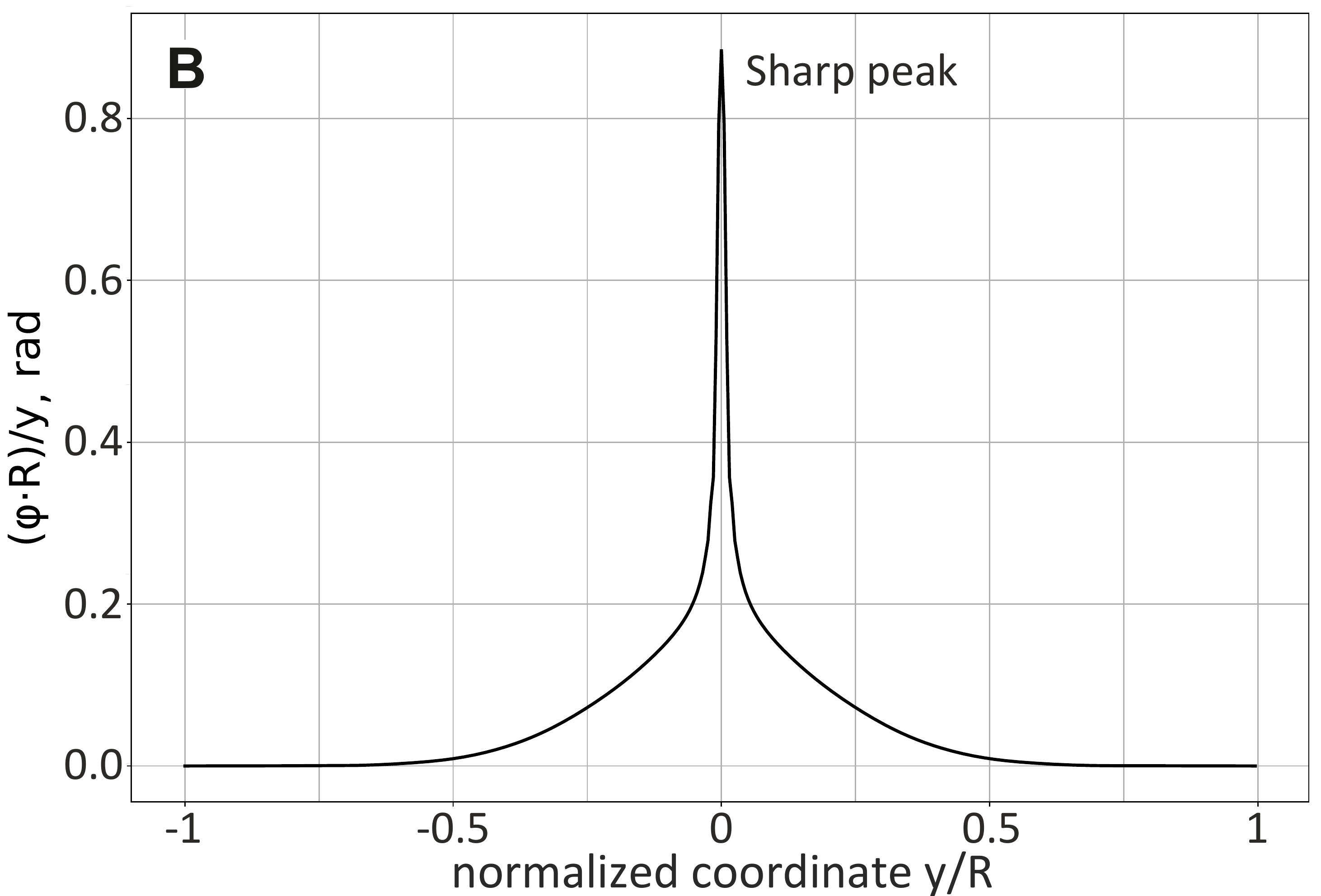}
		\caption{Symmetrized angle in case of axis mismatch (A) and $\varphi/y$ function (B).}
		\label{img:jump}
	\end{center}
\end{figure}

	The reason for such high influence of non-zero  $|y_{plasma}-y_{angle}|$ on calculated SMF behaviour is that symmetrizing angle distribution in respect to wrong axis contradicts the anti-symmetrical behavior of this distribution and will lead to angle(y,z) discontinuity (Fig. \ref{img:jump}A) and non-differentiability at chosen axis. Also, when calculating $F_{\varphi}(y)=\varphi(y)/y$ function it's intuitive that when $\varphi(0)\neq0$ there would be sharp peak at $y=0$, and the peak's value would be limited only by limited resolution of the CCD (Fig. \ref{img:jump}B). Moreover, as shown at equation \ref{eq:abel_inv}, there is a derivative inside the inverse Abel transfrom integral - in case of non-differentiability of the studied function we can most probably face errors performing direct calculation. It was found out, that the sharp peak at (Fig. \ref{img:jump}B) can not be abelized by Gegenbauer polynomial approximation method \cite{Pisarczyk90, Kasper1978} - the required even polynomial degree can not be simulated taking into account width of the abelized function. The Fourier \cite{AgaArt} and Mach-Shardin \cite{Pisarczyk90, Kasper1978} methods work better, however it is known that the latter one is sensitive to data irregularities, it's error is lowest at the boundary and raises to the center of considered function, increasing with the function length increase. Thus all of the data presented in this paper  is calculated using the Fourier method, whose accuracy increases with the number of measured experimental points.

	%%%
	%%%

	According to our evaluation, the most noticeable errors are intensity (background) errors due to irregularities in energy distribution in probe beam and plasma axis uncertainty due to its imperfect symmetry. Avoiding the axes mismatch problem due to the background irregularity requires some modification of measured angle data to make the distribution anti-symmetrical in respect to chosen plasma axis. In other words, the angle value should be set to zero at the $y_{plasma}$. We can think of several ways of modifying the measured data to get $y_{angle}(z)=y_{plasma}(z)$:	
	
	1. Shift each cross-section of angle distribution perpendicular to plasma axis. This is physically unjustified and will influence the angle and density relative positions, which can lead to errors in field distribution in case of sharp changes present in either in density or angle. However, we will test this approach in section devoted to error analysis.

\begin{figure}
	\begin{center}
		\includegraphics[width=0.95\linewidth]{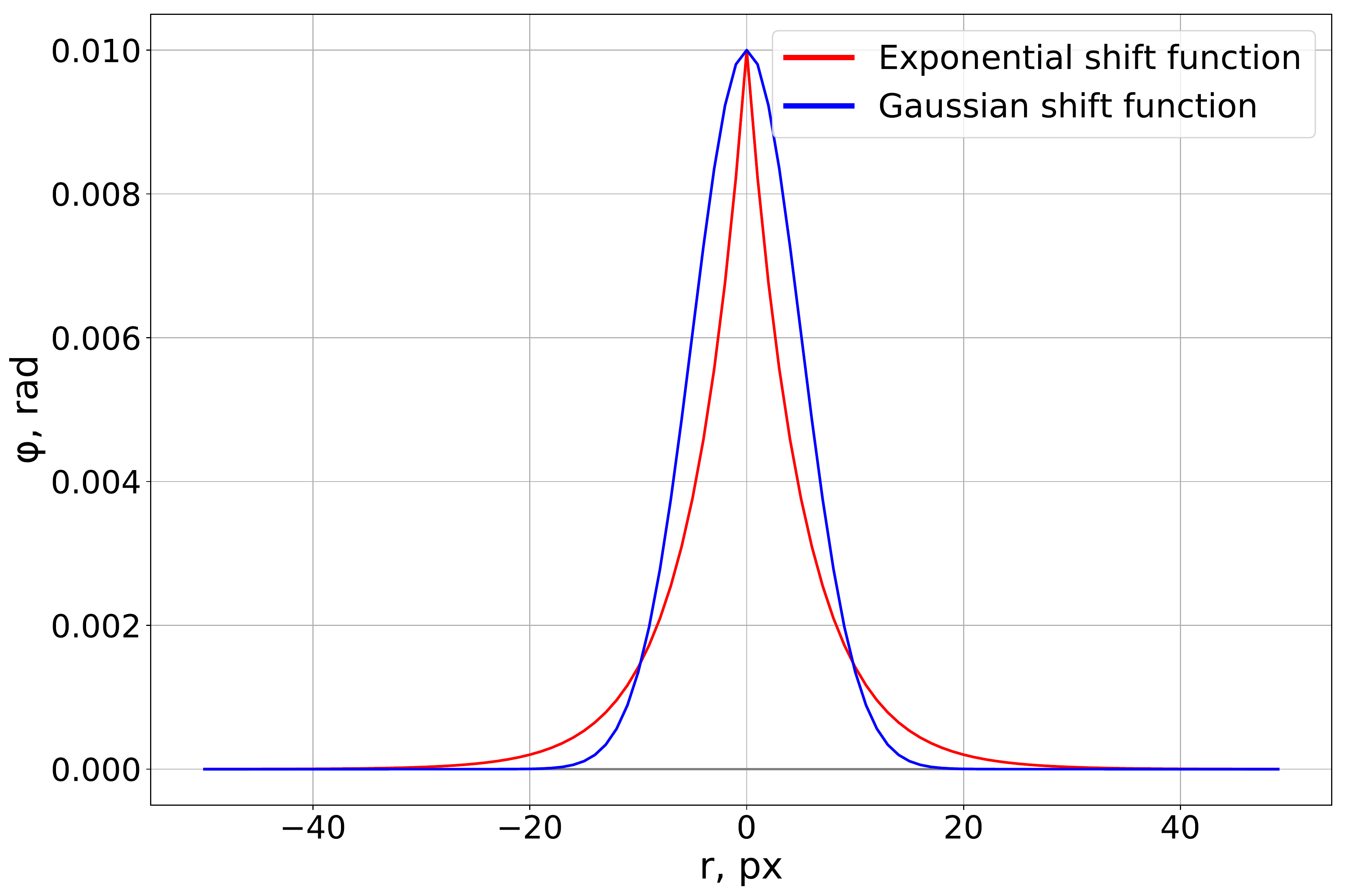}
		\caption{Comparison of tested Gaussian and exponential functions}
		\label{img:Correction Methods}
	\end{center}
\end{figure}

	2. Uniformly shift each angle distribution cross-section intensity-wise, in other words subtract an assumed constant background from each slice. However this will mean that the background is a one-variable (axis z) function, which is physically unreasonable. In case the real background is non-uniform this  method can lead to manyfold increase in error and even change in function sign, especially at the plasma border. We will not test this method in this paper.

	3. Correct a small area close to the plasma axis intensity-wise. This can be seen as a neutralization of the background in a small region near the axis, which is justified by known anti-symmetrical behaviour of the angle and thus known background value at supposed symmetry point. Width of the modified region should be of the same scale as the minimum background non-uniformity period, that can be determined from plasma-free region in the interferogram. Following approaches were tested:
	
	a. Cut the angle distribution cross-section from both sides of the supposed symmetry axis and approximate it to 0 at the axis with linear function. This will lead to constant value of $\varphi/y$  at the corrected region.
	
	b. Subtract a narrow function, centered at the plasma axis, from the angle cross-section. Such function will be further referred to as "shift function". Several shift function profiles were tested, of which the most interesting were Gaussian and exponentially decreasing function (see Fig.\ref{img:Correction Methods}).
	
		\begin{figure*}
		\begin{center}
			\includegraphics[width=0.33\linewidth]{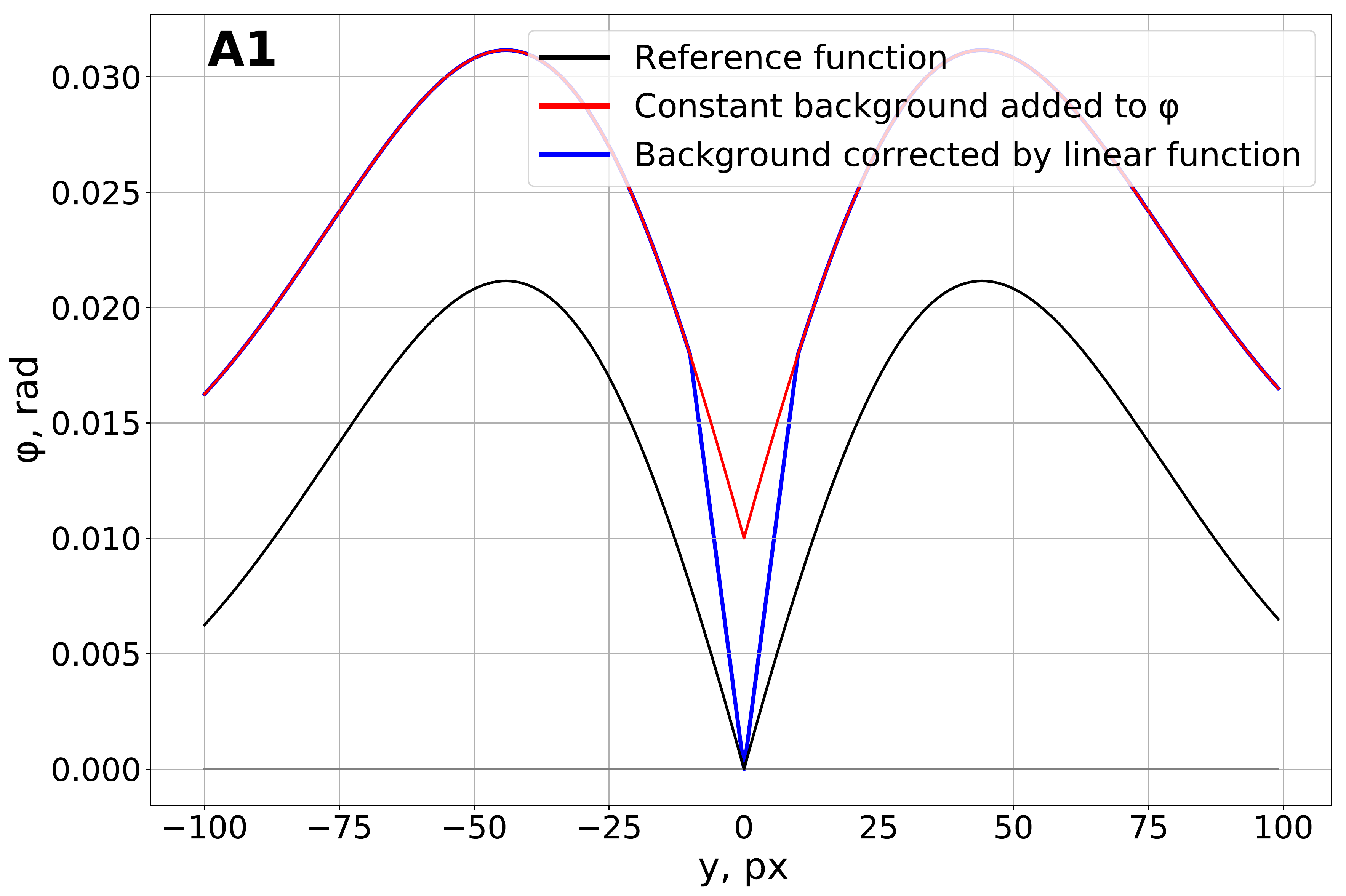}
			\includegraphics[width=0.33\linewidth]{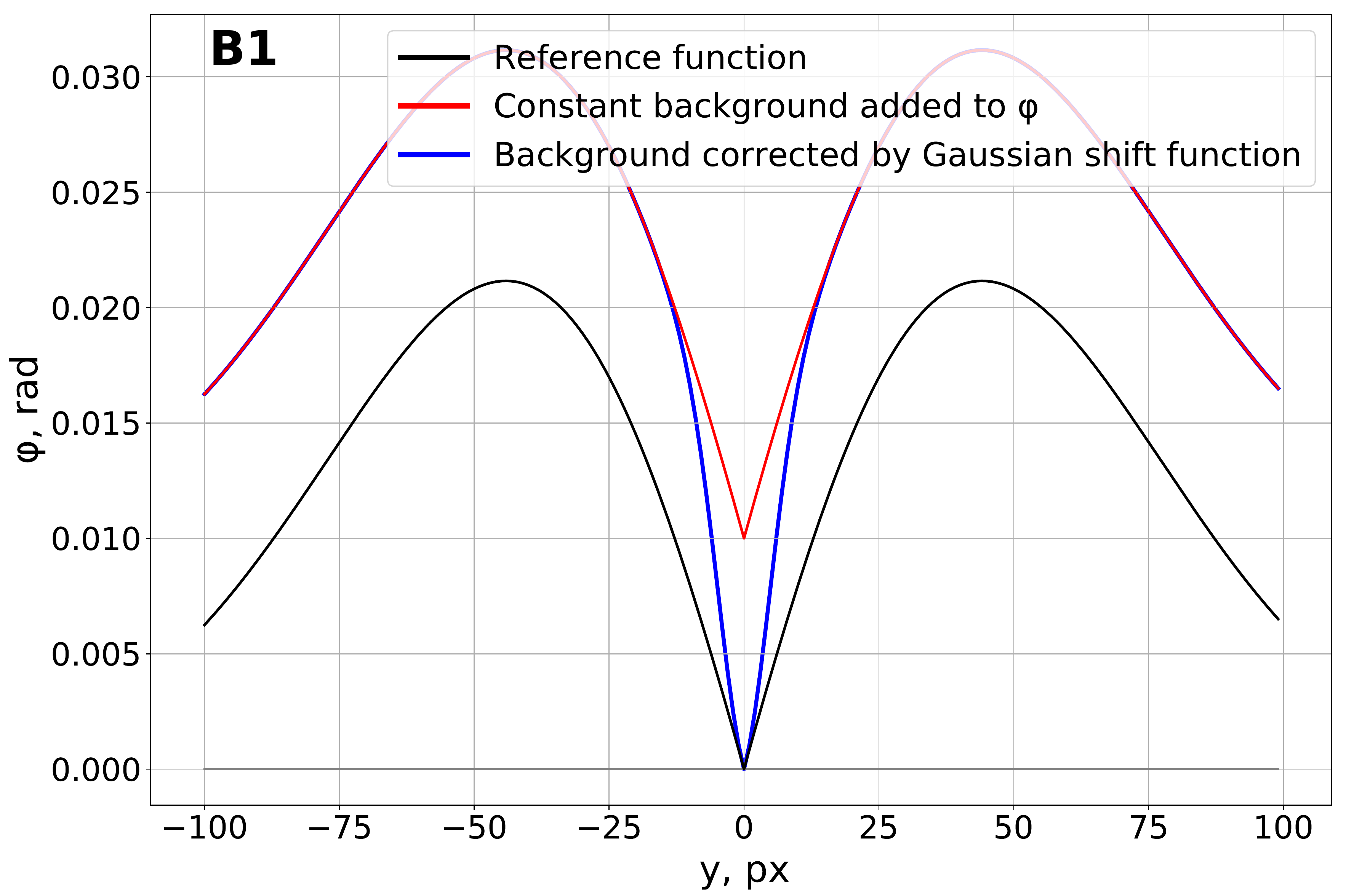}
			\includegraphics[width=0.33\linewidth]{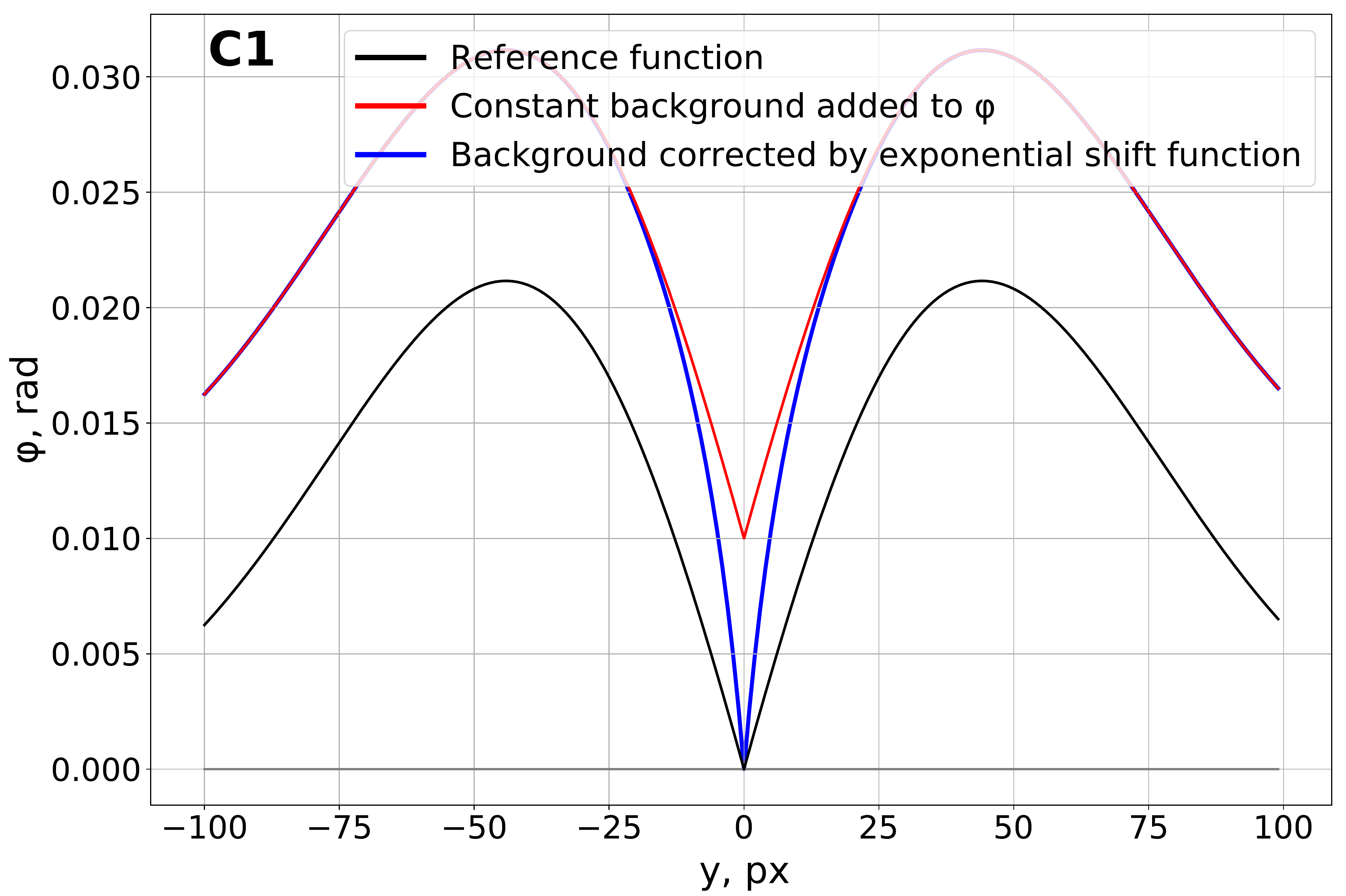}
			\\
			\includegraphics[width=0.33\linewidth]{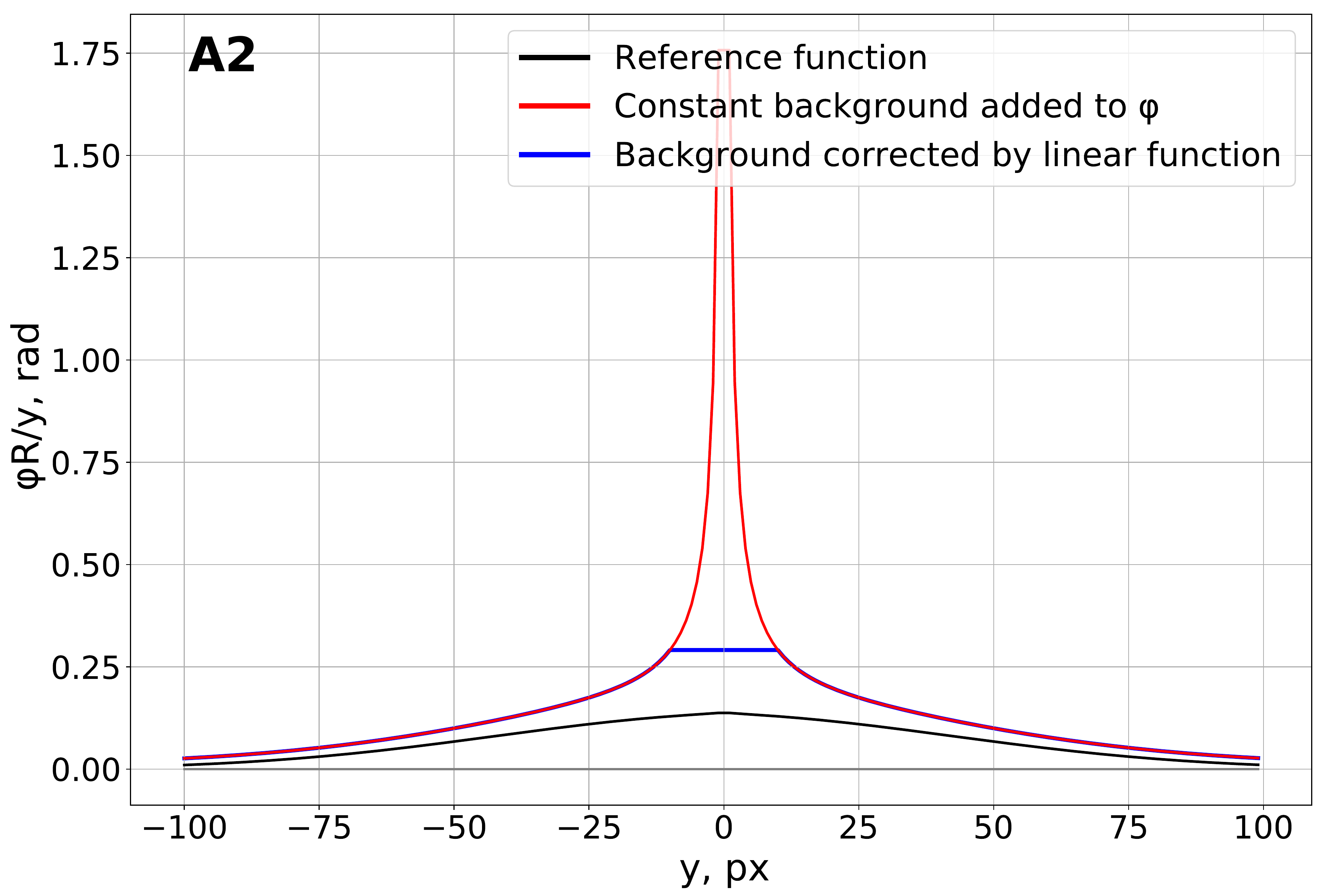}
			\includegraphics[width=0.33\linewidth]{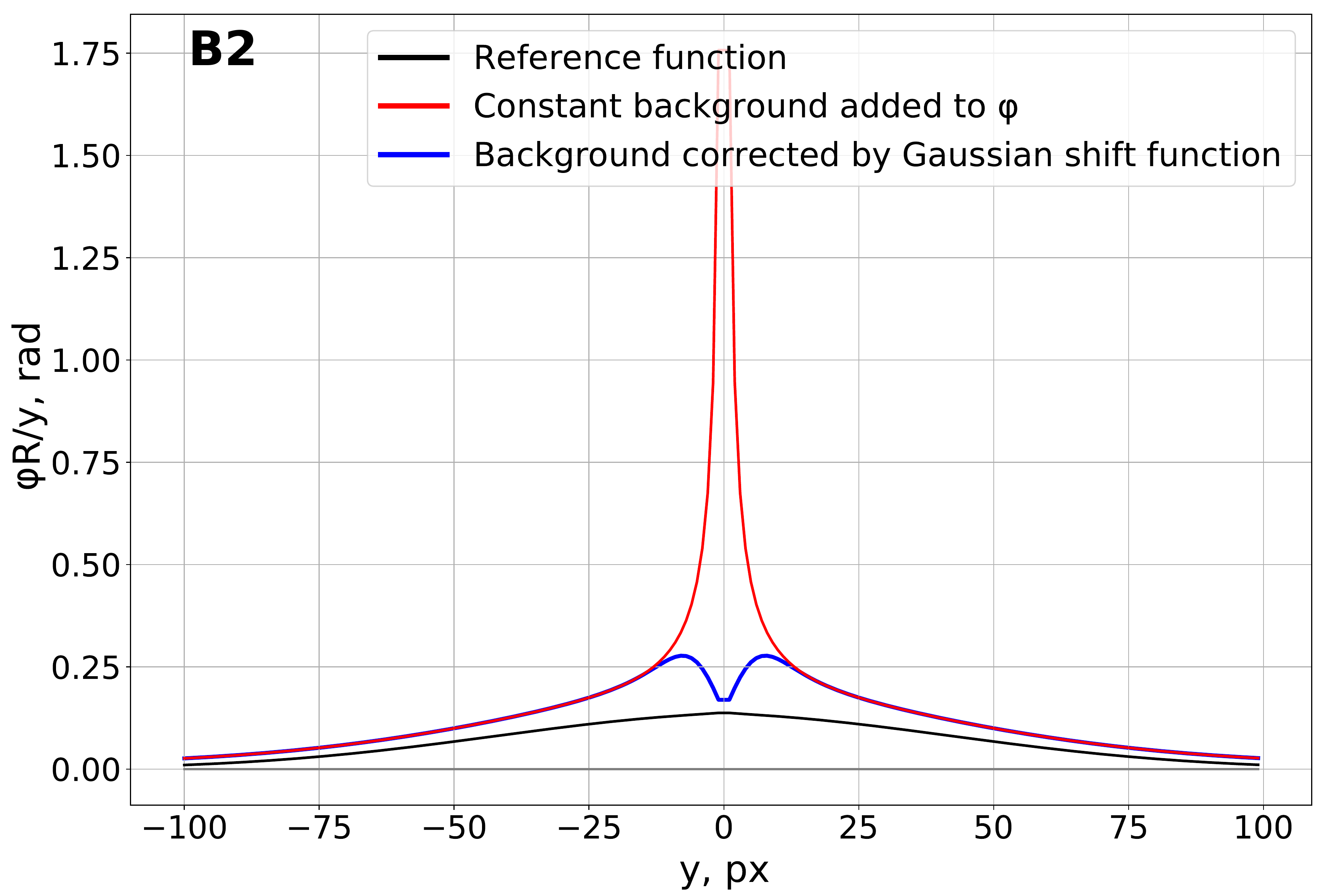}
			\includegraphics[width=0.33\linewidth]{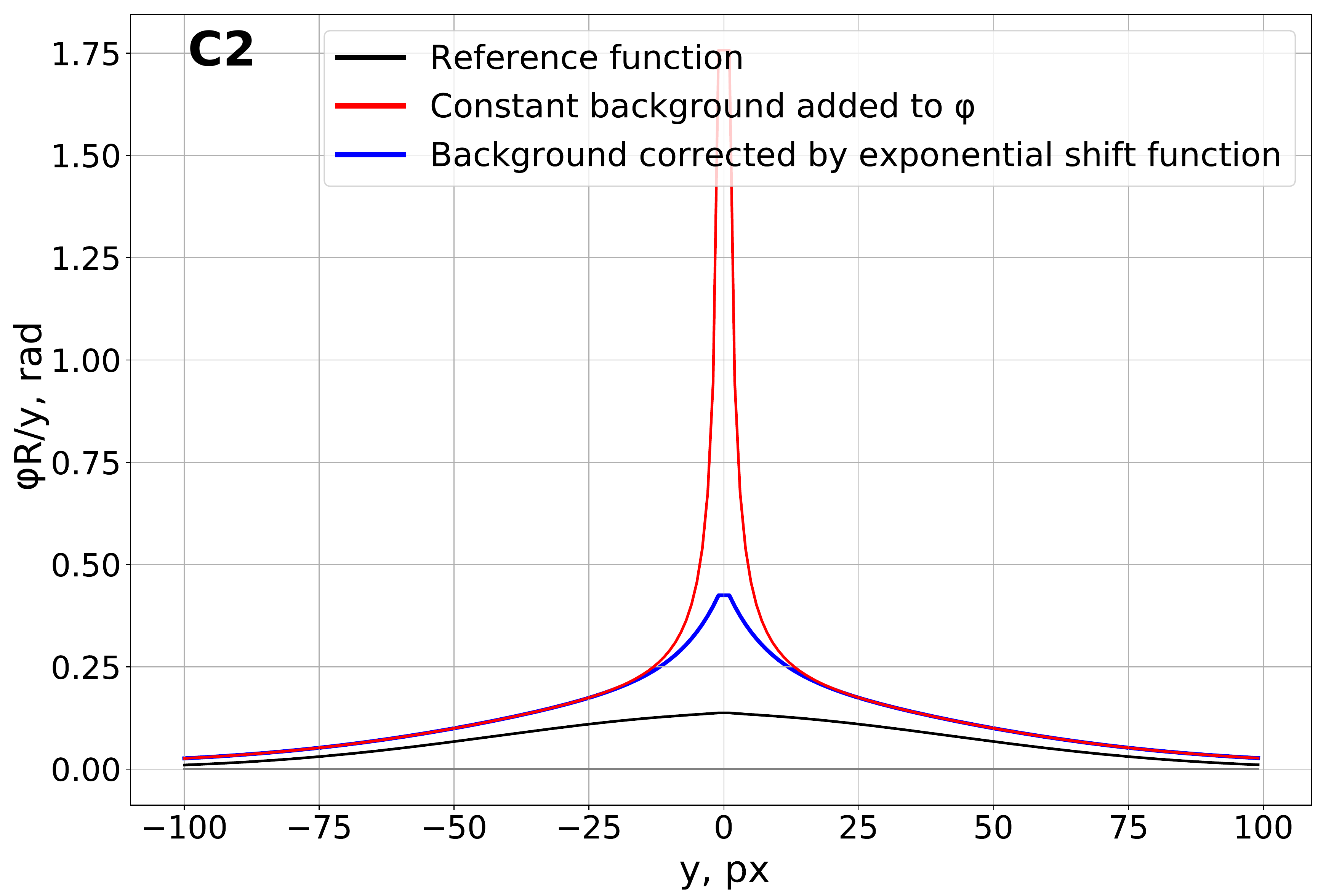}
			\\
			\includegraphics[width=0.33\linewidth]{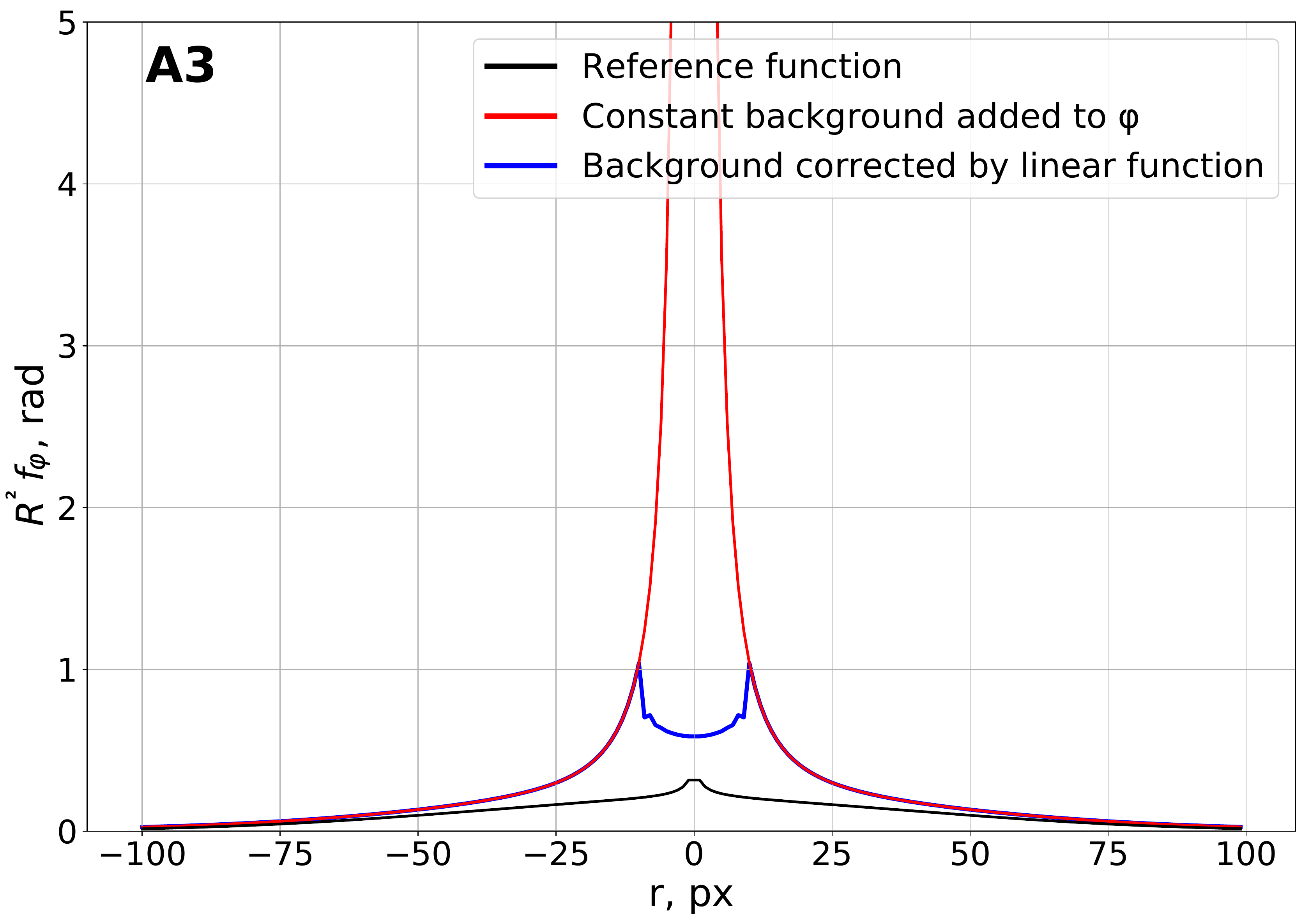}
			\includegraphics[width=0.33\linewidth]{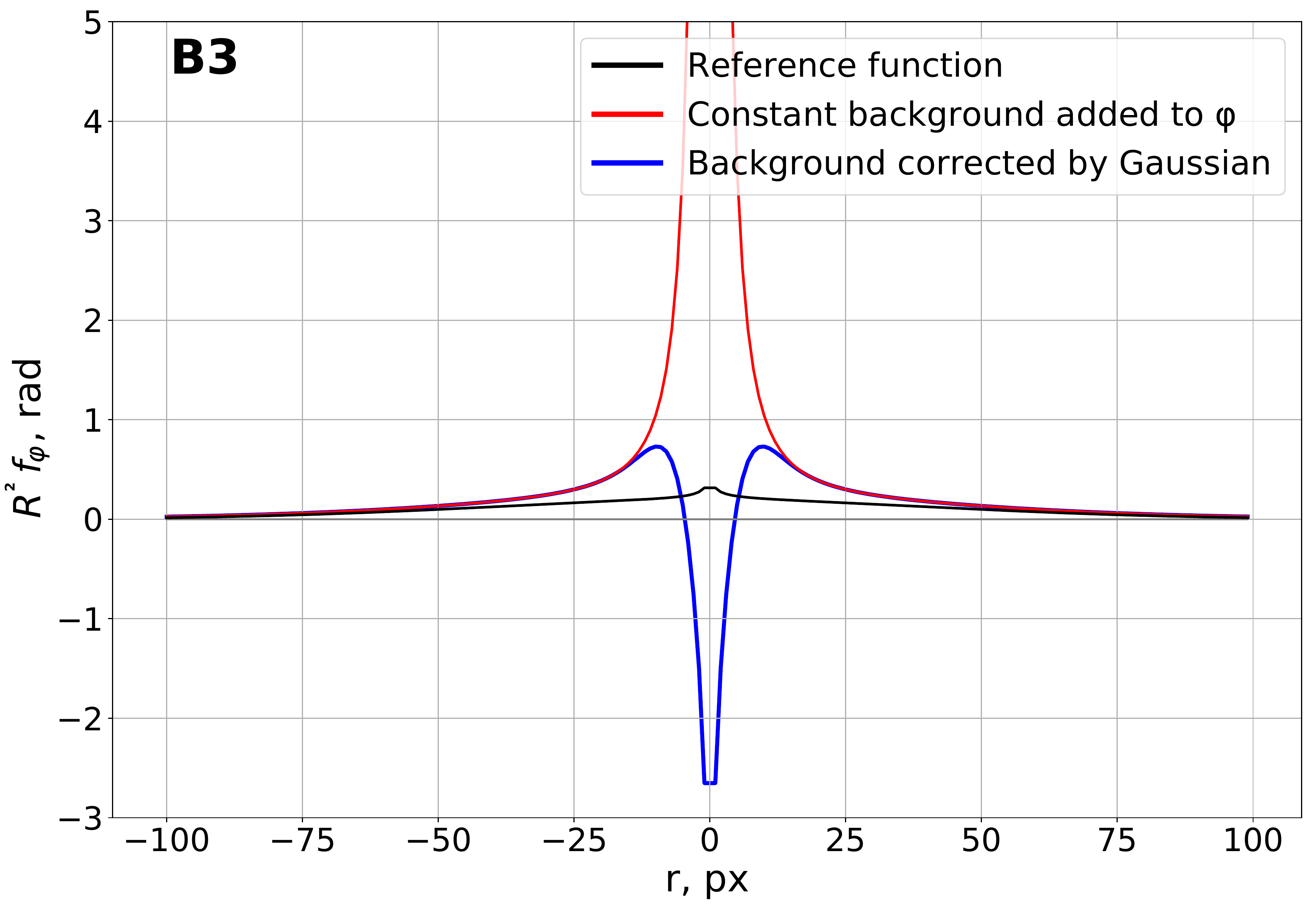}
			\includegraphics[width=0.33\linewidth]{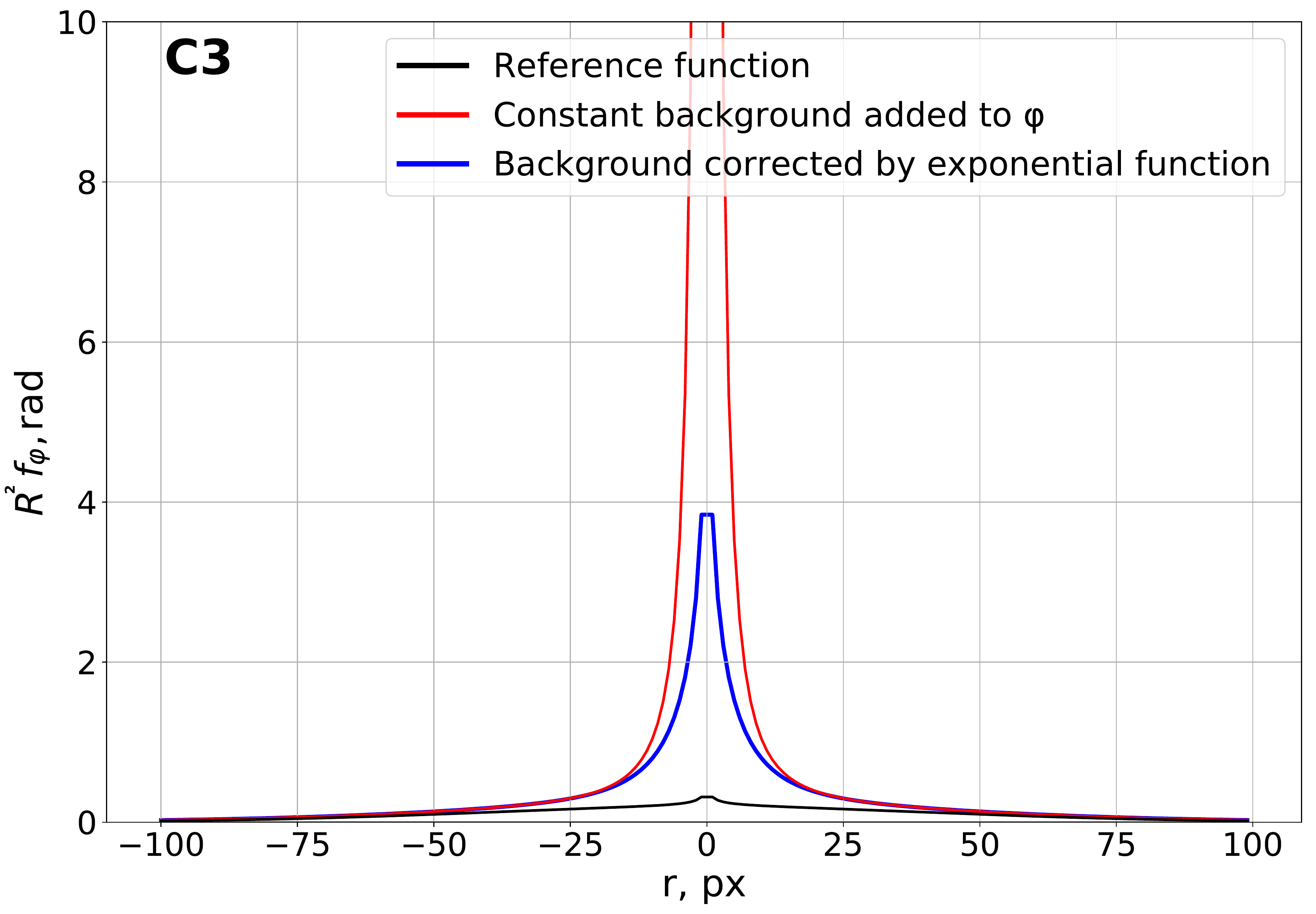}
			\caption{Background correction using linear method, shift by Gaussian and exponential functions (A-C respectively), for angle $\varphi$, normalized angle $\varphi R/y$ and its Abelization $f_{\phi}$ cross-sections (1-3 respectively).}
			\label{img:correction methods}
		\end{center}
	\end{figure*}
	\begin{figure*}
		\begin{center}
			\includegraphics[width=0.7\linewidth]{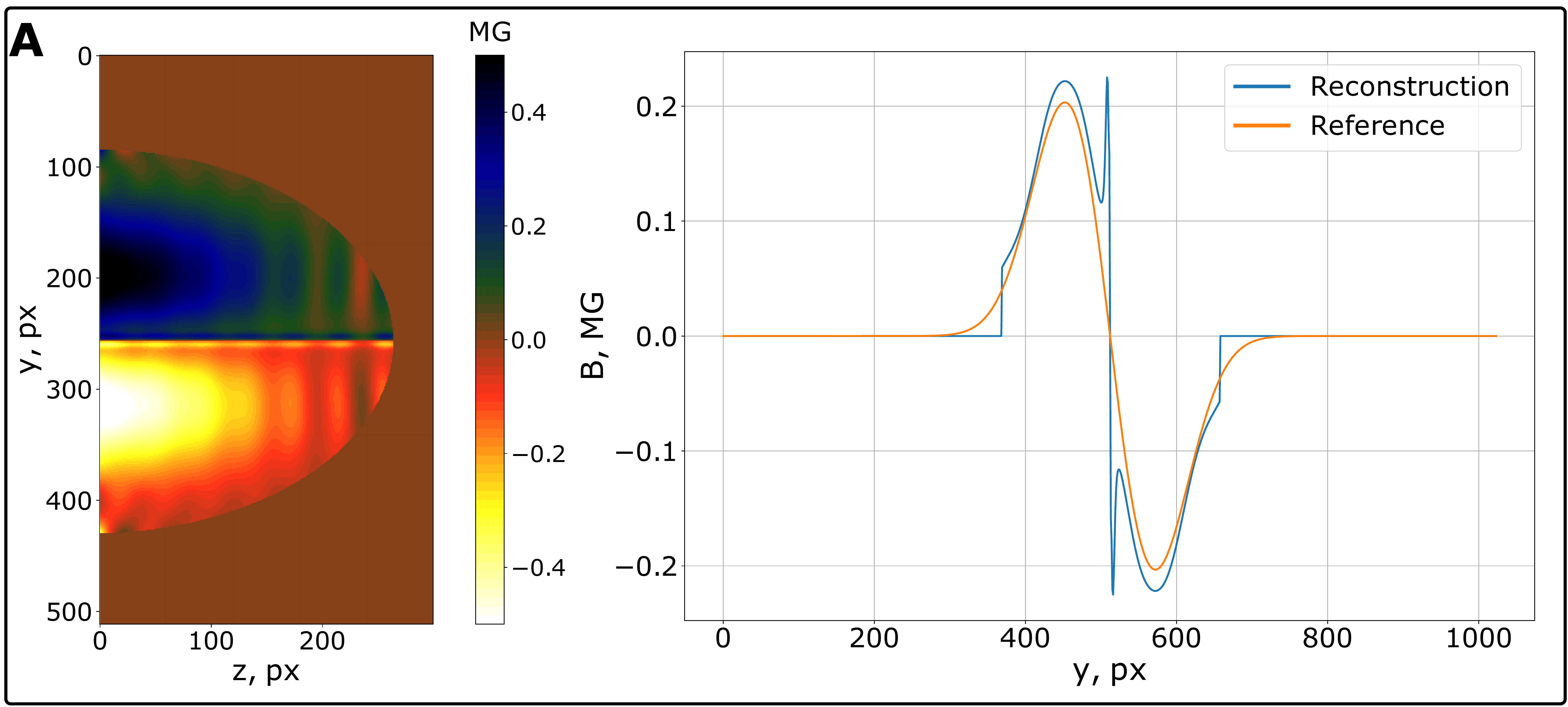}
			\\
			\includegraphics[width=0.7\linewidth]{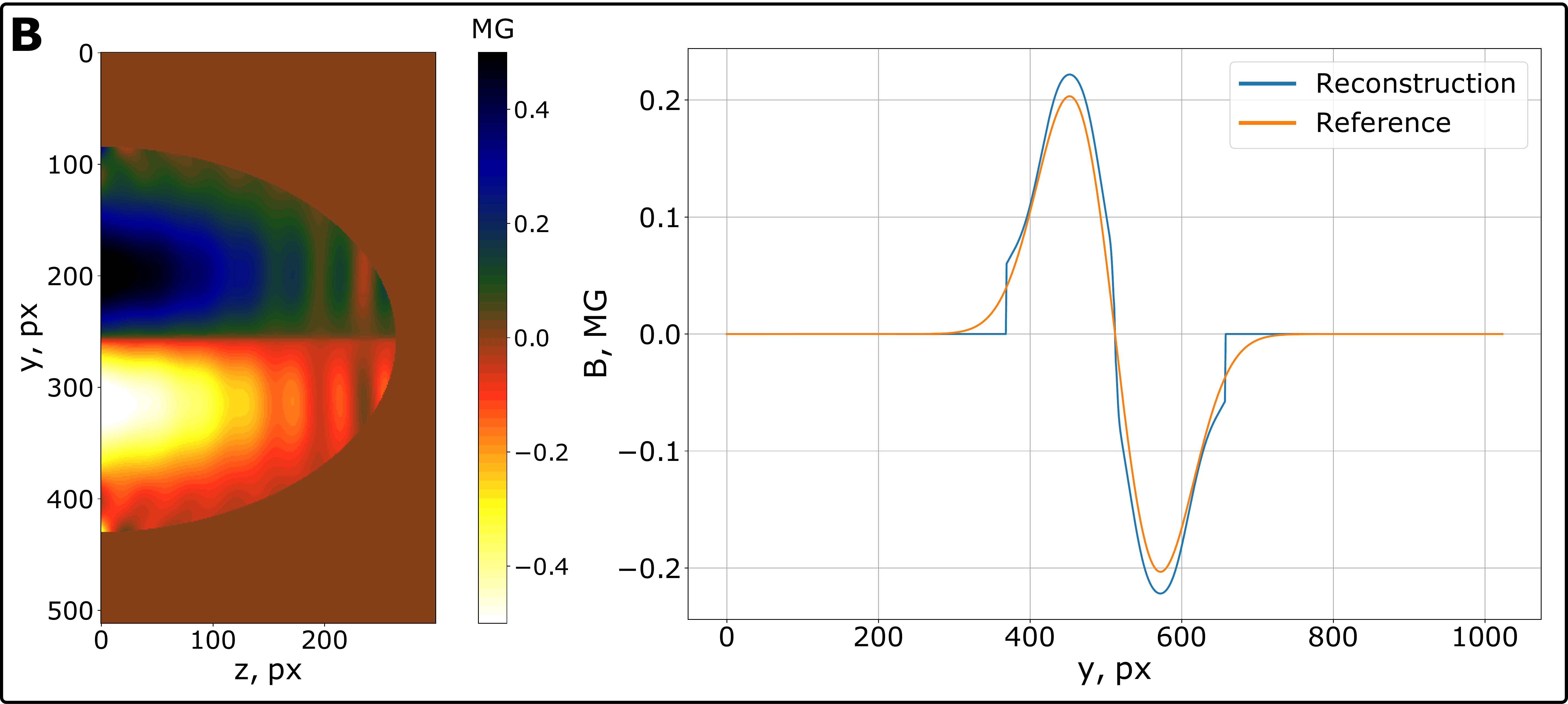}
			\caption{SMF reconstruction in case of $\mathbf{V_{int}}$ error using (A) non-corrected angle distribution, (B) corrected angle distribution }
			\label{img:corr1}
		\end{center}
	\end{figure*}

	\begin{figure*}
	\begin{center}
		\includegraphics[width=0.33\linewidth]{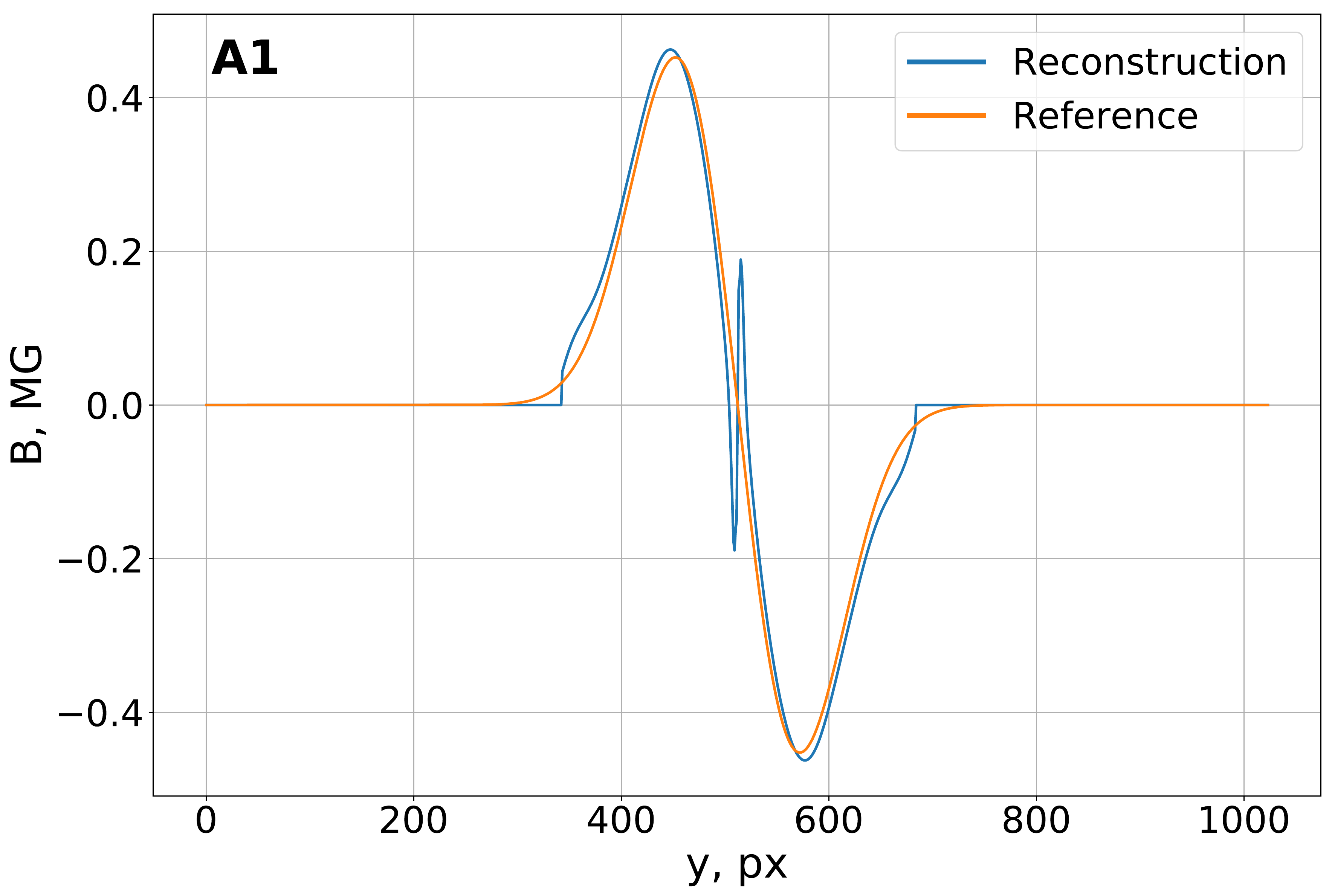}
		\includegraphics[width=0.33\linewidth]{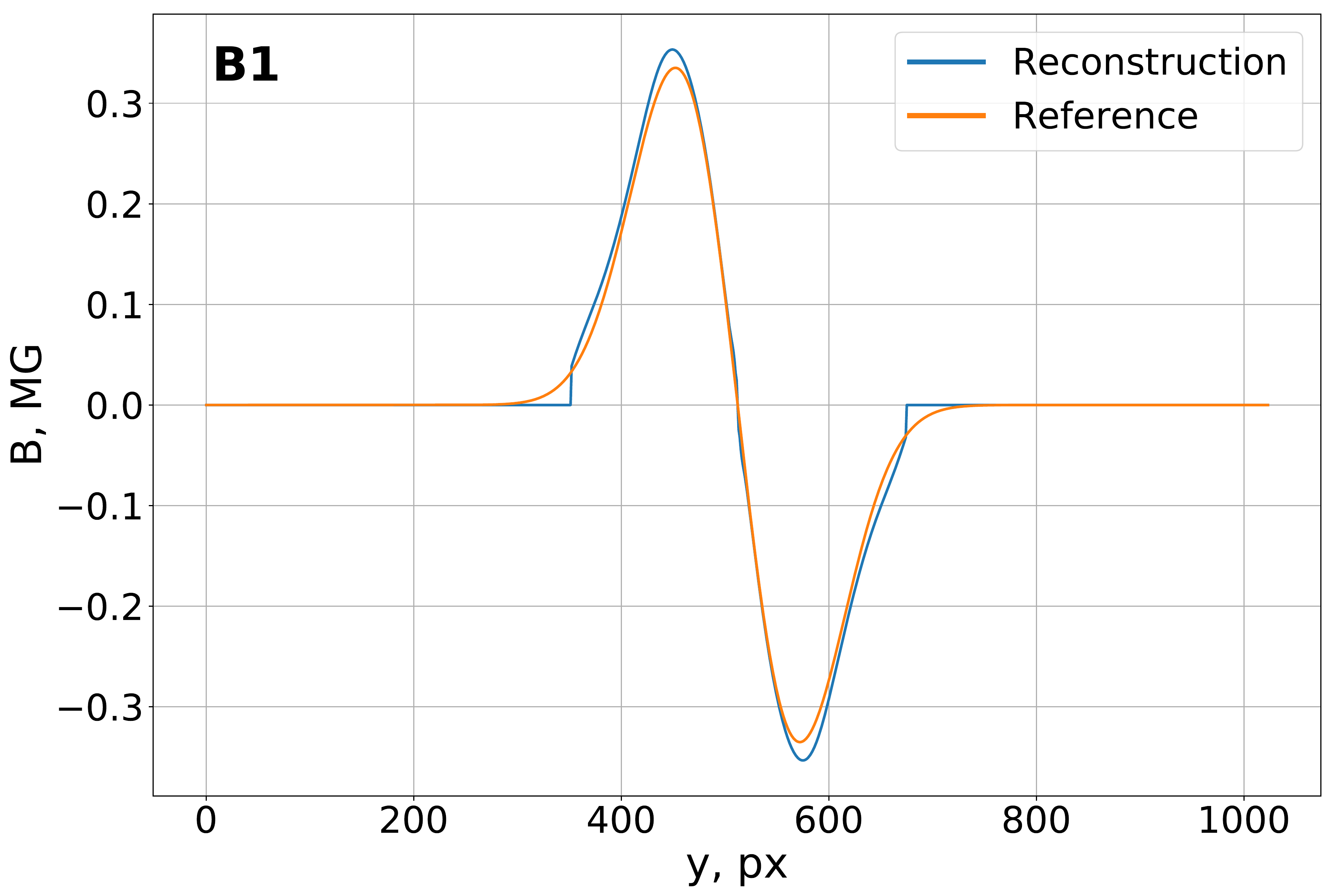}
		\includegraphics[width=0.33\linewidth]{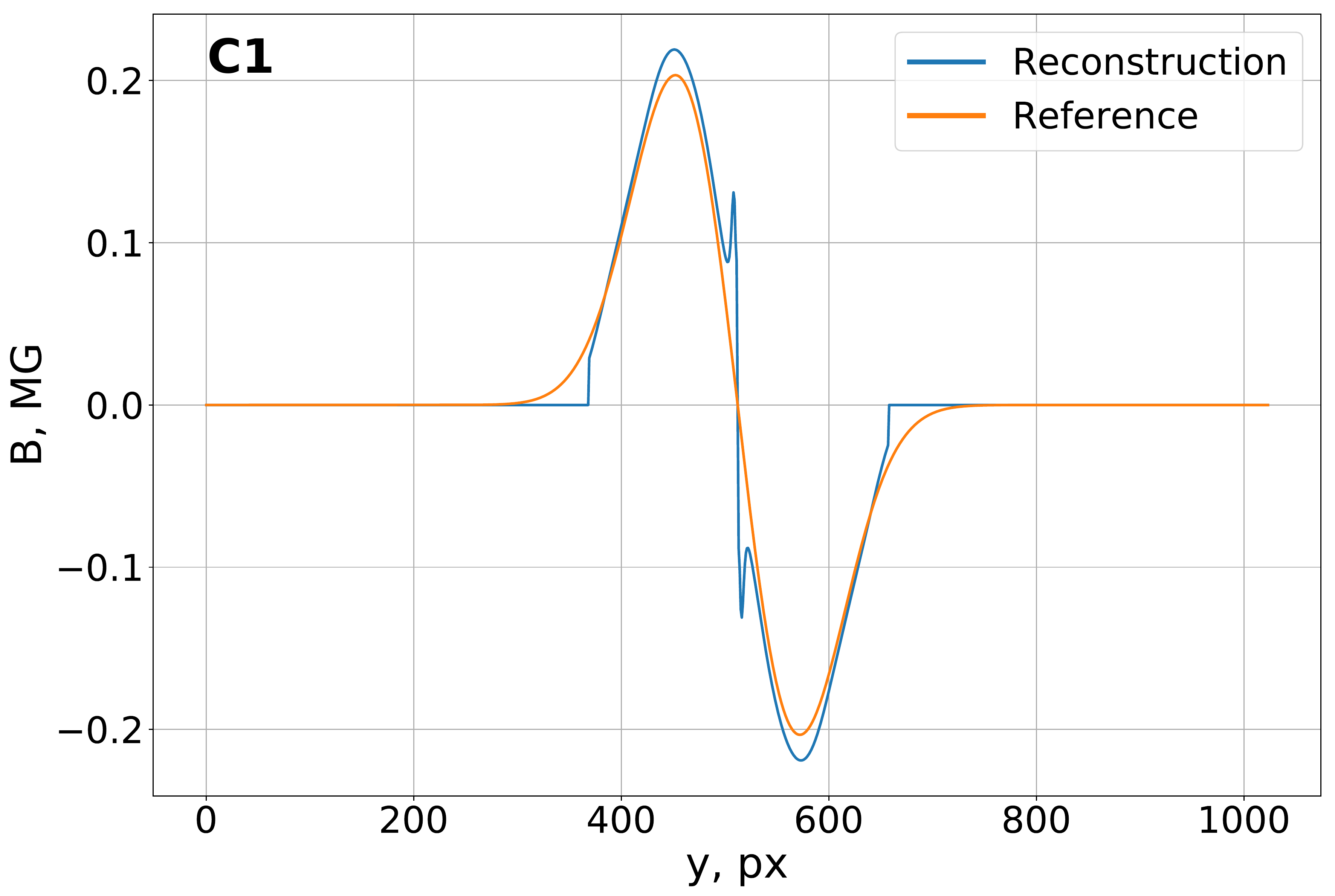}
		\\
		\includegraphics[width=0.33\linewidth]{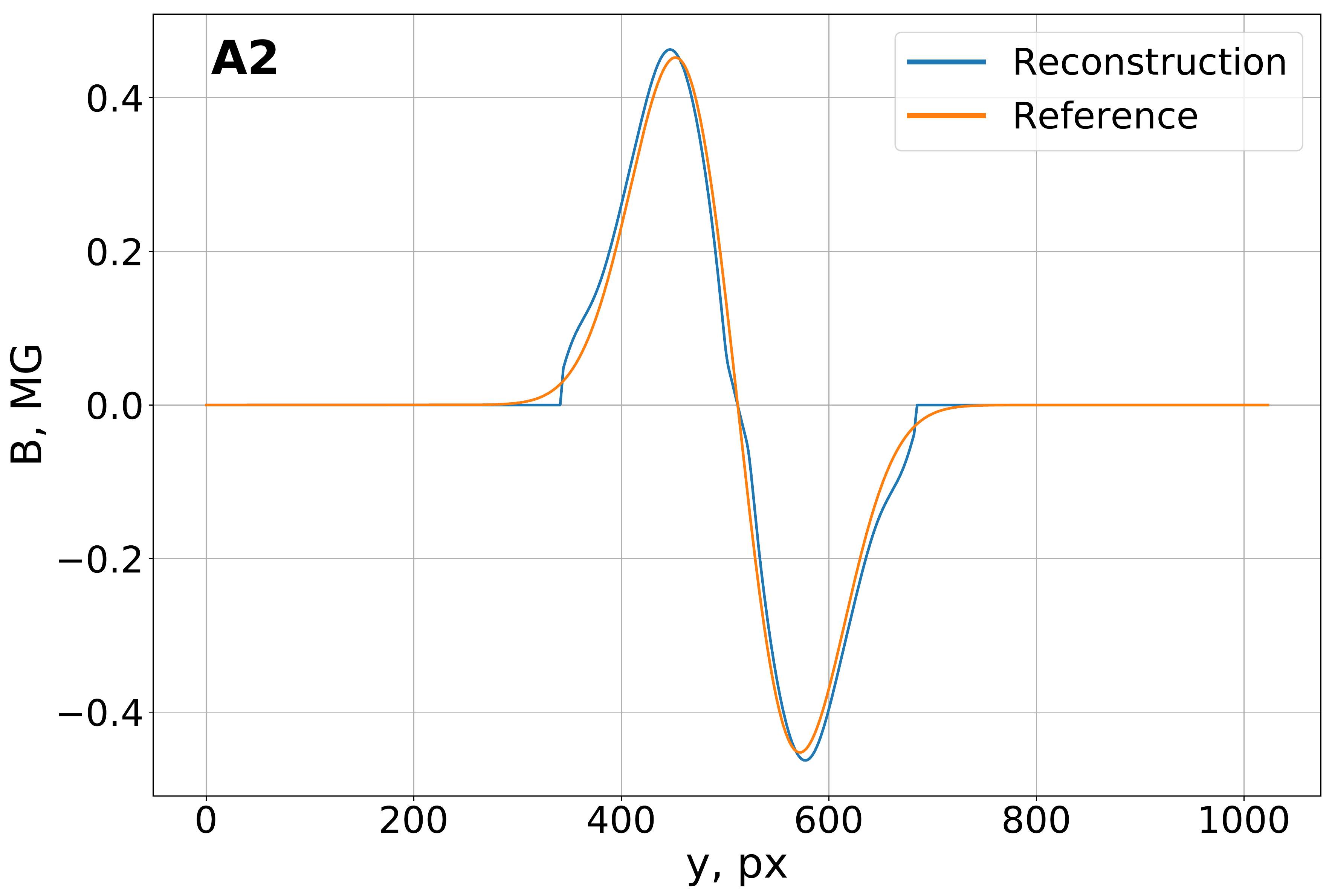}
		\includegraphics[width=0.33\linewidth]{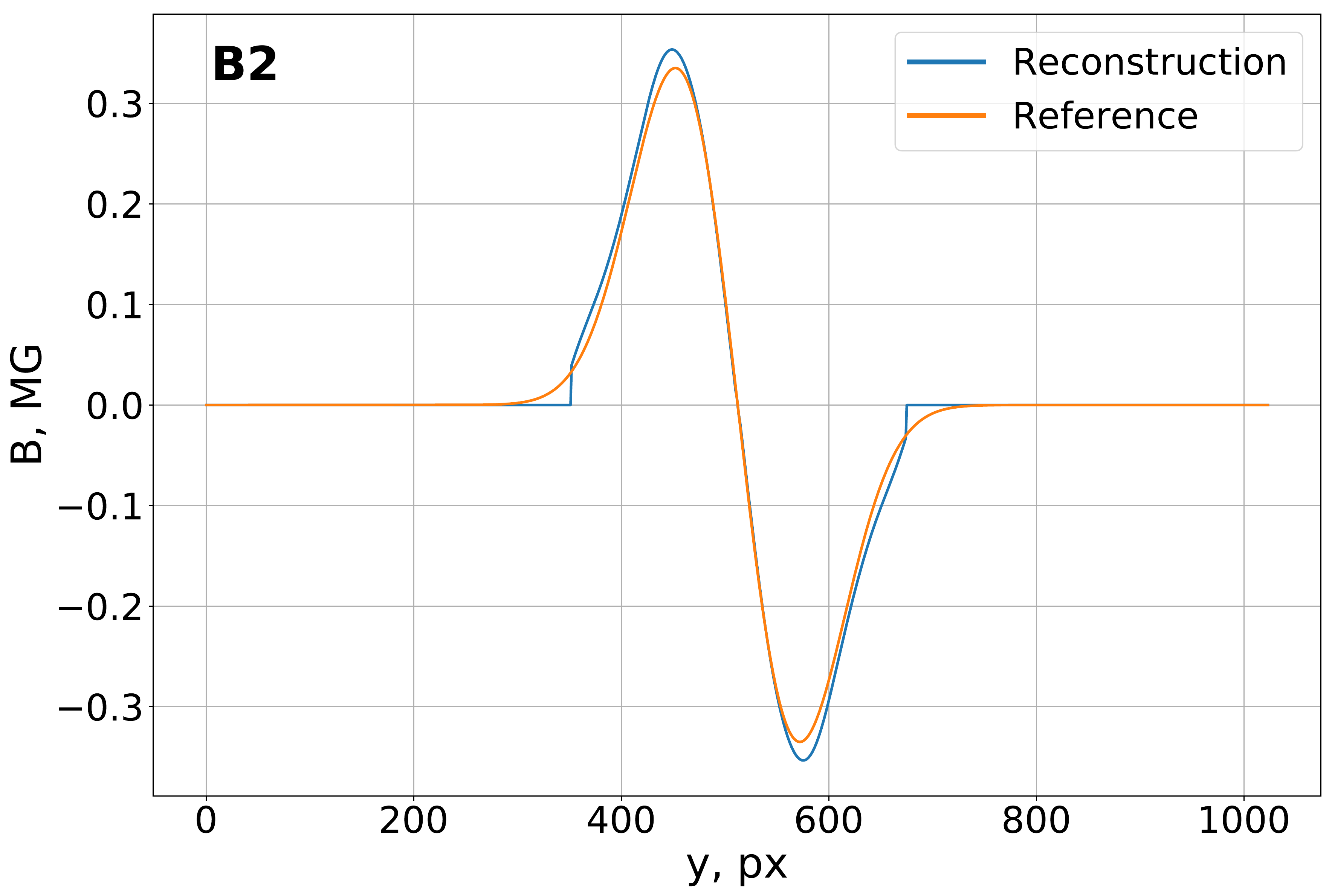}
		\includegraphics[width=0.33\linewidth]{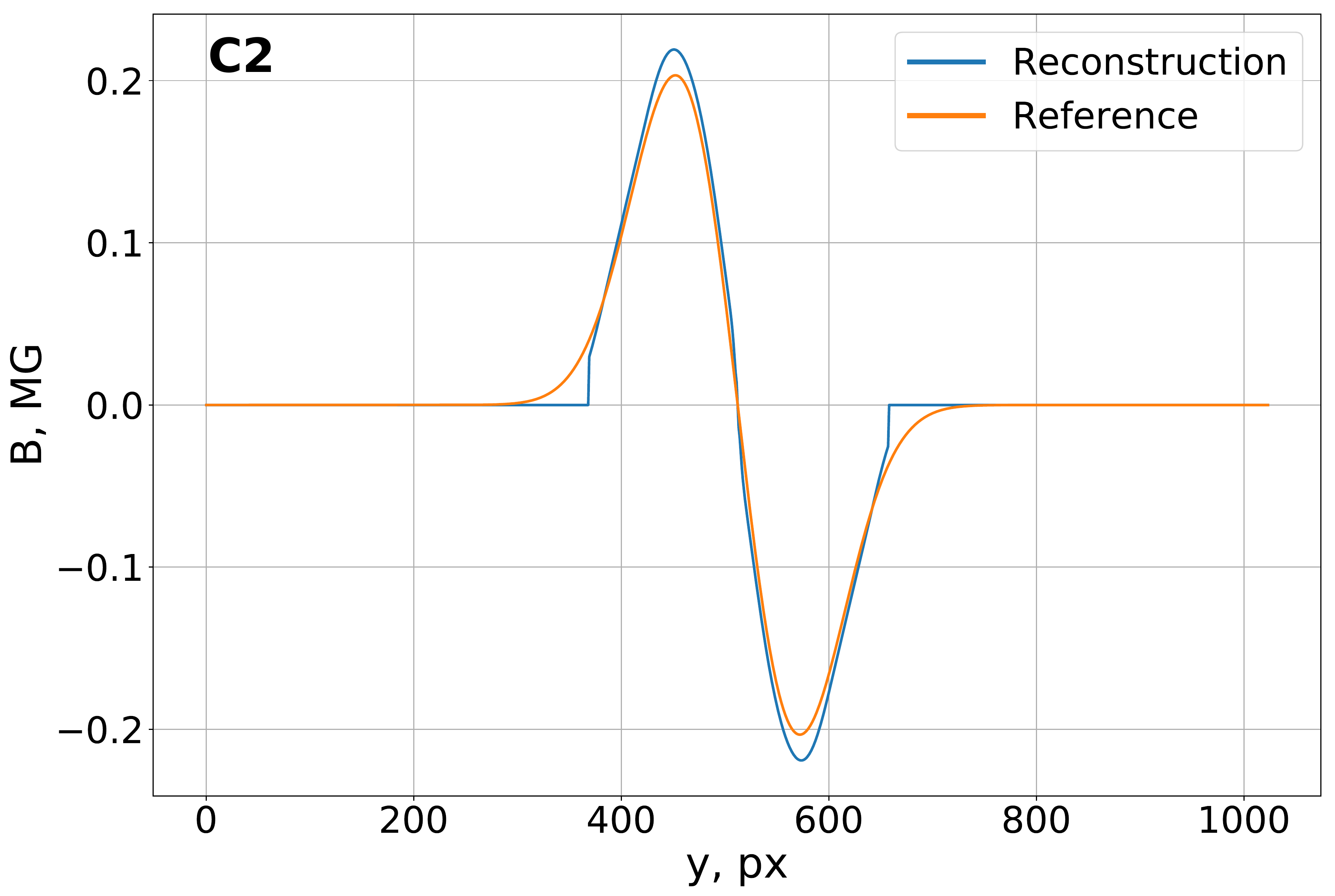}
		
		\caption{SMF reconstruction in presence of $\mathbf{G_{angle}}$ error without and with correction in case $z$=50 px (A1 and A2), $z$=100 px (B1 and B2) and for $z$=150 px (C1 and C2).}
		\label{img:rotcorr}
	\end{center}
\end{figure*}

	\begin{figure*}
	\begin{center}
		\includegraphics[width=0.38\linewidth]{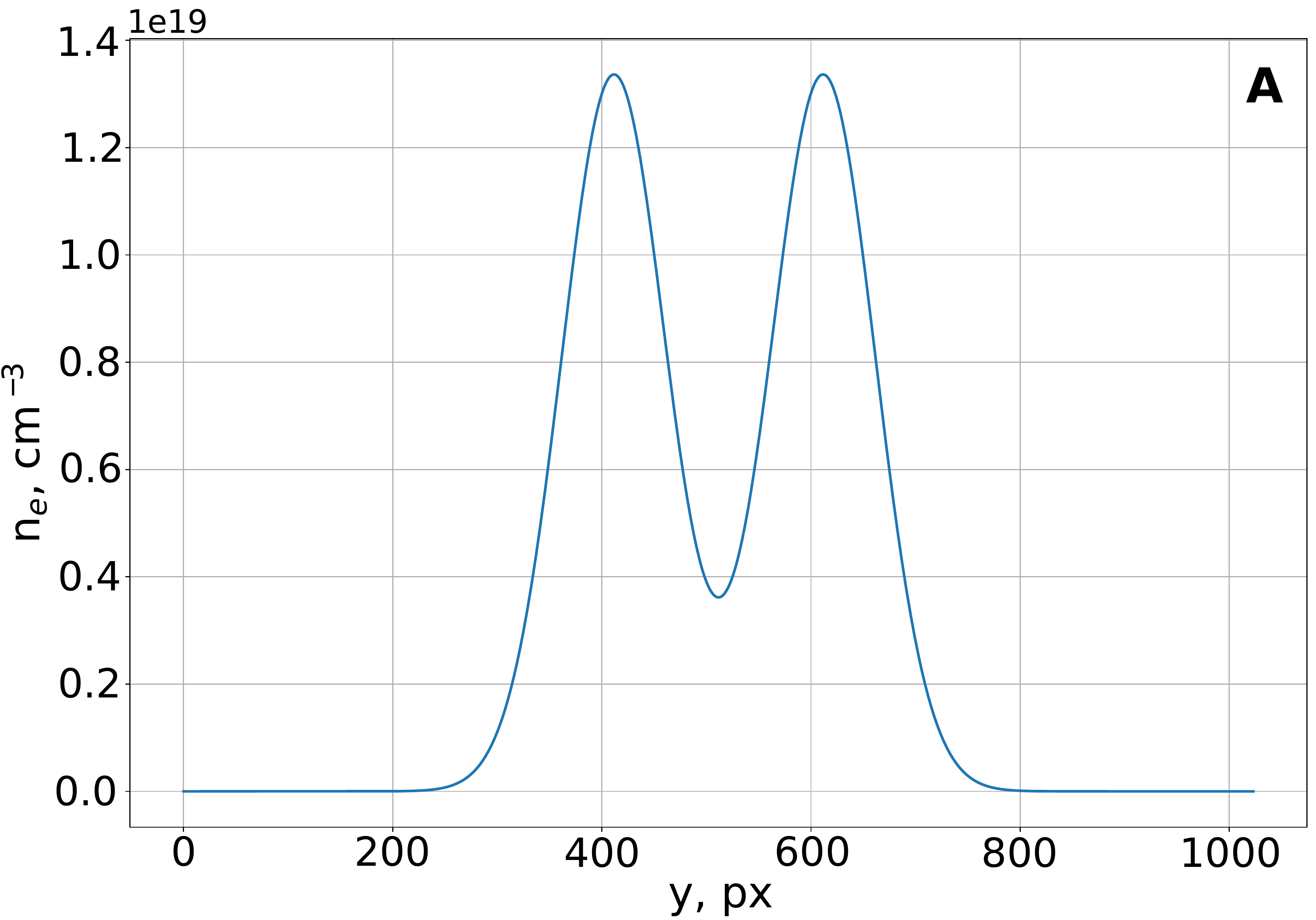}
		\hfill 
		\includegraphics[width=0.38\linewidth]{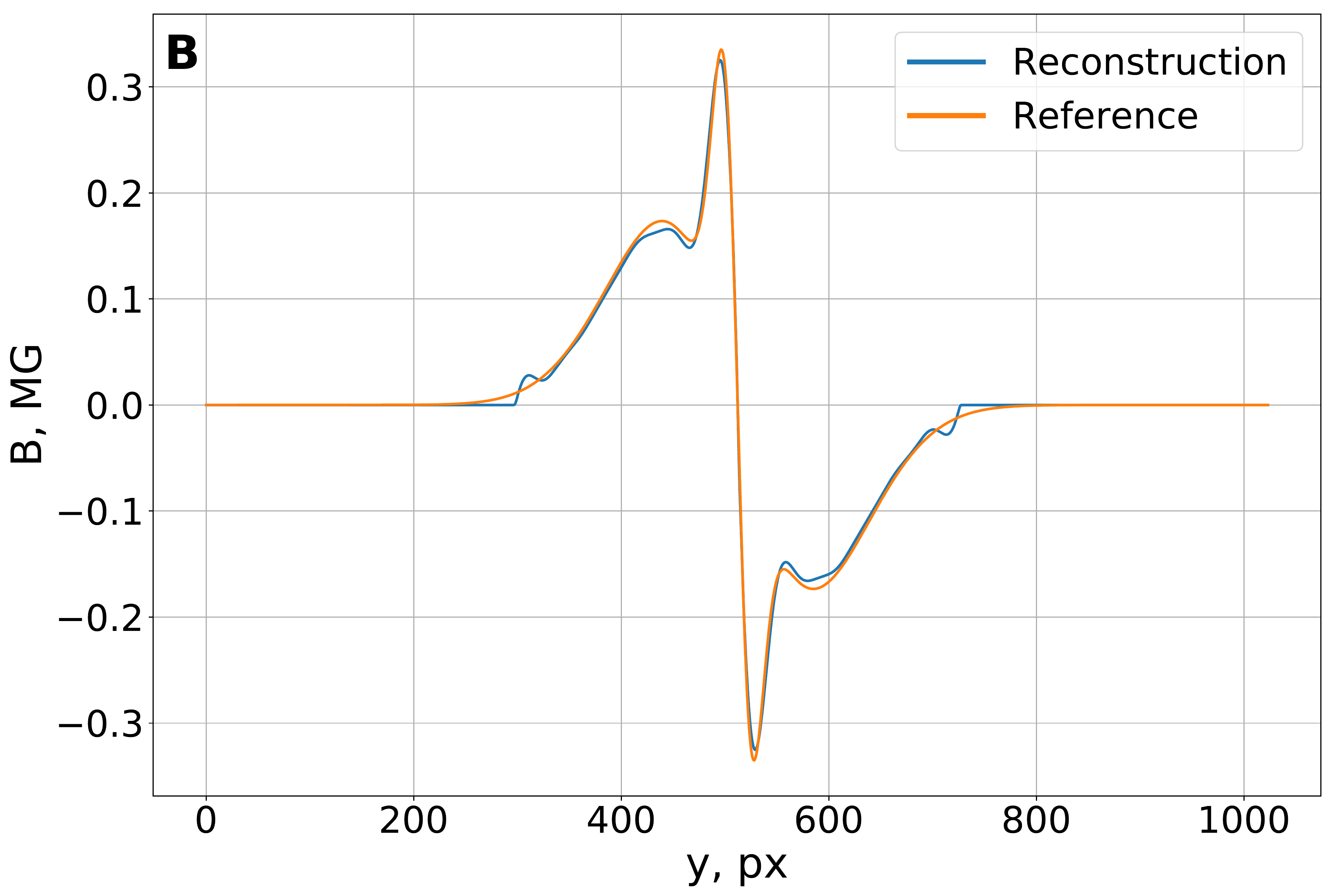}
		\hfill
		\includegraphics[width=0.2\linewidth]{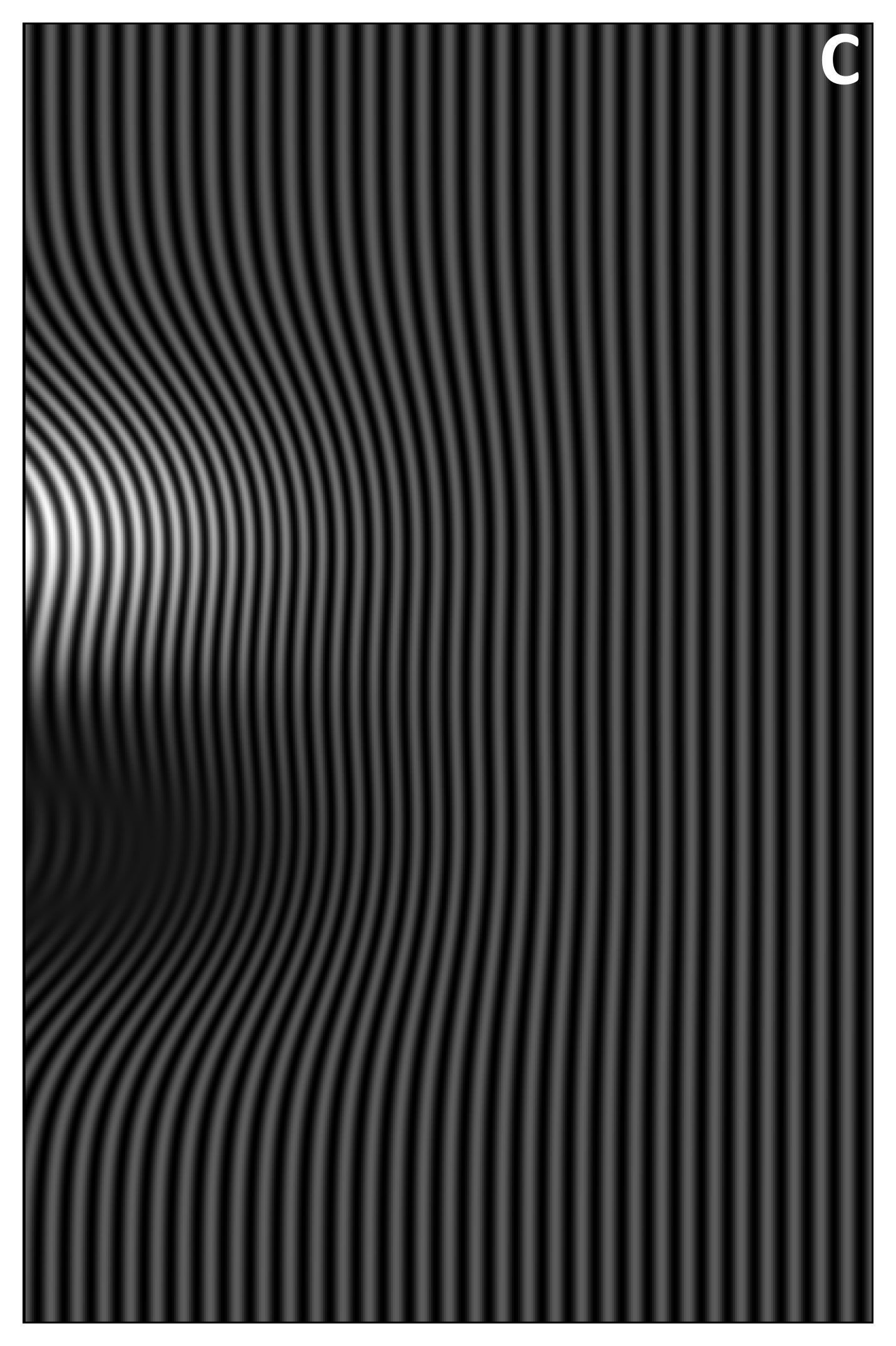}
		\hfill
		\caption{Reference density distribution cross-section for $z$=100 px (A), cross-sections of reference and calculated B-field distributions for $z$=100 px (B), part of synthetic complex interferogram (C).}
		\label{img:SimData}
	\end{center}
\end{figure*}

\begin{figure*}
	\begin{center}
		\begin{subfigure}[b]{0.99\linewidth}	
			\centering		
			\includegraphics[width=0.45\linewidth]{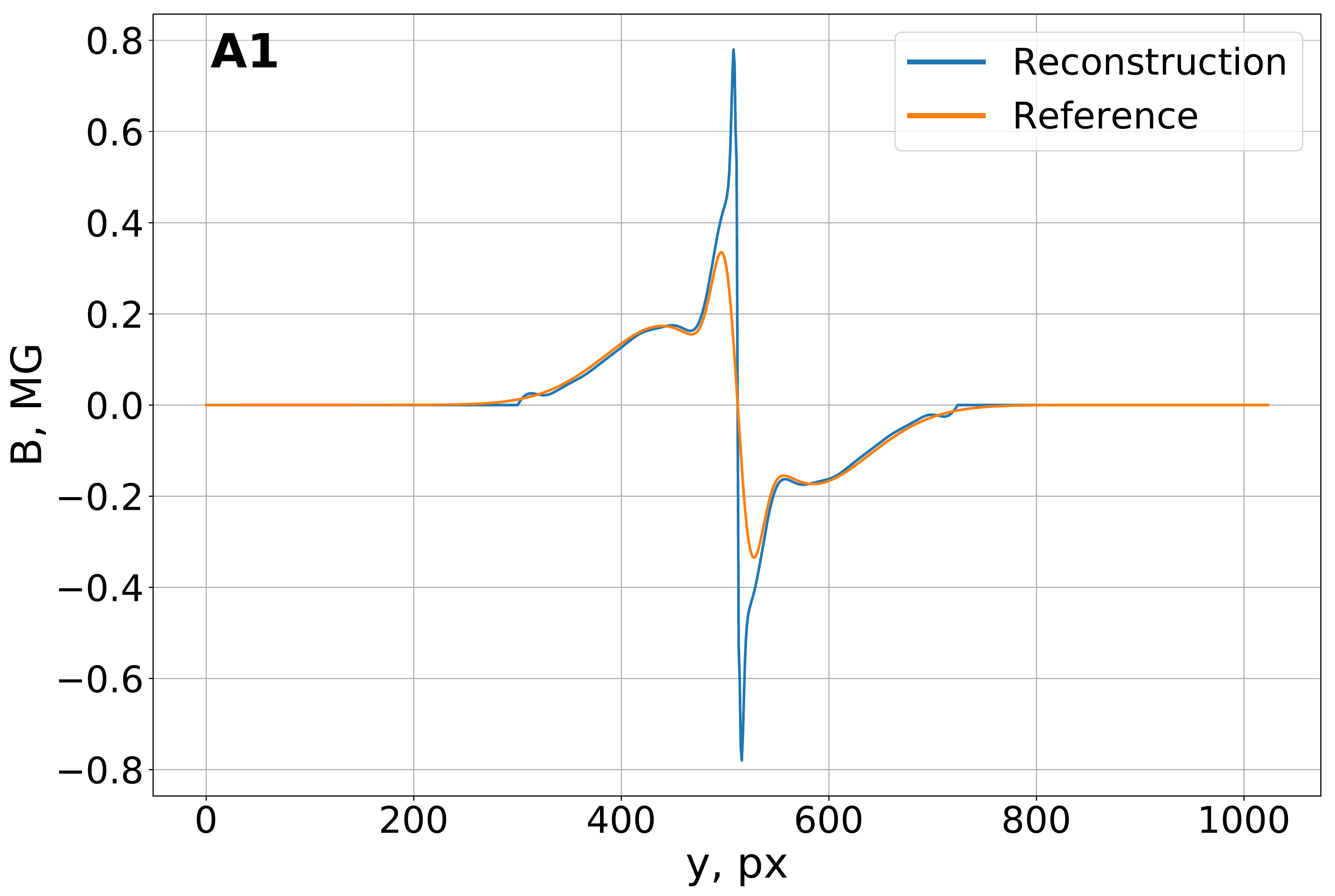}
			\hfil
			\includegraphics[width=0.45\linewidth]{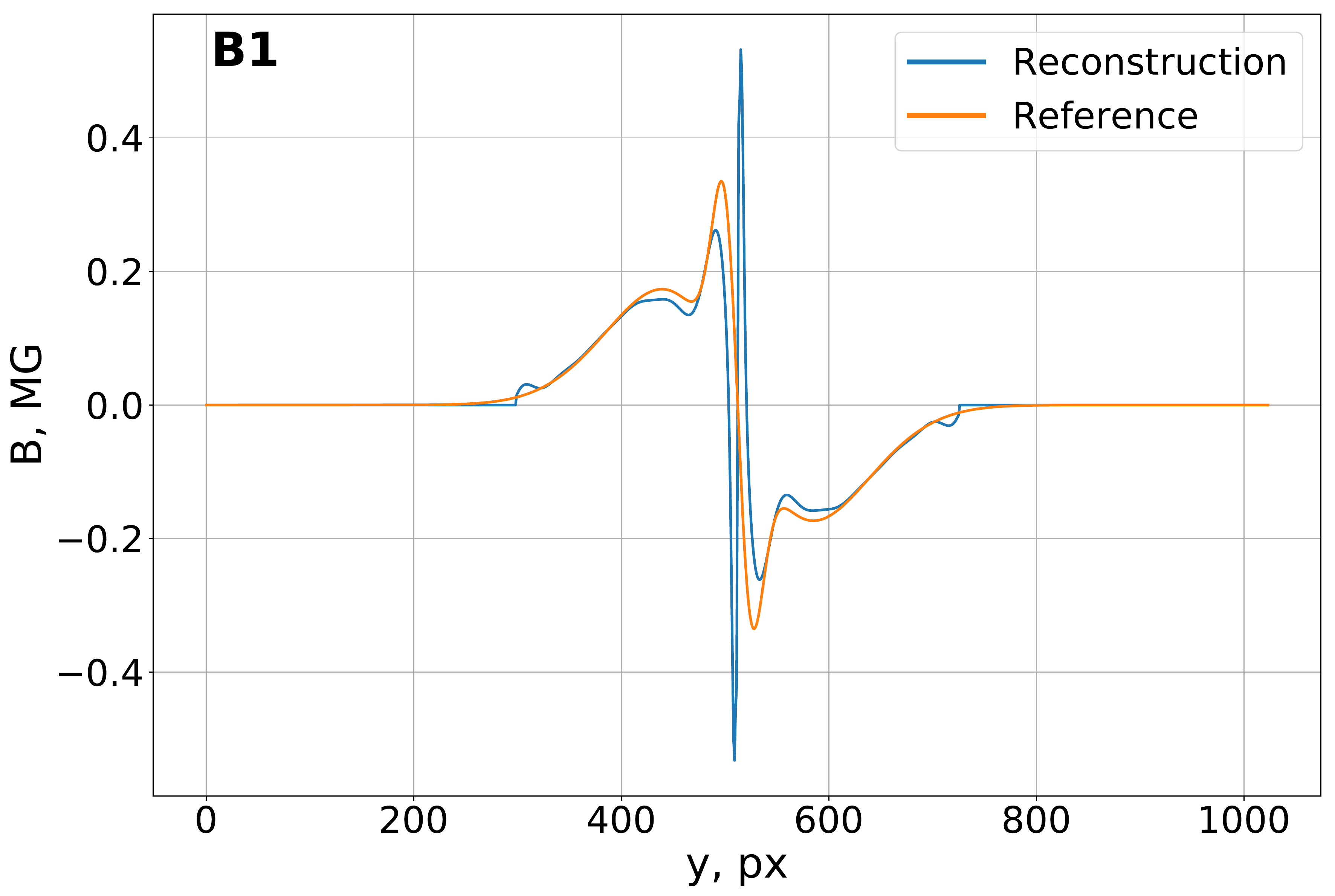}
		\end{subfigure}
		\vspace{1ex}
		\begin{subfigure}[b]{0.99\linewidth}
			\centering		
			\includegraphics[width=0.45\linewidth]{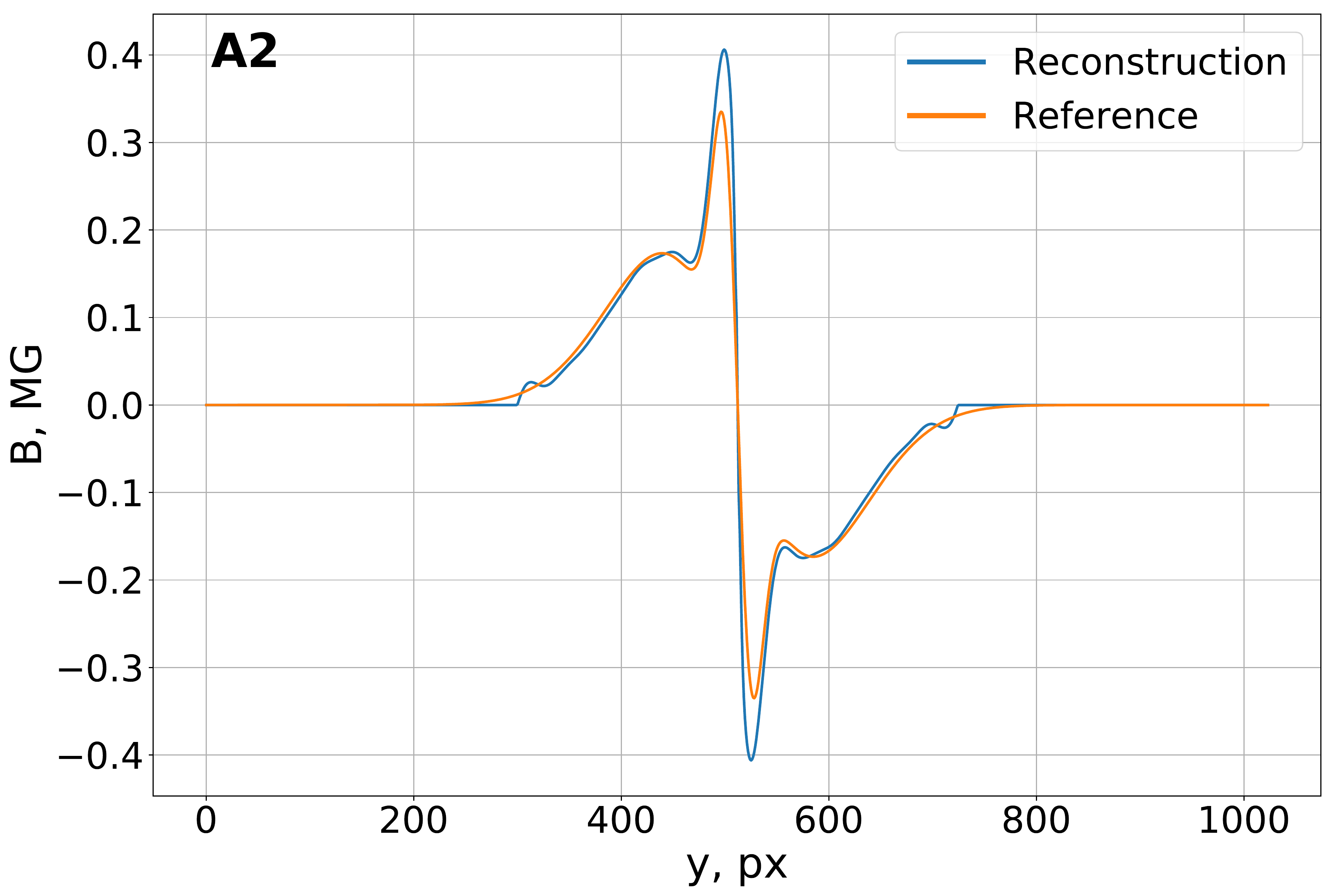}
			\hfil
			\includegraphics[width=0.45\linewidth]{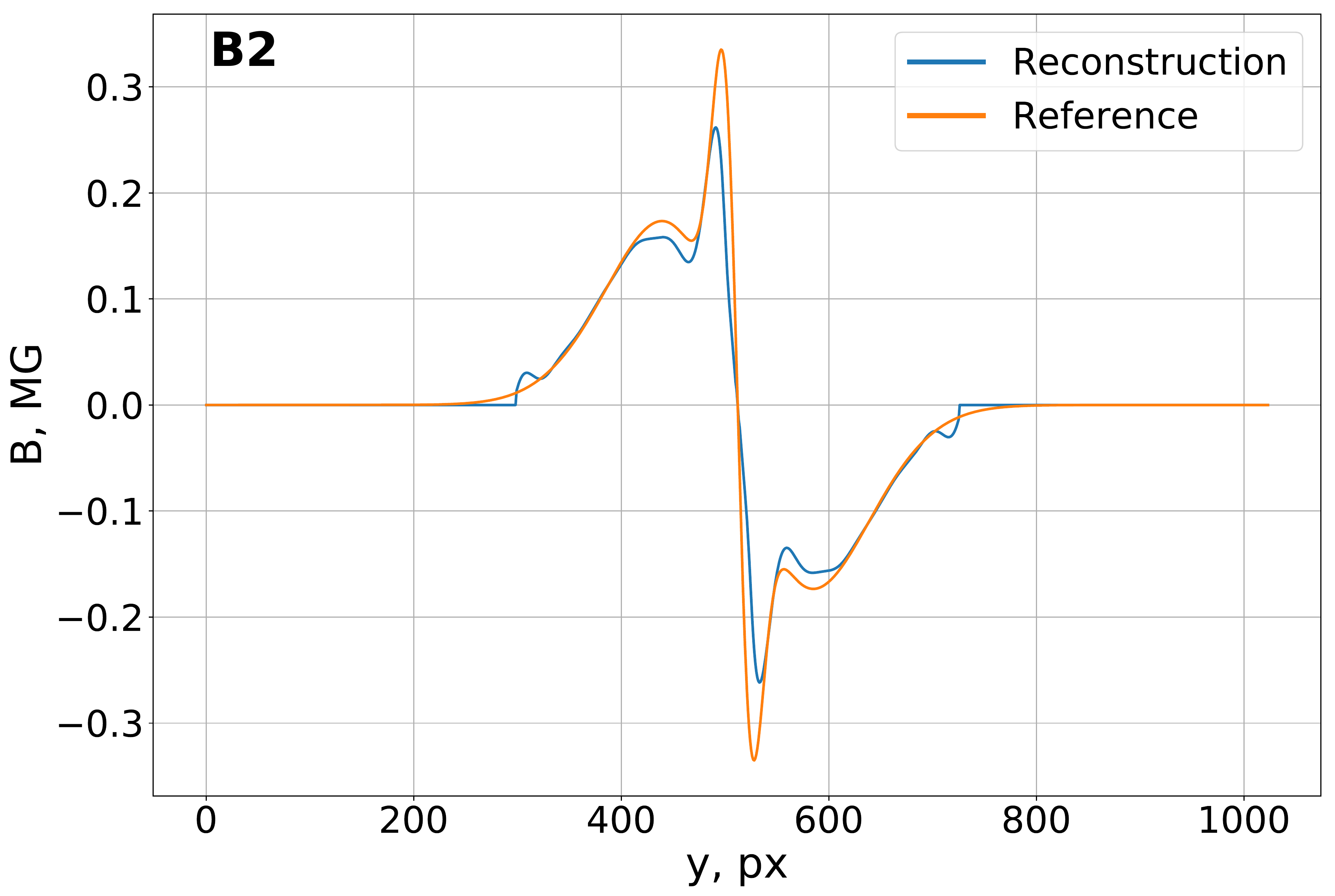}
		\end{subfigure}
		\caption{Calculated B-field with +2 px and -2 px axis offset (A, B), without and with correction (1, 2)}
		\label{img:YesCorr}
	\end{center}
\end{figure*}		

	Choice of the shift function is crucial when using the 3rd option. In case of small axes offset all functions will lead to good results, but it turns out that in more harsh experimental-like conditions the $\varphi$ curve inclination near the axis changes too much, which after $\varphi/y$ and especially the Abel inversion calculations can lead to non-satisfactory results. Linear correction may alter the $f_{\varphi}$, and subsequently the SMF, behaviour near the axis, while using the Gaussian shift function may even change the field sign. Among the considered shift functions the exponential one leads the best SMF reconstruction (see Fig. \ref{img:correction methods}). That is why in all further calculations the exponential shift function was used.

	The plasma axis error (offset from the supposed real axis) can't be corrected due to the nature of the data and will lead to non-zero field error even after any polarization plane rotation angle correction.	Applying chosen method to the cases with  $\mathbf{V_{int}}$,  $\mathbf{V_{noise}}$,  $\mathbf{G_{shift}}$ and  $\mathbf{G_{angle}}$ with $y_{plasma}=y_{real}$ leads to the good reconstruction of the SMF near the symmetry axis(see Fig. \ref{img:corr1} and \ref{img:rotcorr}). 
	
	For more similarity with experimental data, and to check that proposed correction method won't influence the real peak that may be present near the axis, we made a new set of synthetic complex interferograms using analytical density with a gap near the axis,  and magnetic field distribution with peaks near the axis (see Fig. \ref{img:SimData}).

	During simulations we intentionally set an offset of +2 and -2 pixel between $y_{plasma}$ and $y_{real}$ . Measured angle has, as designed, non-zero value at plasma axis, and calculated with such a data magnetic field peak value can be two times bigger than the reference one (Fig. \ref{img:YesCorr} A1) or have an abrupt sign change near the axis(Fig. \ref{img:YesCorr} B1) 
	With angle correction it can be seen that field maximum value became much closer to the reference one. In this case, the consequence of plasma axis error can be clearly seen. Depending on relative position of the axes the 2 pixels offset leads to the reconstructed field being approximately 10\% smaller or greater than intended.

	\section{Error Analysis}
	
	\subsection{Mathematical formulation}
	Here we will estimate the magnetic field calculation error for the simulated data for cases with and without artificial uncertainties.	
	The absolute and relative errors of determining the magnetic field using equation (\ref{eq:B(r)}) is given by
	
	\begin{equation}\label{eq:errorBabs}
	\Delta B=\frac{1.7\cdot10^3}{\lambda}\sqrt{\left(\frac{r\Delta f_{\varphi}}{f_{\delta}}\right)^2 + \left(\frac{r f_{\varphi}\Delta f_{\delta}}{f_{\delta}^2}\right)^2 },
	\end{equation}
	
	\begin{equation}\label{eq:errorB}
	\left(\frac{\Delta B}{B}\right)^2=\left(\frac{\Delta f_{\varphi}}{f_{\varphi}}\right)^2+\left(\frac{\Delta f_{\delta}}{f_{\delta}}\right)^2.  %+\left(\frac{\Delta r}{r}\right)^2+
	\end{equation}
	
	To understand the dependence better we will examine each term of this expression individually. The $f_{\delta}$ and $f_{\varphi}$ represent transformed using Abel inversion $\delta$ and $\frac{\varphi}{y}$ functions respectively. 
	Using equation (\ref{eq:abel_inv}), we can write that error of inverse Abel transformation equals Abel inversion of the error of function under abelization
	
	\begin{equation}\label{eq:errorB1}
	\begin{aligned}
	\Delta f(r)=-\frac{1}{\pi}\int_{r}^{R} \frac{d\Delta F}{dy} \frac{dy}{\sqrt{y^2-r^2}} + \frac{\Delta F(R)}{\pi \sqrt{R^2-r^2}}\\
	= \mathcal{A}^{-1}(\Delta F(y)).
	\end{aligned}
	\end{equation}
	Thus the Abel inversion introduces a modifying coefficient to the resultant  error
	
	\begin{equation}\label{eq:errorBexpand}
	\left(\frac{\Delta B}{B}\right)^2=\left(\frac{\mathcal{A}^{-1}(\Delta\frac{\varphi}{y})}{f_{\varphi}}\right)^2+\left(\frac{\mathcal{A}^{-1}(\Delta\delta)}{f_{\delta}}\right)^2.
	\end{equation}
	
	The phase-shifting interferometry is a very accurate diagnostics. The quality of the interferogram depends primarily on the spatial and temporal coherence of the interfering laser beams and the type of interferometer. The choice of the right probe beam wavelength is also of a great importance. The coherence problem can be solved by using lateral shift interferometry, where the beam interferes with it's shifted copy at small distance from the beam spitting element. The beam shift should be optimal to measure density instead of it's gradient and still have good spatial coherence.
	The errors are also caused by vibrations, laser frequency instability and bad interfringe pattern visibility due to stray reflections, plasma self radiation, and laser intensity instability. Use of ultra-fast probe beam can fix the first problem, good control over laser parameters and additional interferometric filters can reduce contribution of other factors. Note that the implementation of the complex interferometry setup presented in \cite{Pisarczyk2019, AgaArt} corresponds to this situation. Thus the interferometry accuracy is limited mainly by the limited resolution of the CCD and depends on spatial frequency of fringes. Taking half of the pixel value as an absolute error, we can convert it into the  phase uncertainty in units of phase using known width of a interference pattern. As the plasma density gradient increases, the fringes width decreases and the accuracy becomes worse. 
	However, this is not yet the error of the phase shift but error of the phase from one interferogram (reference or shot). As the phase shift is $\delta=\delta_{probe}-\delta_{ref}$, the resultant error can be written as follows  
	\begin{equation}\label{eq:errorPhase}
	\Delta\delta=\Delta(\delta_{ref}-\delta_{probe})=\Delta\delta_{ref}+\Delta\delta_{probe}\approx2\Delta\delta_{probe}.
	\end{equation}
	The relative phase error can be then calculated by division of $\Delta\delta$ by the measured phase shift. 
	
	In approximation of constant interference pattern width the the absolute value of phase uncertainty is constant over the entire interferogram area, and we can easily calculate the error of it's Abel inversion:
	\begin{equation}\label{eq:errorPhase_abel}
	\Delta f_{\delta}(r)=\frac{\Delta\delta}{\pi R \sqrt{1-(r/R)^2}},
	\end{equation}
	where $r/R$ is a radius, normalized to maximum calculation radius R.

	Considering the absolute error of $\varphi/y$ ratio is a bit more challenging.
	To estimate it, consider three axes: $y_{real}$, $y_{plasma}$ and $y_{angle}$. The first one can not be determined experimentally and contributes to the unavoidable error, the second one is calculated from plasma phase distribution but due to imperfect symmetry of plasma there is some uncertainty in it's position, and the last one, extracted from polarization plane rotation angle distribution, is influenced by a set of inherent factors and is therefore determined with an error. Even a small mismatch of these axes can lead to noticeable error in SMF calculations.

\begin{figure}[H]
		\centering
		\begin{subfigure}[b]{0.99\linewidth}
			\centering
			\includegraphics[width=0.99\textwidth]{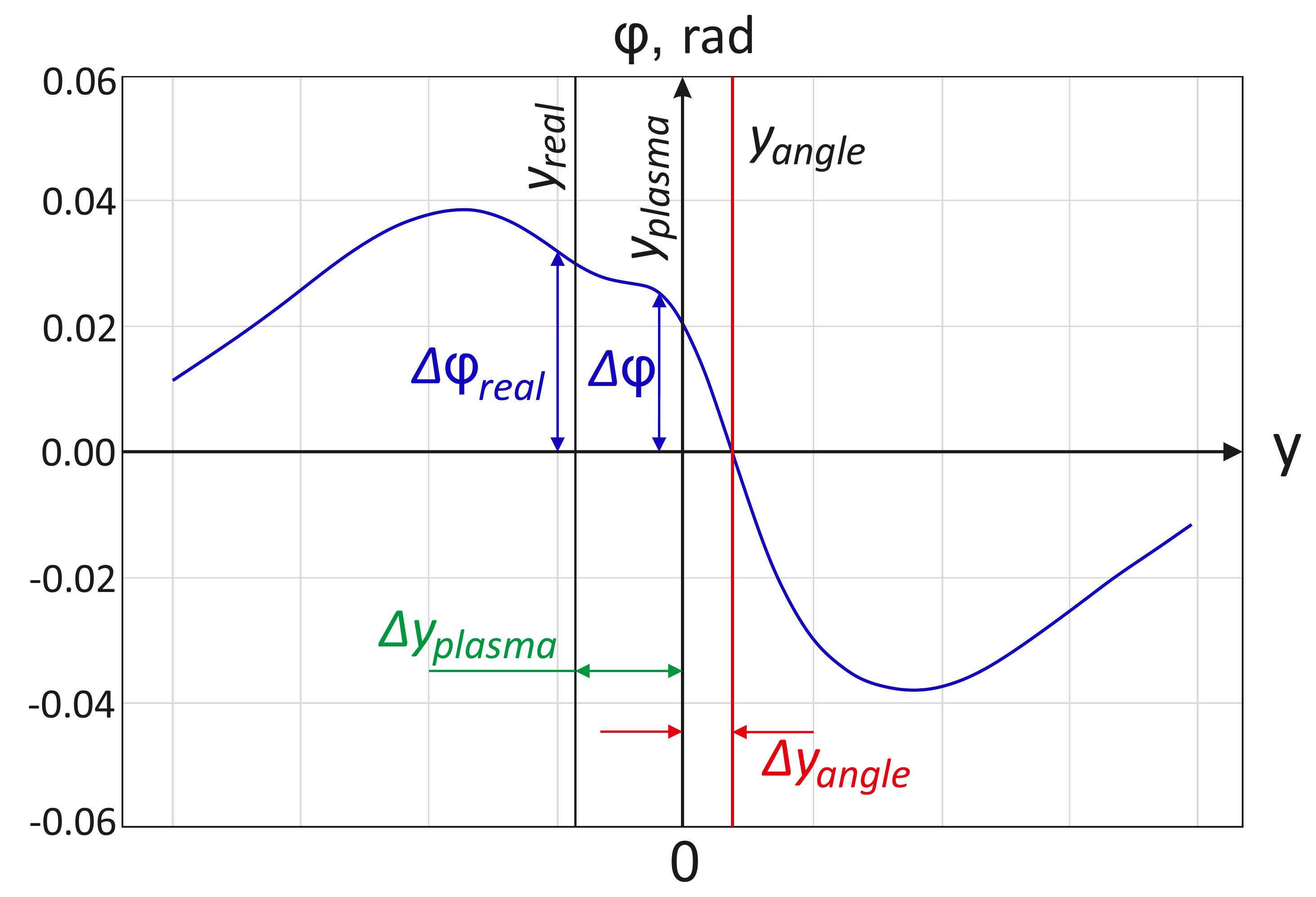}\\[2mm]  
		\end{subfigure} 
		\caption{Mismatched axes - experiment-like case (illustrative).}
		\label{img:all cases}
	\end{figure}

	Next let us consider the most general case of relative axes positions - mismatched axes (Fig. \ref{img:all cases}) and calculate the error of $\varphi/y$. Here $\Delta y_{plasma}$ is plasma axis offset from the real one,  $\Delta\varphi$ and $\Delta\varphi_{real}$ are the angle values at the plasma and at the real axis positions respectively, $\Delta y_{angle}$ is an offset between plasma axis $y_{plasma}$ and angle axis $y_{angle}$. Note that the error will be calculated in assumption that there are no other uncertainties during the calculations, i.e. the angle is ideally recreated from synthetic interferograms, the fringe contrast is good enough for accurate phase detection. 
	
	The direct calculation of $\varphi(y)/y$ in respect to the chosen plasma axis, will lead to big error near this axis as was previously showed in this paper. The correct calculation way (impossible for real experimental data) would be to subtract background, resulting in $\varphi$=0 at real axis position, and shift plasma axis to the real axis. The correct formula reads:
	\begin{equation}\label{eq:error5}
	F(y)=\frac{\varphi(y)-\varphi_{bg}(y)}{y+\Delta y_{plasma}},
	\end{equation}
	where the $\varphi_{bg}(y)$ is an function disturbing the measured angle due to all previously the discussed effects. 	
	
	The obtained function $F(y)$ is shifted in respect to the directly calculated $\varphi(y)/y$, thus make a comparison and calculate the error we need to shift it back:
	
	\begin{equation}\label{eq:error5_1}
	\begin{aligned}
	\Delta \frac{\varphi(y)}{y} = \frac{\varphi(y)}{y}-F(y-\Delta y_{plasma})\\
	=\frac{\varphi(y) - \varphi(y-\Delta y_{plasma})+\varphi_{bg}(y-\Delta y_{plasma})}{y},
	\end{aligned}
	\end{equation}
	and here we know $\varphi_{bg}(-\Delta y_{plasma})=\varphi(-\Delta y_{plasma})=\Delta\varphi_{real}$ and $\varphi(0)=\Delta\varphi$. From this equation it can be seen, that the error on chosen plasma axis is equal to 
	
	\begin{equation}\label{eq:error_inf}
	\Delta \frac{\varphi(y)}{y}[y=0] = \lim_{y\to 0}\frac{\Delta\varphi(y)}{y},
	\end{equation}
	
	This means that the error near the calculation axis in general tends to infinity, and only in case of coincident $y_{plasma}$ and $y_{angle}$ is limited by some constant value. This  limited error value has a direct relationship with plasma axis determination error and tends to zero when measured axes tends to the real axis. 
	
	To lower the error in case of mismatched axes, a correction of the angle distribution can be performed. This will force $\varphi(0)$ and $\Delta\varphi(0)$ to be zero, superposing plasma and angle axes and therefore turning infinity-like error into limited one. However, as the real axis position is a mathematical abstraction and can not be determined, some finite error will always remain. 
	
\begin{figure}[H]
		\centering
		\begin{subfigure}[b]{0.99\linewidth}
			\centering
			\includegraphics[width=0.99\linewidth]{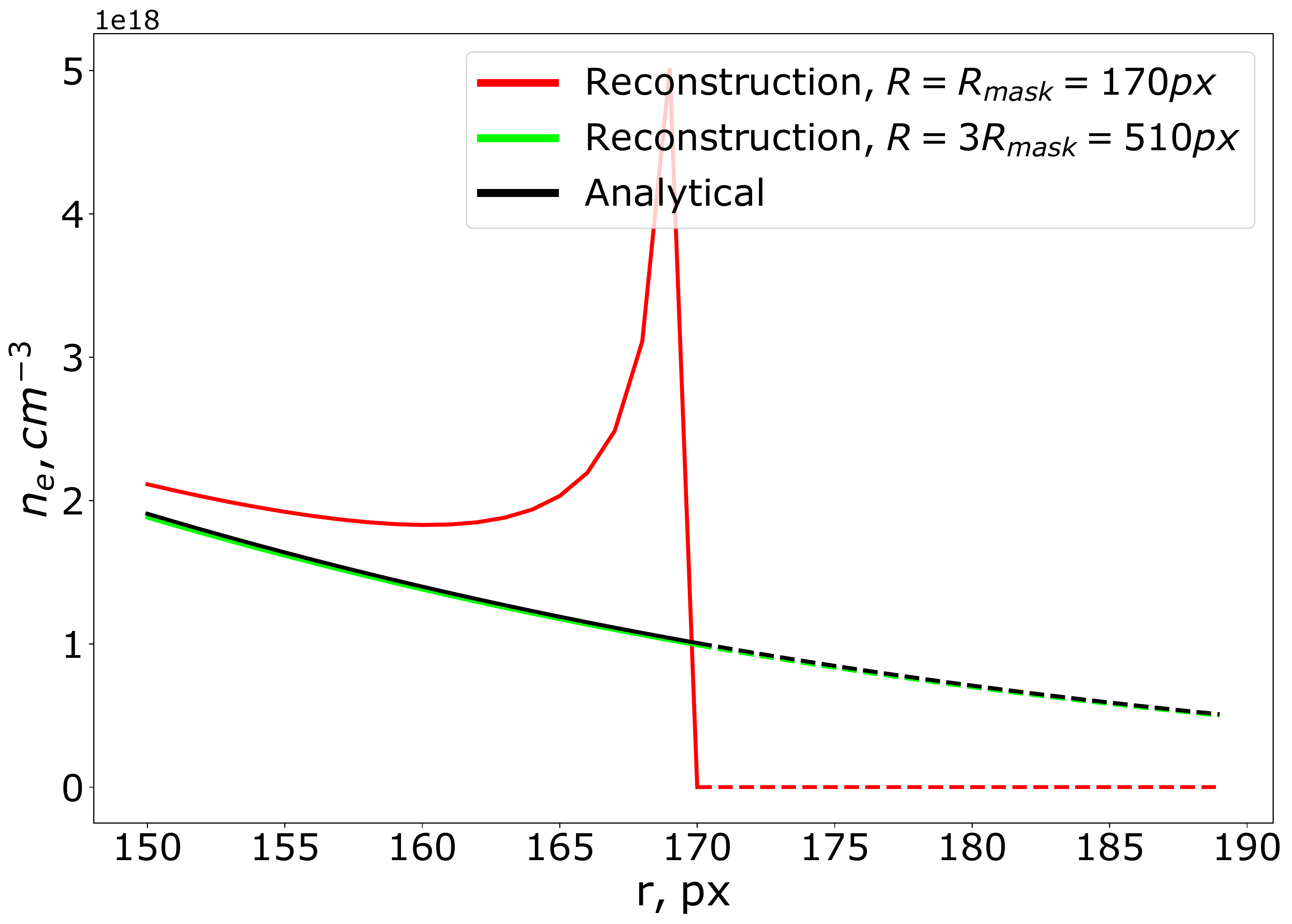}\\[2mm]  
		\end{subfigure} 
		\caption{Density reconstruction using different maximum radius R.}
		\label{img:Abel_radius}
	\end{figure}

	From equation (\ref{eq:errorB1}) it is seen that the error of an Abelized function depends primarily on the error distribution. Considering the constant background $\Delta\varphi$, used to create artificial complex interferograms, we can write:
	
	\begin{equation}\label{eq:errorAngle_abel}
	\begin{aligned}
	\Delta f_{\varphi}=
	\frac{1}{\pi}\int_{r}^{R} \frac{\Delta\varphi}{y^2} \frac{dy}{\sqrt{y^2-r^2}} + \frac{\Delta\varphi}{\pi R \sqrt{R^2-r^2}}\\
	=\frac{R\Delta\varphi}{\pi r^2 \sqrt{R^2-r^2}}=\frac{\Delta\varphi}{\pi R^2 \tilde{r}^2 \sqrt{1-\tilde{r}^2}},
	\end{aligned}
	\end{equation}
	where $\tilde{r}=r/R$ is the normalized radius. This equation is valid only for constant background and in case of introduced correction the inverse Abel of $\Delta(\varphi/y)$ should be calculated to receive the error distribution.
	In experimental case the error behaviour may change significantly however usage of the proposed approach can help us to evaluate the error in the best-case scenario.
	From Eq. \ref{eq:errorPhase_abel} and \ref{eq:errorAngle_abel} it follows that in case of constant measured data absolute error in order to minimize the absolute error after the Abelization, the plasma radius R should be maximized. On the other hand, even choosing the maximum possible R value the relative error in some areas may still be too big for the data to be representative. Thus the border  $R_{mask}$ should be applied after all of the calculations to cover areas with an excessive error. The example of the Abelization error for $R=R_{mask}$ compared to $R=3R_{mask}$ is presented at Fig. \ref{img:Abel_radius}.
	
	\subsection{Application to used synthetic data}
	The interference period, needed for error evaluation, is a function of probe and reference beams convergence angle and wavelength. In our simulations it depends on the spatial frequencies which were introduced in Eq. \ref{eq:I} and set to be 12 px.	
	Taking half of the pixel value as an absolute error, we convert it into the phase uncertainty of 0.04 fringes (here and after the phase will be presented in number of fringes, which equals phase[rad]/2$\pi$). This is valid for both cases with and without artificial uncertainties. It is obvious that the phase shift relative error increases in low-density regions, where $\delta={\delta_{ref}-\delta_{probe}}$ tends to zero, therefore during calculations the density and phase data should be abridged by using the  $R_{mask}$ condition. In the  presented calculations, the phase uncertainty is chosen on the level of 0.04 fringes, leading to $\Delta\delta=0.08$ fringes. The plasma axis $y_{plasma}$ determined using the assumed symmetry of plasma distribution is accurate for the synthetic data, but in the experiments due to geometrical error $\mathbf{G_{angle}}$ has an uncertainty on the order of a few pixels.
	
	The error analysis was carried out at one chosen cross-section ($z$=50 px). All calculations were carried out for maximum radius R=512 pixels laying at the border of the image, while for this cross-section the  $R_{mask}$, corresponding to electron density value $n_e=10^{18} cm^{-3}$, is 170 px. Calculating the  $f_{\delta}$ relative error taking use of Eq. \ref{eq:errorPhase_abel}
	we see that it does not exceed 5\% at $r=R_{mask}$ (Fig. \ref{img:Phase and phi Error}A).
	
	In the experiments, however, due to sharp plasma density gradients the interference period can be as small as a few pixels, which leads to much higher local phase uncertainty. This is one of the reasons why application of the presented here error analysis methods to real interferograms may lead to error underestimation. The other reason is that the angle can't be perfectly extracted from the complex interferometry data due to Fourier filtering method limitations and filtering window choice uncertainty. Using the synthetic data, scanning though all the possible parameters and calculating the mean squared error between the extracted angle and analytical one the optimal window (34x44 pixels) for the studied cross-section ($z$=50 px) was found. The polarization plane rotation angle error near the axis reached 10\% (see Fig. \ref{img:Phase and phi Error}B), decreased to the minimum of 0.85\% at $0.3R_{mask}$, that corresponds to the angle extrema position, and then started to rapidly increase after $0.8R_{mask}$, approaching 100\%. The non-ideality of the extracted angle at the axis plays an important role as it may be magnified after division by the coordinate and subsequent Abelization.
	
	We will consider following cases: 	
	\begin{enumerate}

		\item Without artificial uncertainties. The only error here, the origin of which is non-ideal data extraction from complex interferogram ($y_{angle}\approx y_{plasma}=y_{real}$), may be increased during post-processing.
		\item With constant non-corrected background of 0.01 rad (about 20\% of maximum angle value) and error in plasma axis determination. Here we  distinguish two sub-cases: 
		\begin{enumerate} 
			\item The angle and plasma determination errors compensated and the axes overlapped, being shifted from the real axis ($y_{angle} = y_{plasma}\neq y_{real}$). The axes offset resulting from the chosen background level is $\Delta y_{plasma}=\Delta y_{angle}$=7 px (see Fig.  \ref{img:all cases}).
			\item All axes are mismatched	($y_{angle}\neq y_{plasma}\neq y_{real}$). Here $\Delta y_{angle}$=7 px, $\Delta y_{plasma}$=2 px. 
		\end{enumerate} 
	\begin{figure}[H]
		\begin{center}
			\includegraphics[width=0.90\linewidth]{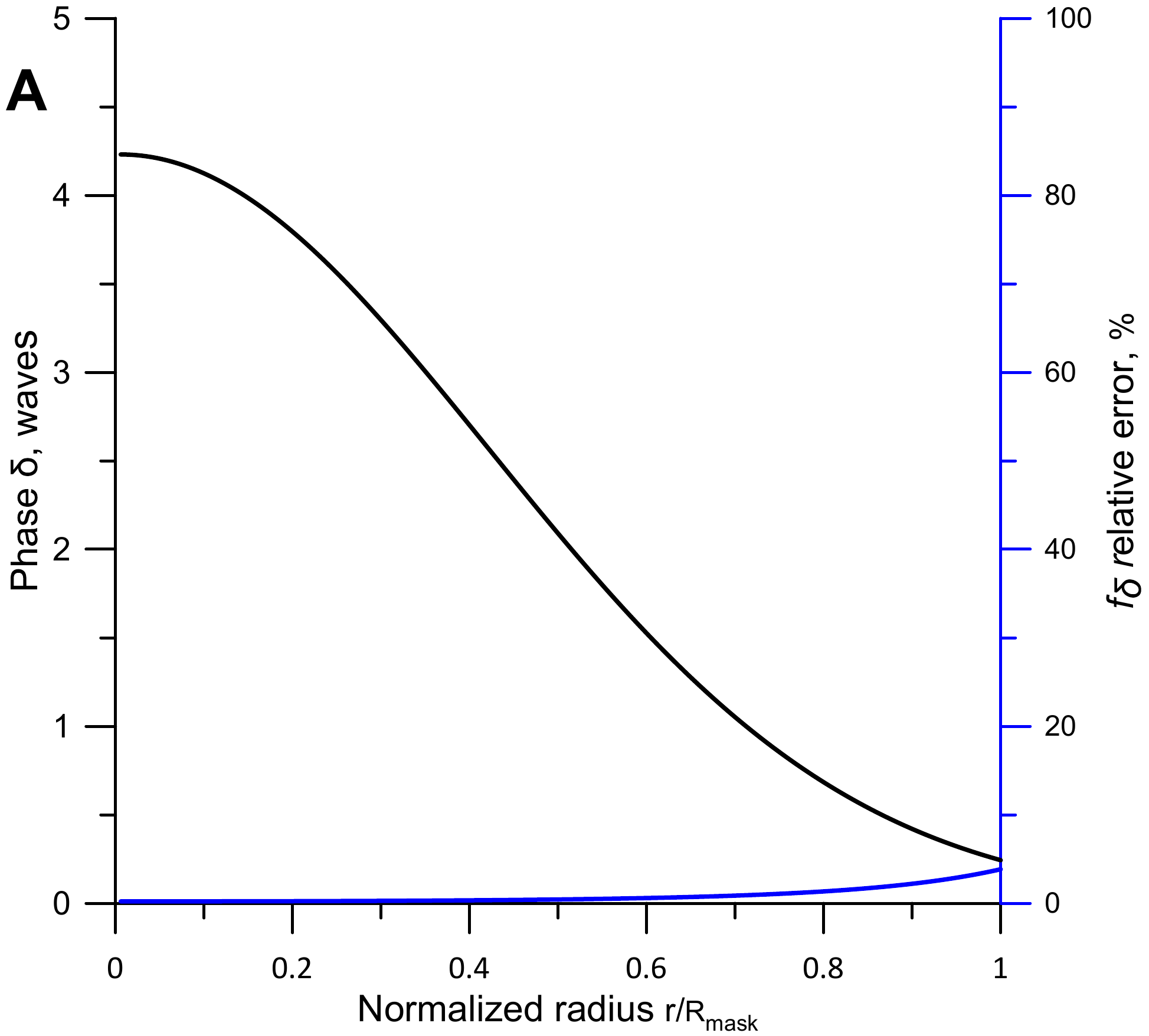}
			\\
			\includegraphics[width=0.90\linewidth]{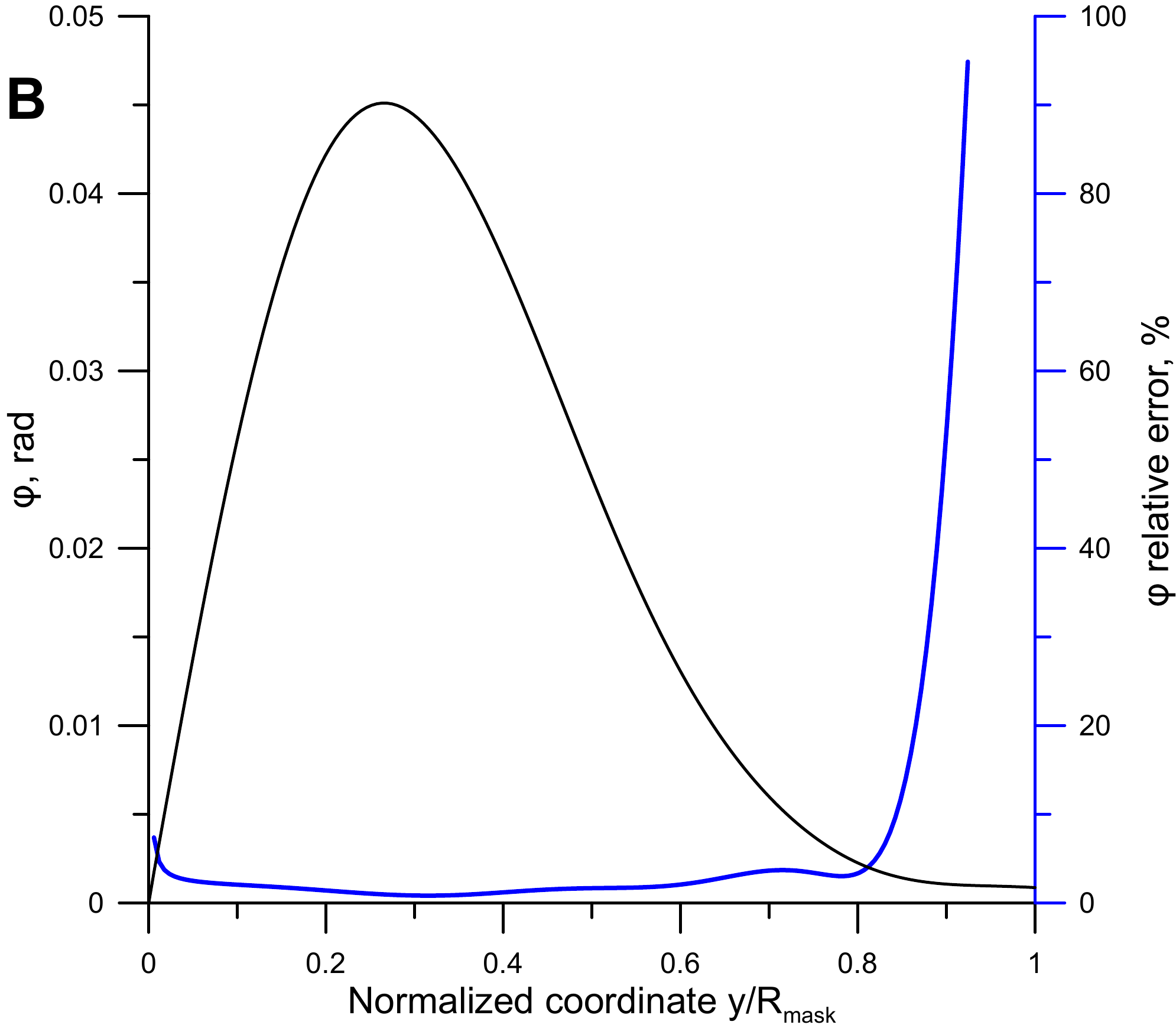}
			\caption{Extracted from synthetic complex interferogram phase and the relative error of its abelization (A); extracted polarization plane rotation angle and it's relative error (B) }
			\label{img:Phase and phi Error}
		\end{center}
	\end{figure}	

\begin{figure*}
	\begin{center}
		\begin{subfigure}[b]{0.9\linewidth}	
			\centering		
			\includegraphics[width=0.43\linewidth]{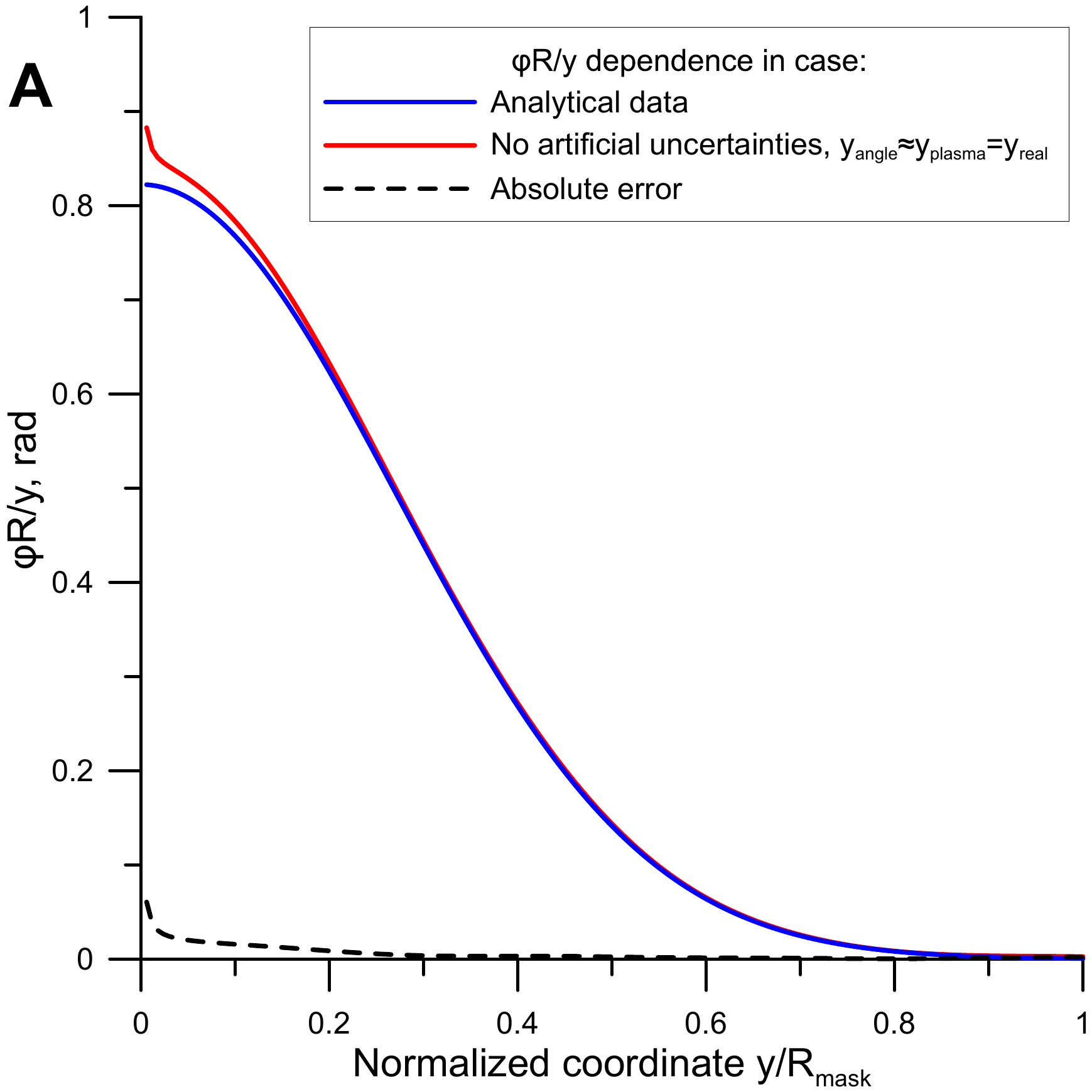}
			\hfil
			\includegraphics[width=0.43\linewidth]{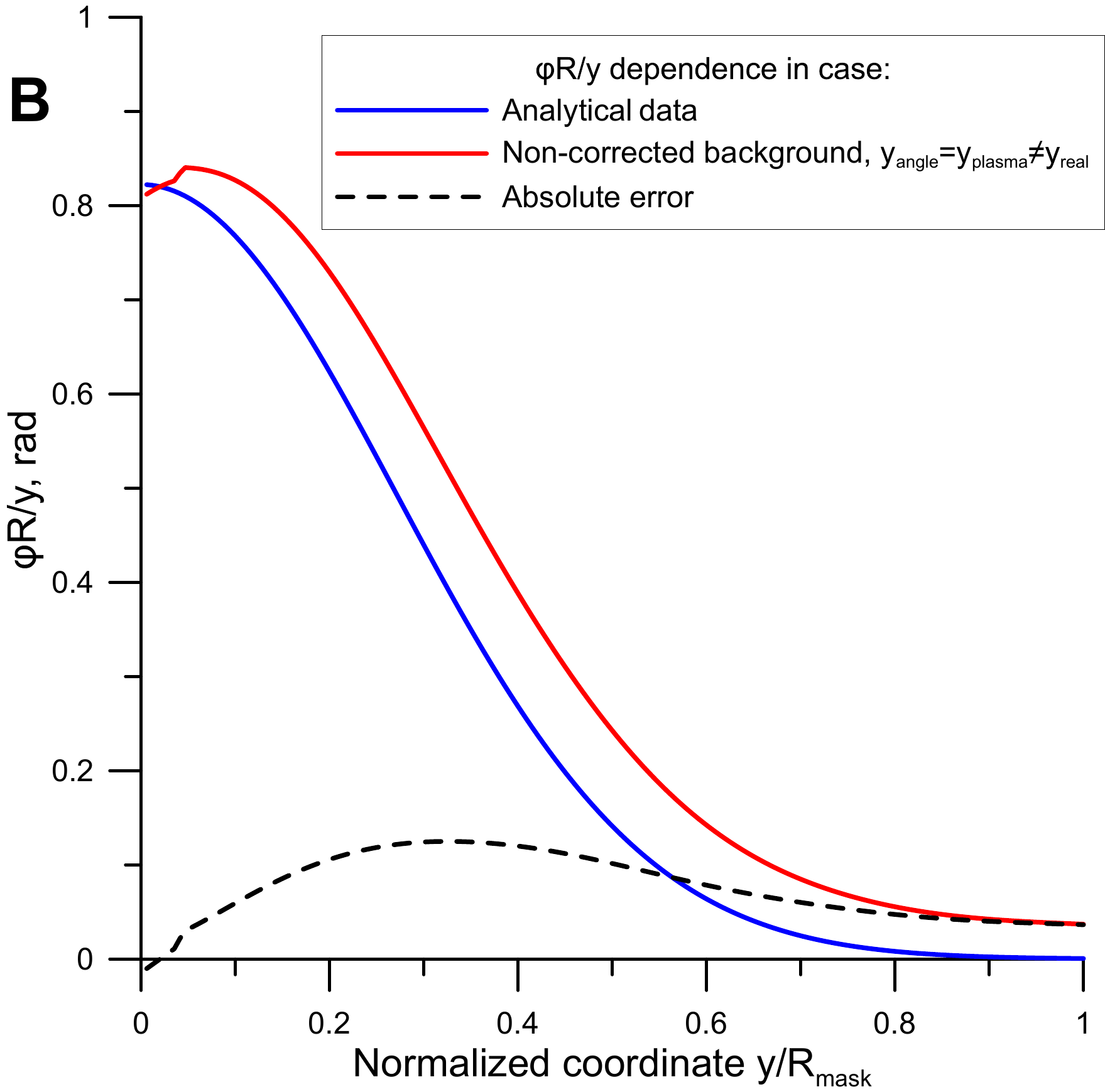}
		\end{subfigure}
		\vspace{1ex}
		\begin{subfigure}[b]{0.9\linewidth}
			\centering		
			\includegraphics[width=0.43\linewidth]{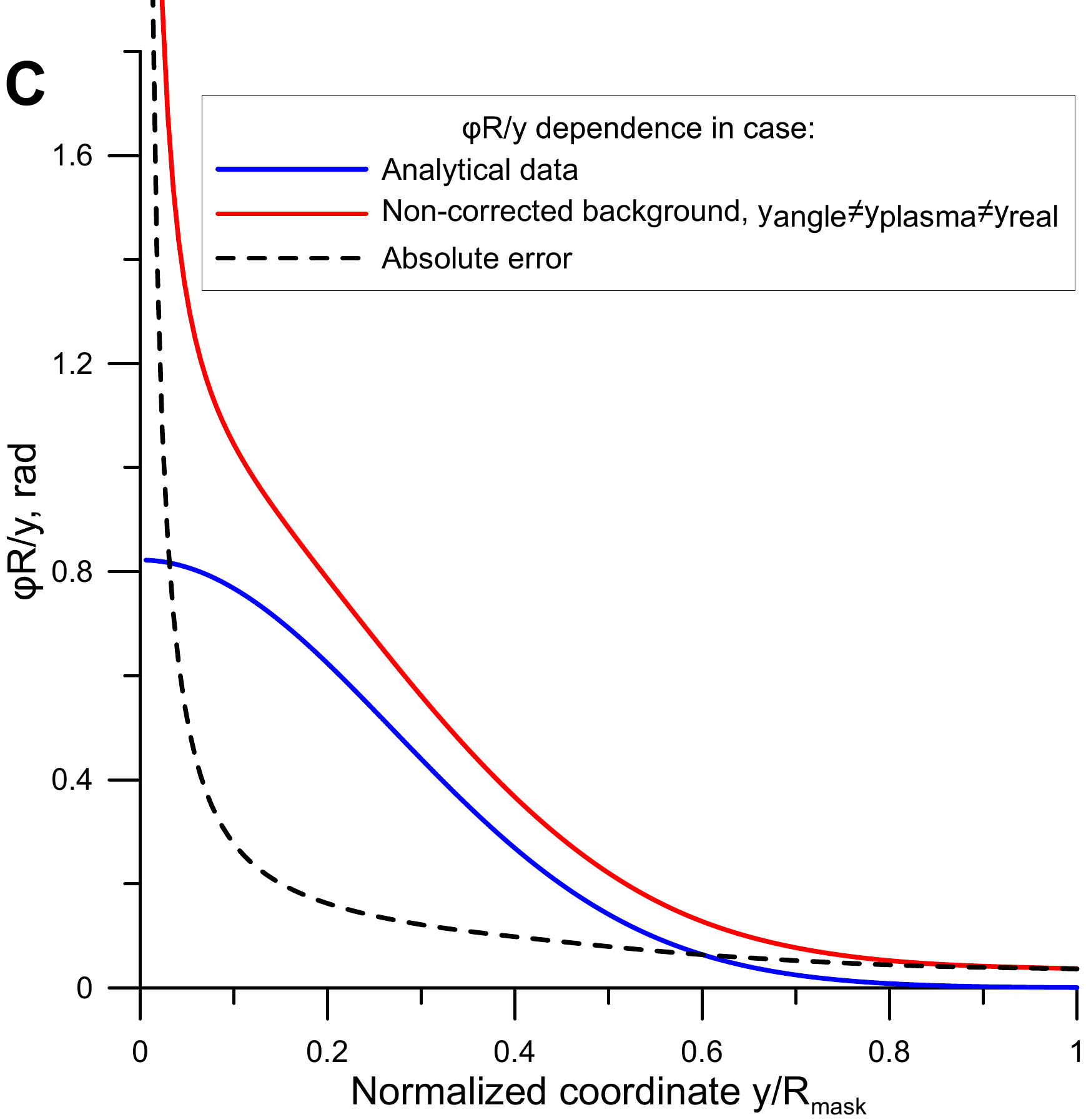}
			\hfil
			\includegraphics[width=0.43\linewidth]{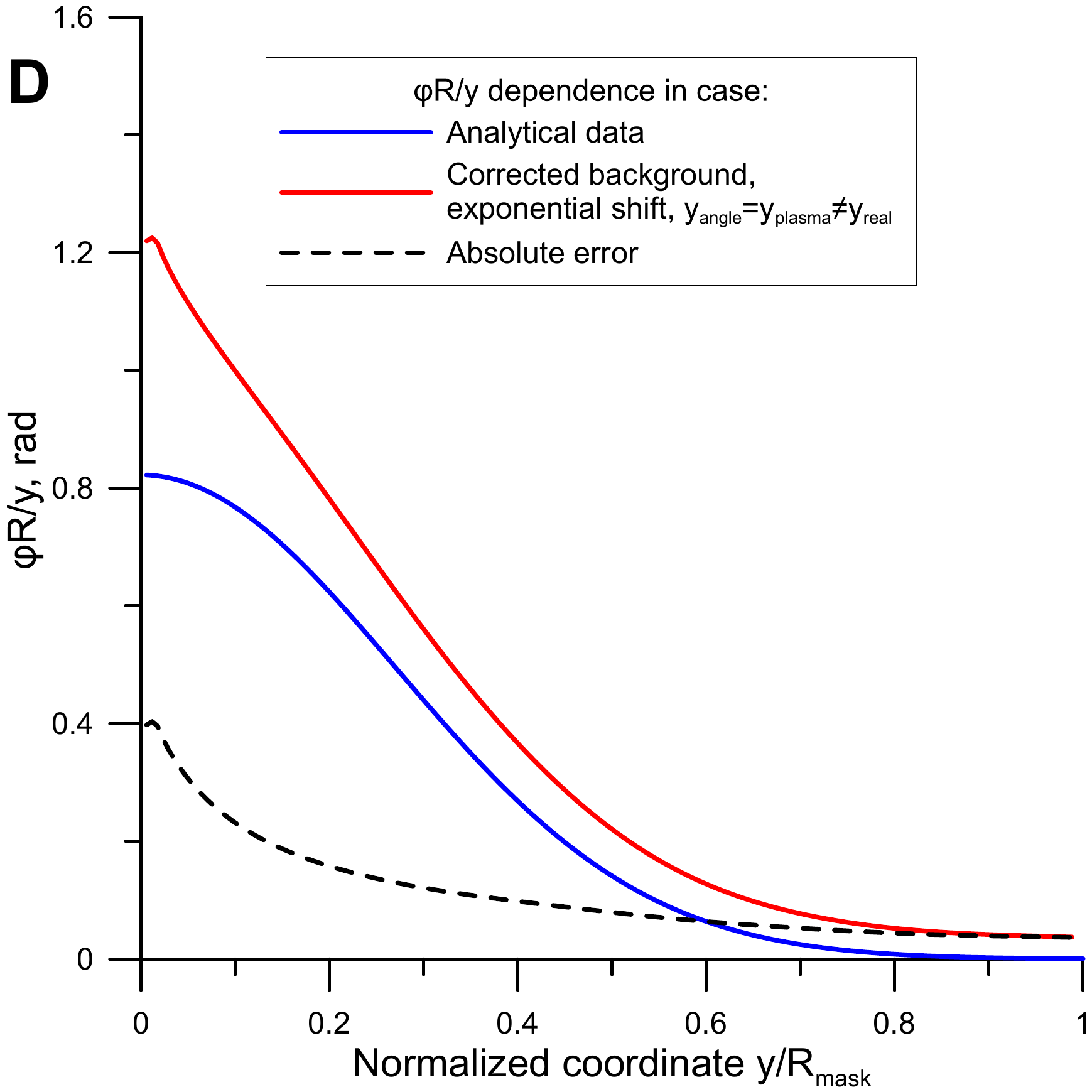}
		\end{subfigure}
		\caption{Axes relative positioning options: extracted from complex interferogram without any artificial uncertainties (A), with background, plasma axis is shifted from the real one to the angle axis position (B), with background, all axes mismatched (C), with background corrected using exponential shift function (D)}
		\label{img:all cases error}
	\end{center}
\end{figure*}

	\begin{figure*}
	\begin{center}
		\includegraphics[width=0.32\linewidth]{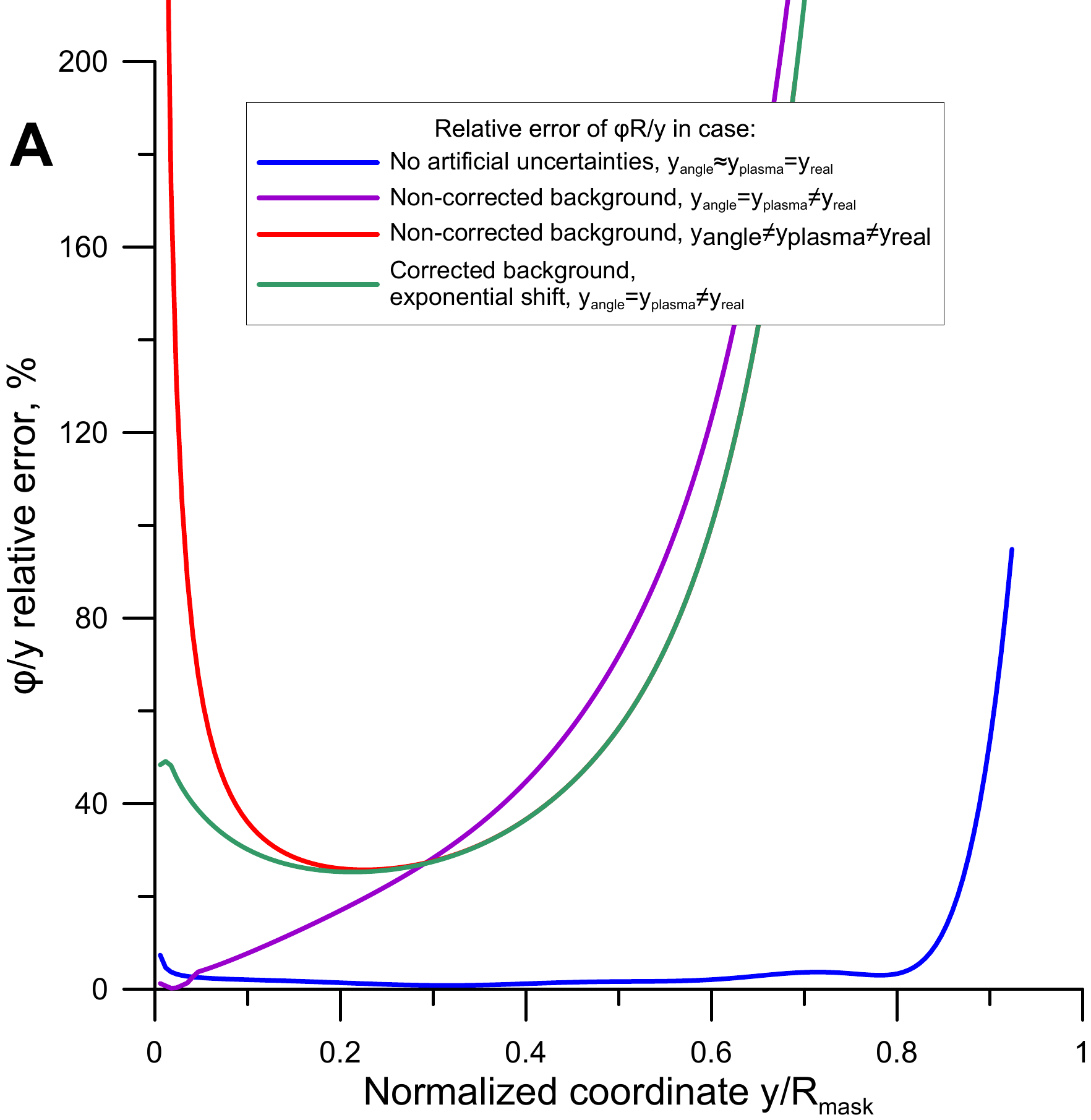}
		\hfil
		\includegraphics[width=0.32\linewidth]{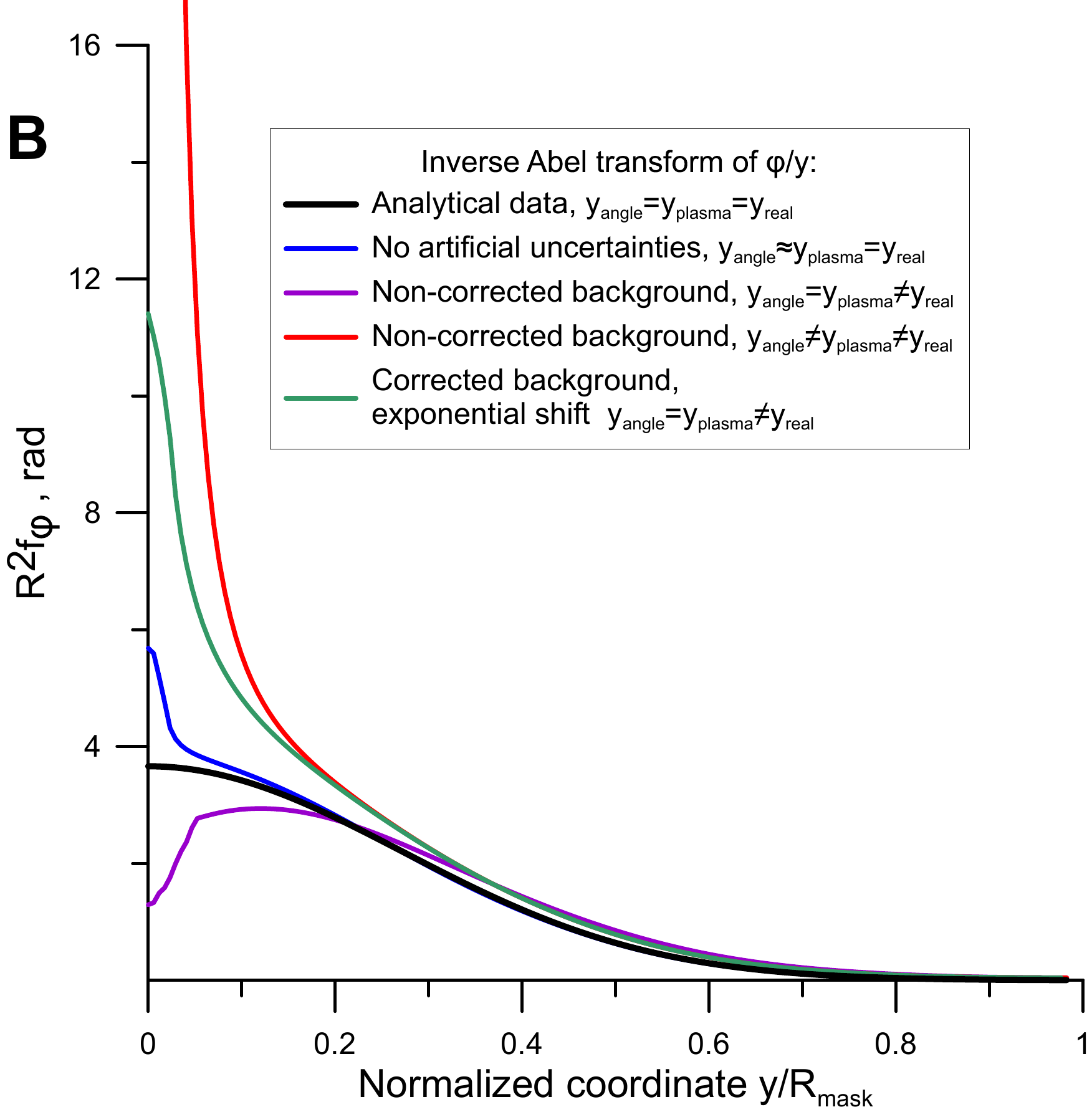}
		\hfil
		\includegraphics[width=0.32\linewidth]{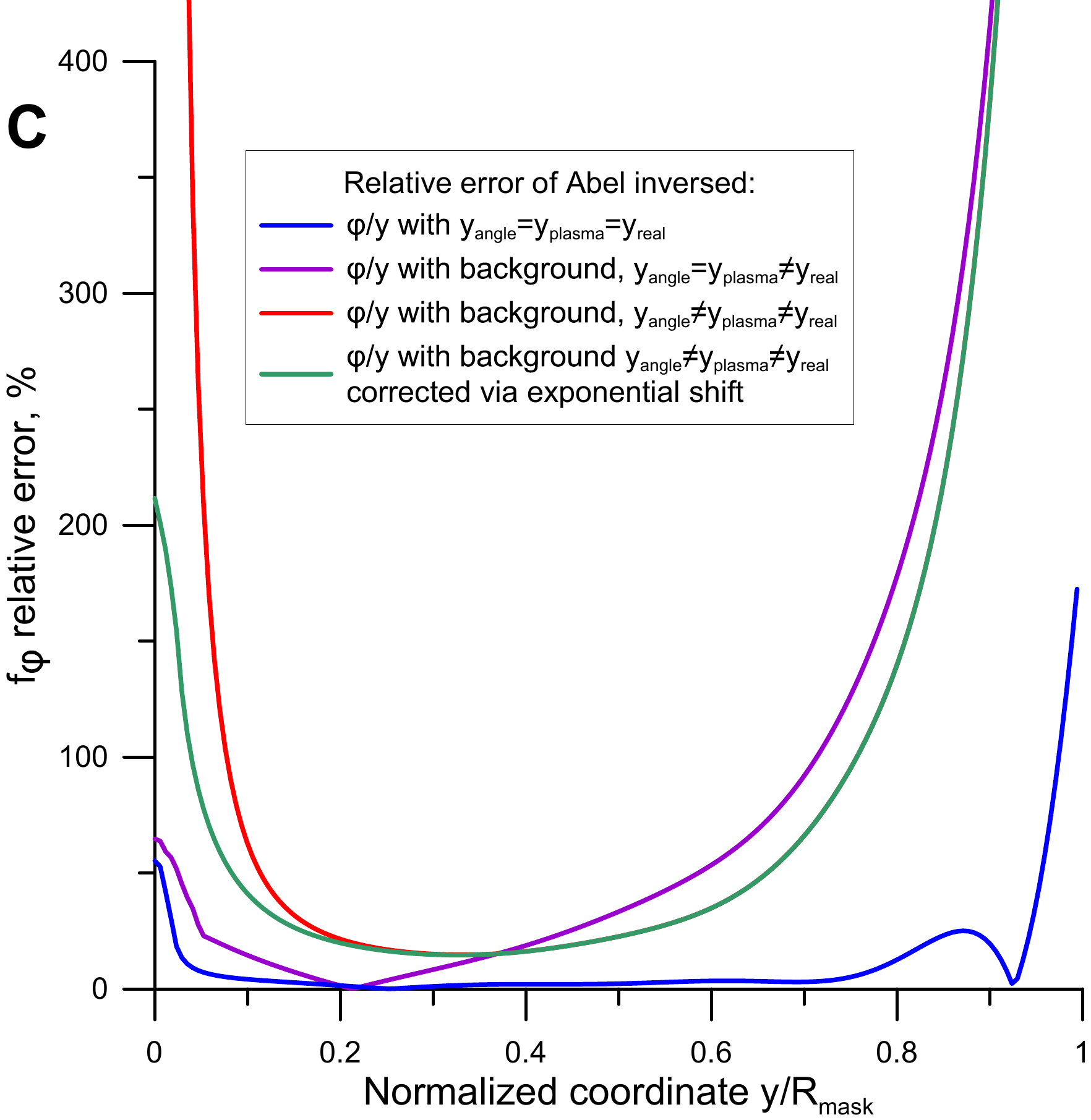}
		\caption{(A) Relative error of $\varphi/y$ functions; (B) Abelized $\varphi/y$ functions and (C) their relative error}
		\label{img:absolute error abel}
	\end{center}
\end{figure*}

		\item  Corrected background case ($y_{angle} = y_{plasma}\neq y_{real}$).	As we are considering Faraday effect and axial magnetic field only, we know that the angle value on the axis should be zero, thus the proposed correction method (described in the previous section) using narrow exponential shift function makes $\Delta\varphi=0$ at the chosen plasma axis. Here $\Delta y_{plasma}$=2 px, the angle is corrected from initial  $\Delta y_{angle}$=7 px to 0 px. The width of the exponential shift function should be less than the minimum measured angle irregularity period (natural or due to laser instabilities), and in here was chosen to be 14 pixels FWHM. 
	\end{enumerate}

	The $\varphi/y$ function behaviour in the region near the axis in all cases differs from the reference. This error is connected with the angle extraction error (Fig. \ref{img:all cases error}A), with the background presence and offset between real, plasma and angle axes (Fig. \ref{img:all cases error}B, C) and with the shift function influence (Fig. \ref{img:all cases error}D).

	The Abelization procedure, itself having a fairly good accuracy tested on analytical functions, increases all already existing errors (compare \ref{img:absolute error abel}A and C) in region adjacent to plasma axis. For example, extracted angle error of 10\% near the axis, which remained constant for $\varphi/y$ calculation, increases to almost 60\% after the Abel inversion; small drop in $\varphi/y$ function due to plasma axis offset from real one (Fig. \ref{img:all cases error}B) leads to Abelized function intensified decrease near the axis (purple line at Fig.\ref{img:absolute error abel}B). The error of non-corrected $f_{\varphi}$ distribution is extremely large close to the axis (several thousands of percent for our resolution), but can be limited to maximum value of about 200\% by exponential correction. 
	On the other hand the relative error for region with $y>0.1R_{mask}$ seems to be only mildly affected by the Abel inversion.  For example, both $\varphi/y$ and $f_{\varphi}$ functions error in case of ideal data reaches drops down to $\approx$2\% in range from 0.2 to 0.8 of $R_{mask}$. 
	
\begin{figure}[H]
		\begin{center}
			\includegraphics[width=0.99\linewidth]{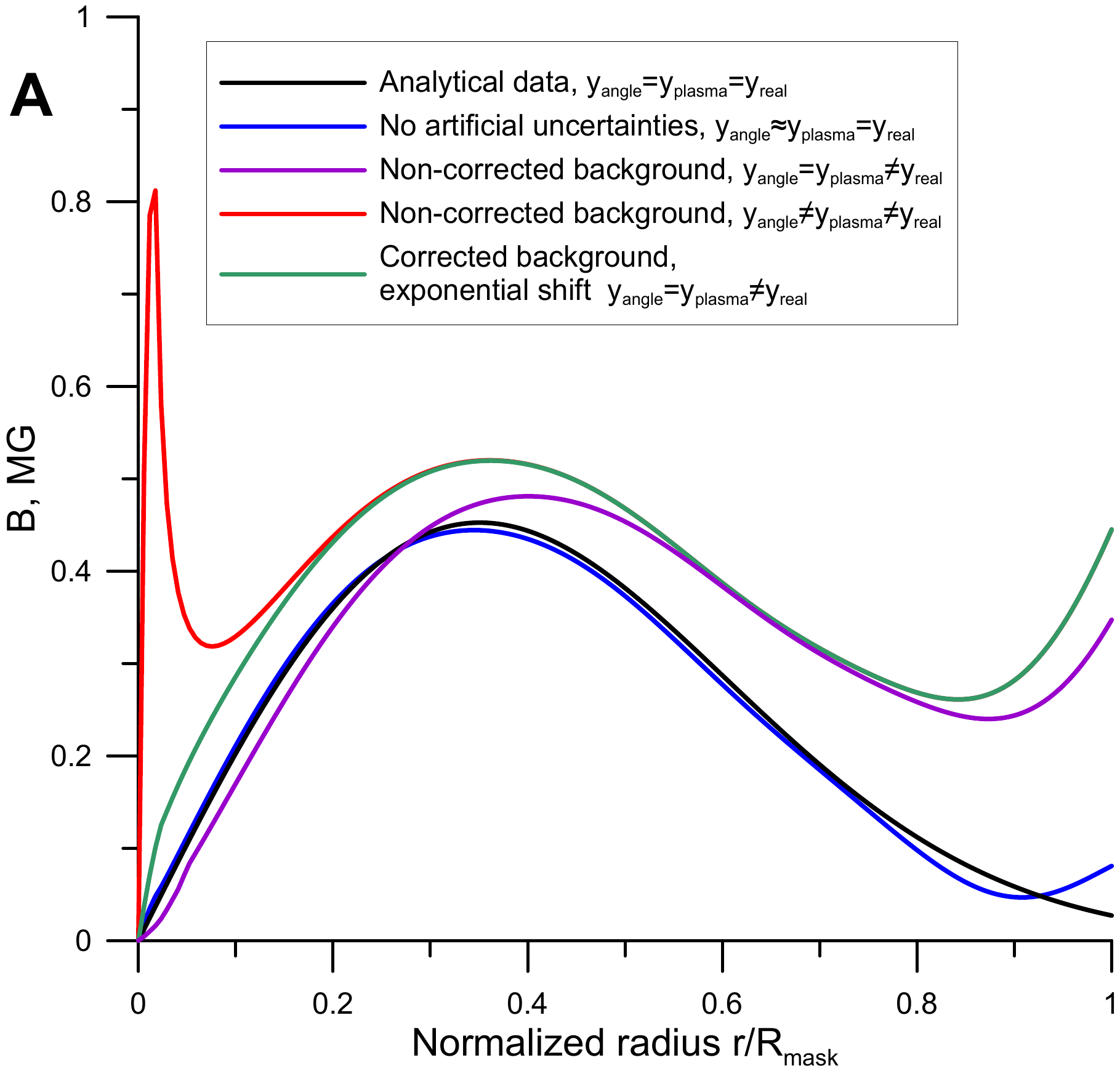} 
			\\
			\includegraphics[width=0.99\linewidth]{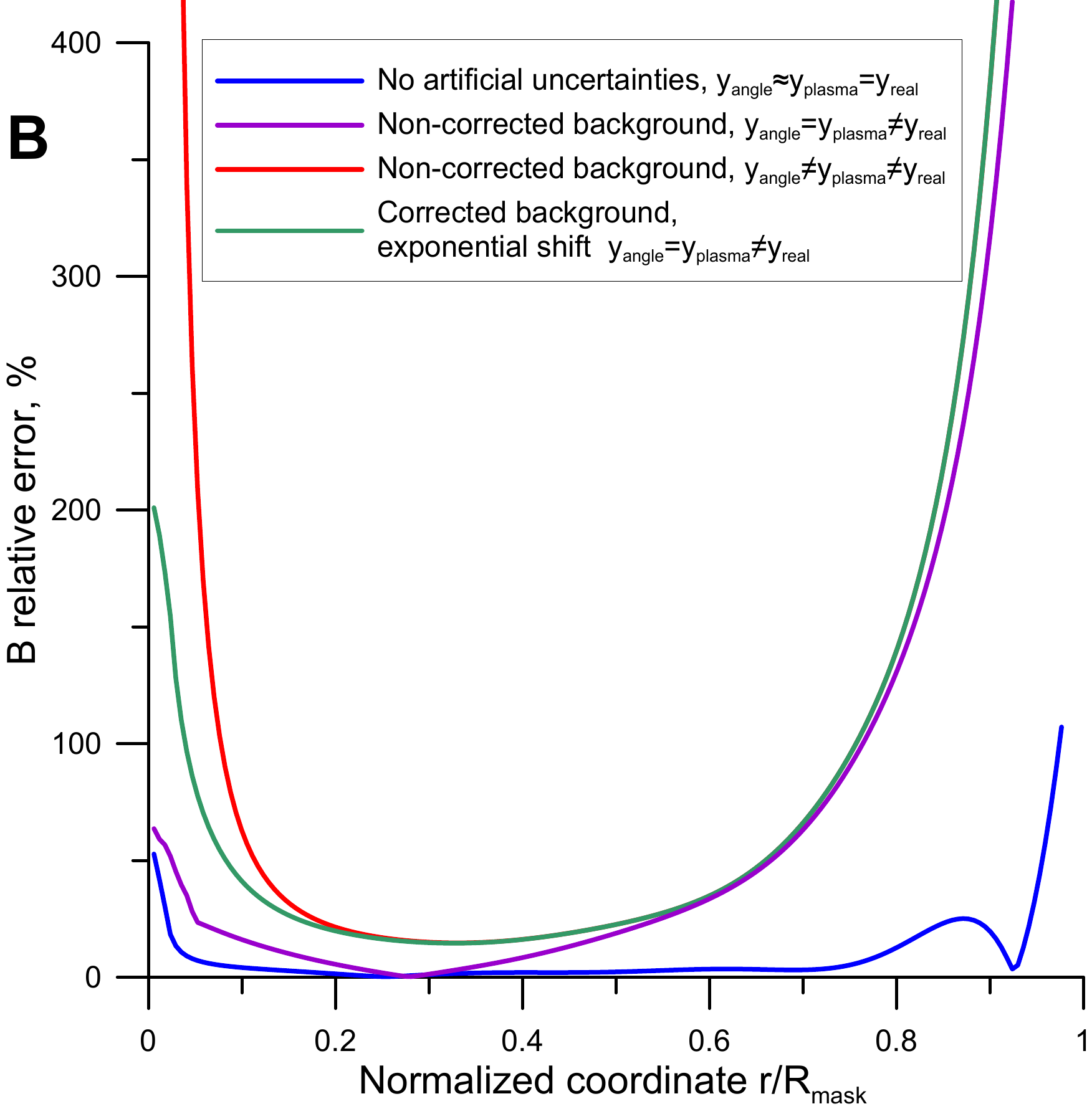} 
			\caption{Magnetic fields distribution (A) and relative error (B) comparison.}
			\label{img:B_abs_value}
		\end{center}
	\end{figure}

	The reference and calculated magnetic fields are presented at Fig. \ref{img:B_abs_value}. It should be noted, that as the magnetic field tends to zero near the axis, formidable value of relative error may leads to only small absolute error.  It can be seen that in the ideal case of absence of any artificial uncertainties the reconstructed field distribution almost coincide with the analytical one near the axis in spite of relative error of 70\%. The field reconstruction error then drops to $\approx$2\% at $r=0.3R_{mask}$, which corresponds to the position of maximum angle value, and starts to gradually increase. The visible separation appears only at $r>0.8R_{mask}$, where the relative error starts to exceed 10\%.

	The field reconstructed from simulated interferograms with constant background, which value was chosen to be similar to maximum of that of experimental images (0.01 rad) demonstrates much higher error near the plasma border and this can not be fixed without modifications of diagnostics system for measurement of the background value. The error has minimum value of 15\% at the same position as in background-free case, $r=0.3R_{mask}$. In case of all mismatched axes the error tends to infinity near the axis, however utilizing the proposed in this paper correction via the exponential shift function the error can be limited to $\approx$200\%.  In this case the field absolute error value near the axis is more than an order of magnitude smaller compared to uncorrected case - from a few MG down to hundreds of kG.

	The field reconstructed from simulated interferograms with constant background with coincident angle and plasma axes shifted from real one, demonstrates surprisingly good field reconstruction accuracy.  The error close to the axis is comparable to that of ideal case, and only after $r=0.5R_{mask}$ reaches the value of fully mismatched axes case. On the other hand, the calculations were carried out for smooth distribution without any distinctive features. Shift of the plasma axis will influence relative position of such features in angle and density distributions thus reducing useful information, that can be extracted from experimental data. However, in case of smooth density distribution, obtained in the experiment, use of such an approach is cogent and will provide more accurate reconstruction near the axis.

	\section{Conclusions}
	
	The complex interferometry came a long way to become a powerful diagnostic tool. Being properly used, it may provide accurate magnetic field distribution measurements in laser plasmas with parameters permitting phase shift and polarization plane rotation detection. However, with the diagnostics universality comes the data processing sophistication and need to evaluate all factors that can affect the measured values and their accuracy. 
	
	In this work, geometrical factors, connected to measured distributions deviations from symmetry, and value factors, connected to various additional effects changing the light intensity on the sensor, were taken into account. It was found out that all the considered factors lead to one common consequence - the main error is located near the assumed symmetry axis and originates from mismatch in measured plasma symmetry axis $y_{plasma}$ and polarization plane rotation angle antisymmetry axis $y_{angle}$. Using synthetic complex interferograms, it was shown that even in absence of any artificial uncertainties due to small antisymmetry axis oscillations because of Fourier image filtering, the calculated field near the axis has some irregularities. In case of added artificial uncertainties, the error drastically increases and can reach several hundred of percent.
	
	To solve this problem, several approaches were proposed and analysed. The first one consists in the plasma axis shift to the angle axis, which will however increase the reconstruction error in case narrow distinctive features are present in  density or angle distributions.  	 
	The second introduced approach is polarization plane rotation angle modification technique.
	As the angle distribution behaviour is considered to be anti-symmetrical, it's value on the axis is supposed to be zero. Thus the non-zero value on the axis can be treated like the background and can be subtracted near the axis area. Such angle correction removes the plasma and angle axes mismatch, therefore decreasing the analysed error. The substructed shift function should be wide enough to avoid sharp changes in function derivative, but smaller than minimal background irregularity width, that can be roughly estimated in plasma-free region. The shift function shape was found to be of a great importance as well, as the corrected angle may become non-physical and lead to change of function sign after Abelization. Using a narrow exponential-like function was found to be the most effective.
	
	To evaluate the error and the introduced correction method's influence on it, the calculations for a simplified case of a constant background added to the synthetic angle data, were carried out. It was shown, that interferometric part's error depends in limited CCD resolution and always increases in low-phase (low-density) regions. To ignore areas with the excessive error, a new border condition $R_{mask}$ was introduced, and all data beyond this border was cut.  The polarimetric's part depends on angle extraction error due to Fourier method limitations and ambiguity in filtering window size; on the background level and on the relative real, plasma and angle axes positions. 
	It was found out that in the experiment-like case the absolute error of magnetic field calculation near the axis can reach up to 10 times the reference field maximum value, which leads to an enormous values of relative error at that region. 
	With the applied correction, the angle value becomes zero at the plasma axis, and due to the correction function narrowness - everything else remains the same thus only small part of the field distribution is changed. The correction leads to decrease of previously infinity-like relative error near the axis to the level of $\approx$200\%, lowering the field absolute error from MG values down to hundreds of kGs. At the same time, the field reconstruction error in the region $0.2R_{mask}<r<0.8R_{mask}$ has the mean value of 25\% with the minimum of 15\% at the maximum measured angle point. The plasma axis shift approach was also tested and provided even better field reconstruction, comparable to that of the background-free case. However this method should be used only when one of the studied distributions is wide and smooth function, otherwise all significant field details would be lost. 
	
	Summing up, the relative error is smallest in the area from 0.2 to 0.8 of radius $R_{mask}$ and in background-free case is about 2\% there, caused by angle, phase extraction and Abelization errors. Upon addition of experimental like background (0.01 rad which equals to 20\% of maximum angle value in the studied cross-section) to the angle distribution this error raises to 25\%.  The background shifts the $y_{angle}$ position in respect to the calculation axis $y_{plasma}$, which leads to infinity-like error near the axis. Shift of the plasma axis reduces the error to about 70\% , which is comparable with background-free case, but leads to loss of details. The proposed correction technique via angle modification decreases error near the axis to the level of about 200\% leaving all the details intact.
	Utilizing the presented correction techniques provides more correct information on the magnetic field distribution.	 This makes the complex interferometry a more accurate and reliable diagnostic tool and allows more precise measurements of space-time distributions of SMF and the electron density in ablative plasma, which opens new perspectives in laser plasma studies related to ICF and laboratory astrophysics.
	
	\section*{Acknowledgments}
		This research was supported by the Access to the PALS RI under the EU LASERLAB IV (Grant Agreement No. 654148) and LASERLAB AISBL (Grant Agreement No. 871124) projects; by the Ministry of Science and Higher Education, Republic of Poland (Decision   NR 5084/PALS/2020/0), by the Ministry of Education, Youth and Sports of the Czech Republic (projects No. LM2015083 (PALS RI), LTT17015 and No.EF16\_013/0001552) and by the Czech Science Foundation (Grant No. 19-24619S). The work was partially supported by the MEPhI Academic Excellence Project (contract No. 02.a03.21.0005, 27.08.2013) and by the project \# FSWU-2020-0035 (Ministry of Science and Higher Education of the Russian Federation). We acknowledge resources of NRNU MEPhI High-Performance Computing Center. We thank the Joint SuperComputer Center of the Russian Academy of Sciences for providing its computational resources.
	
	\section*{Appendix: Application of the discussed technique to an example real experimental data.}
	
	In the simulation it is impossible to perform all of the different effects that present in real interferogram. To test the proposed correction method, we used the 2 frames of data obtained during one of the experiments at PALS laser facility. At first, the interferograms were processed using the Fourier method to extract phase information \cite{AgaArt}, the density distribution was calculated and the plasma axis position was defined.

\begin{figure}[H]
	\begin{center}
		\includegraphics[width=0.6\linewidth]{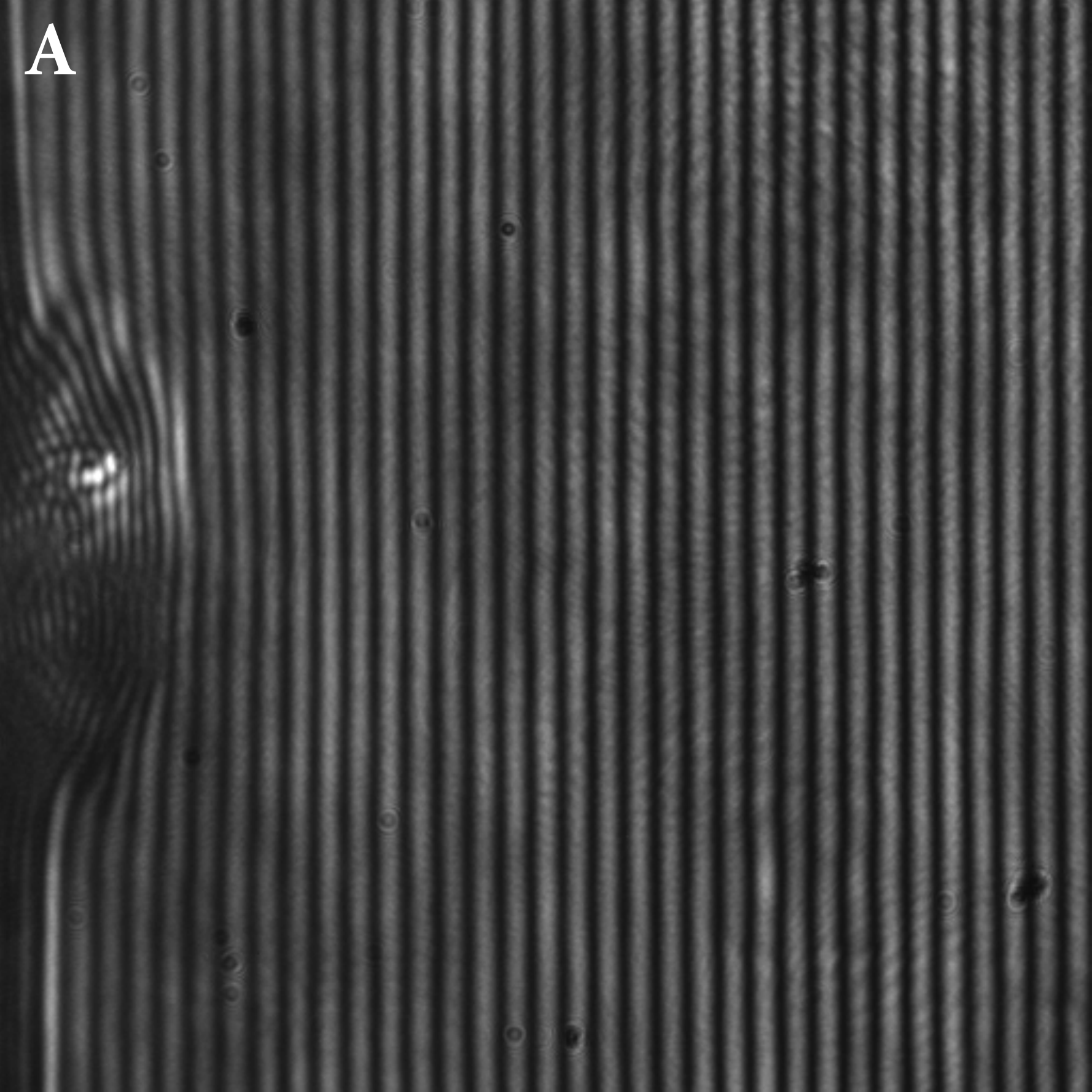}
		\\[2mm]
		\includegraphics[width=0.99\linewidth]{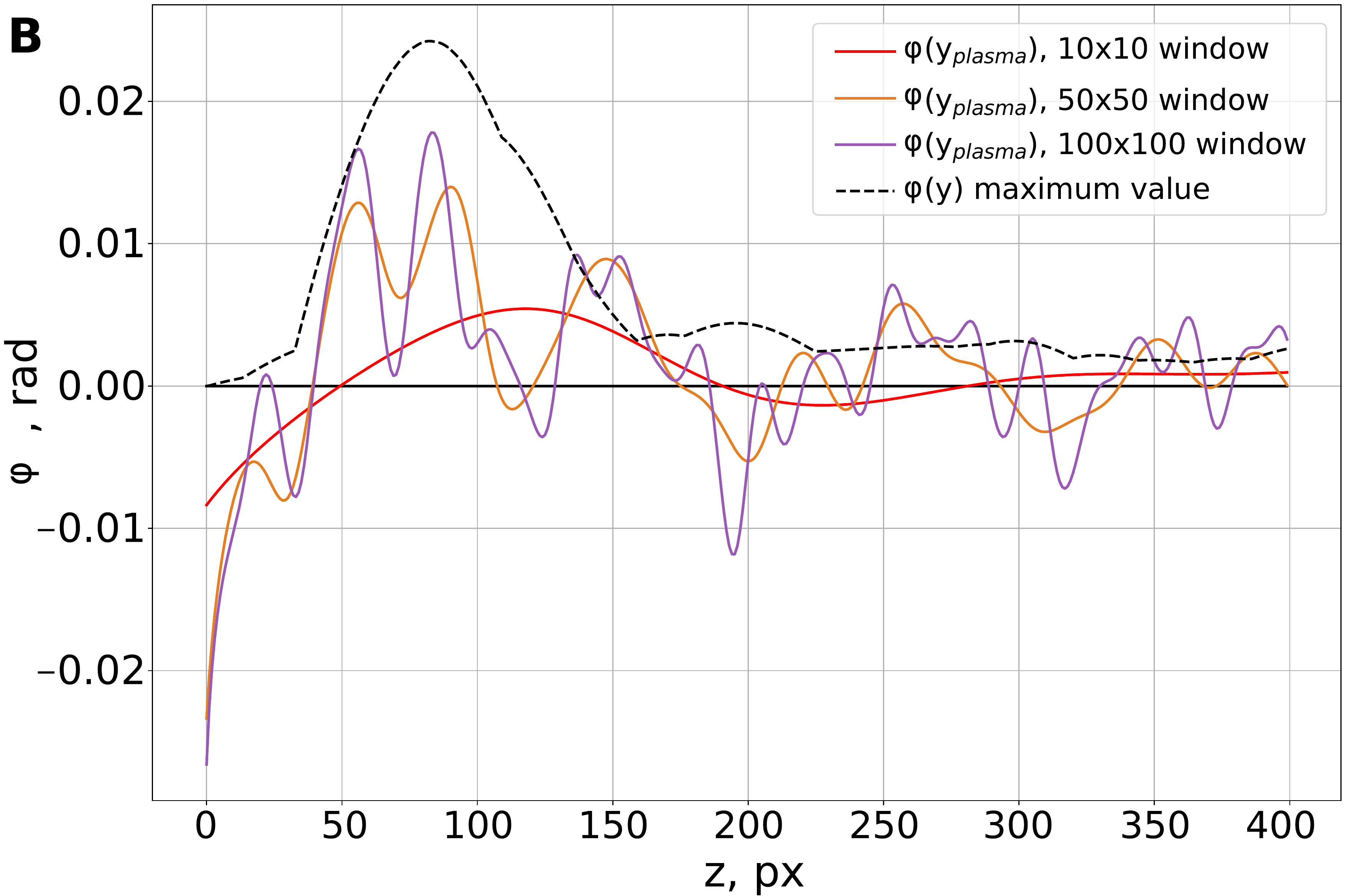}			
		\caption{ (A) Part of the complex interferogram for the shot \#52991$\_$1; (B) Calculated angle value at the plasma axis (solid lines) and maximum angle value at the corresponding cross-section (dashed line) for this interferogram}
		\label{img:real oscill}
	\end{center}
\end{figure}

	Investigating the extracted angle data, we can see that it's value at the found plasma axis is often far from zero.	 As can be seen at Fig. \ref{img:real oscill} for the first  of experimental data (shot 52992 frame 1) - the angle value at determined plasma symmetry axis $y_{plasma}$ is highly unstable and shows oscillations with a much higher amplitude than caused by Fourier filtering. This leads to the measured anti-symmetry point for each z coordinate to be shifted by up to 10 pixels from plasma symmetry axis. The angle value itself also doesn't drop down to zero in plasma-free density region, which can help us to estimate the effective background level and the error in angle determination. Unfortunately this noise can't be completely removed, but can be reduced if special measures were taken while conducting the experiment (e.g. measurement of additional information - intensity distributions in the reference and probe beams \cite{Kalal_2016}). 
	
\begin{figure*}
		\begin{center}
			\begin{subfigure}[b]{0.99\linewidth}
				\centering
				\includegraphics[width=0.45\linewidth]{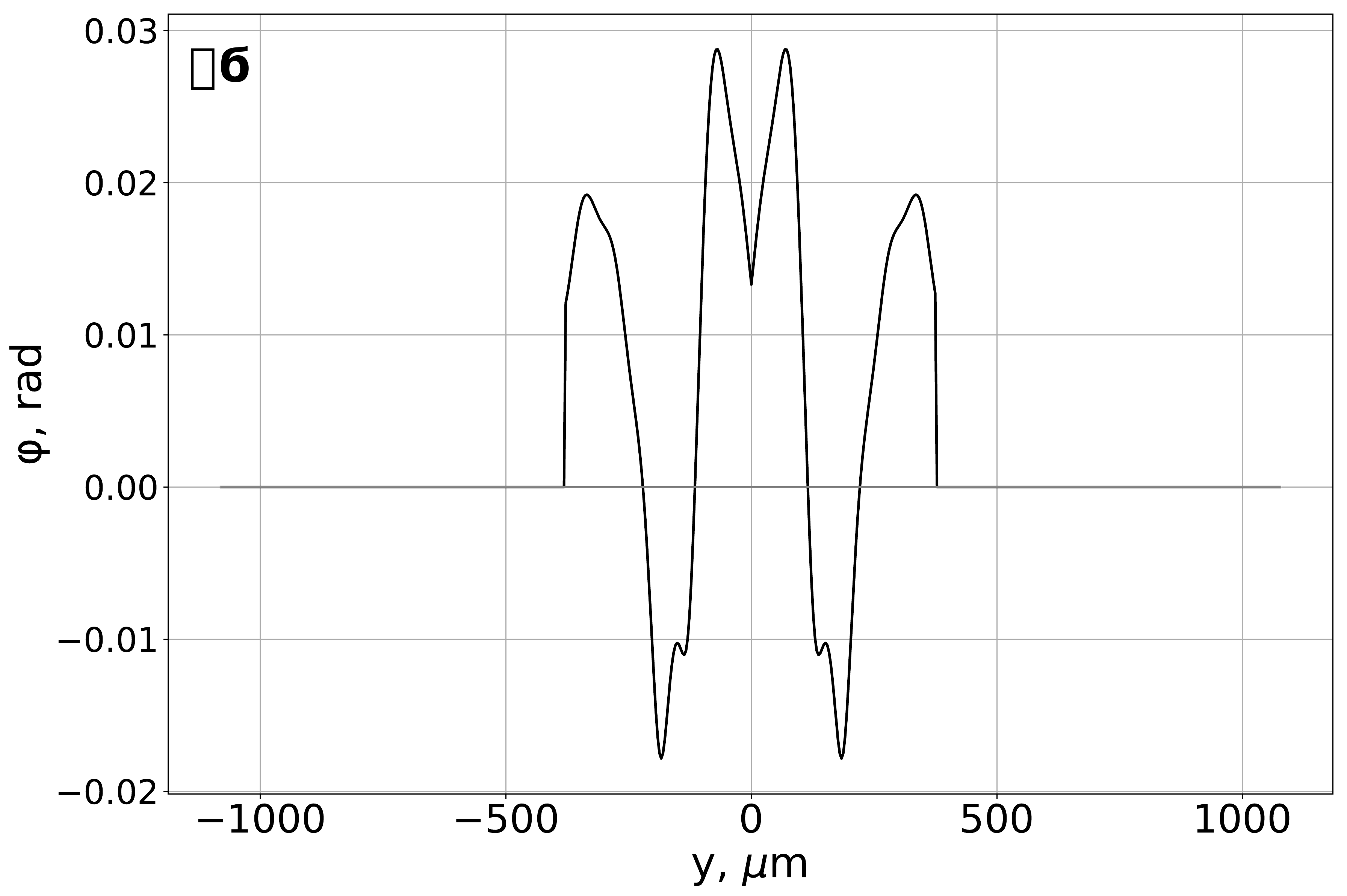}
				\hfil
				\includegraphics[width=0.45\linewidth]{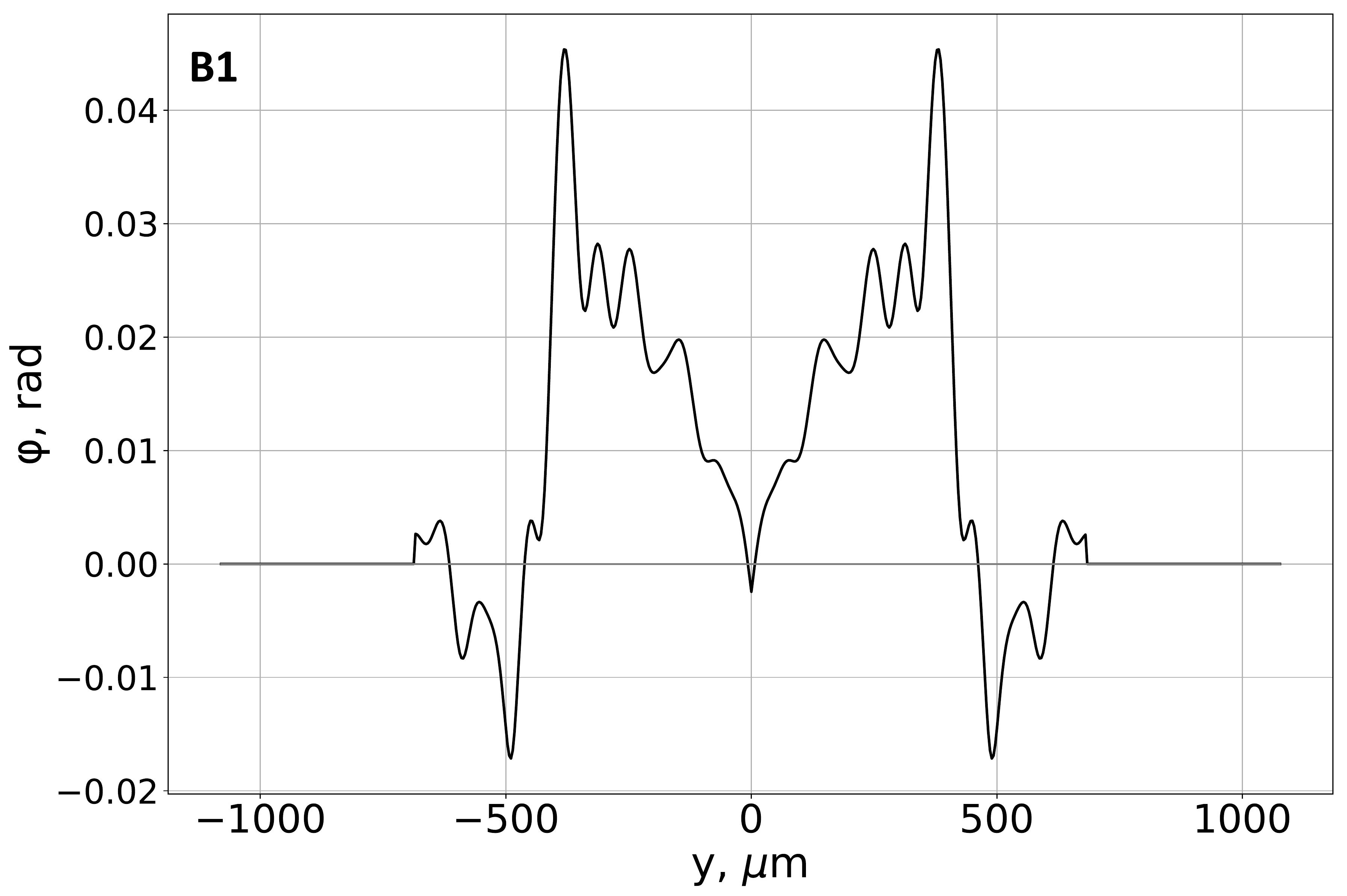}
				%\hfil
				%\includegraphics[width=0.3\linewidth]{3_Sleft_520_nc.pdf}
			\end{subfigure}
			\vspace{1ex}
			\begin{subfigure}[b]{0.99\linewidth}
				\centering
				\includegraphics[width=0.45\linewidth]{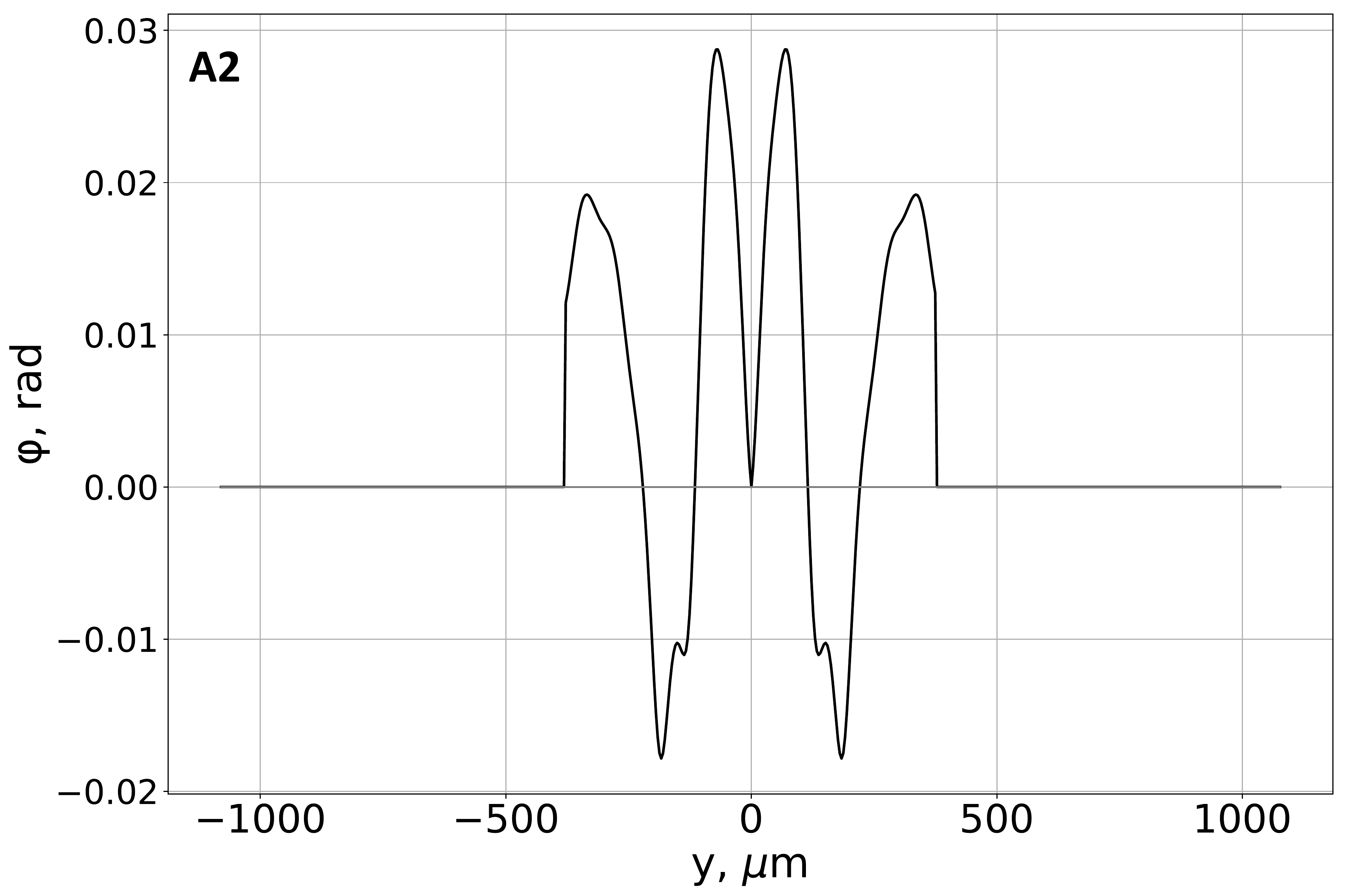}
				\hfil
				\includegraphics[width=0.45\linewidth]{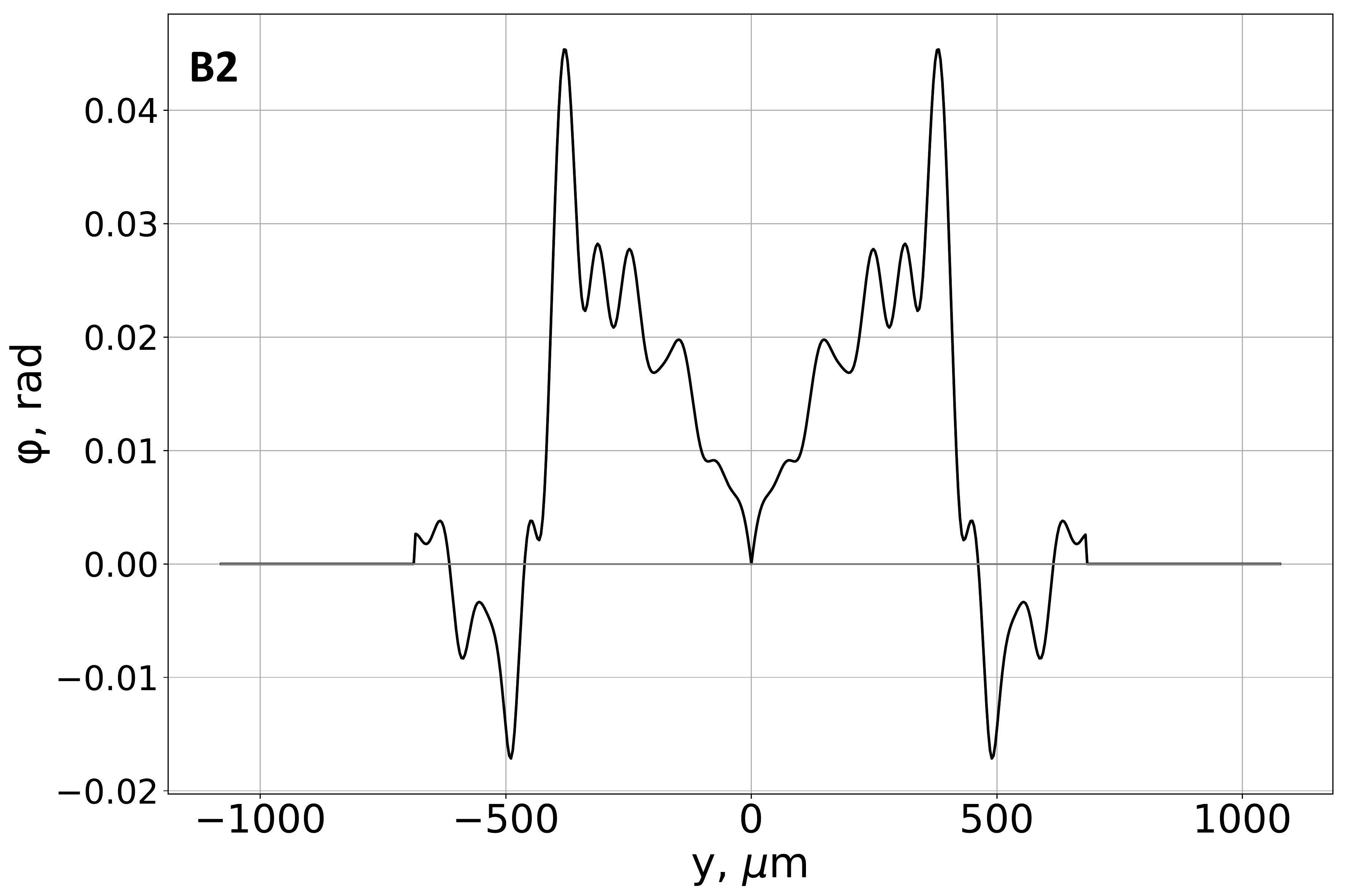}
				%\hfil
				%\includegraphics[width=0.3\linewidth]{3_Sleft_520_c.pdf}
			\end{subfigure}
			\caption{Applying correction to 2 frames of experimental angle data (shot \#52992). Upper row - uncorrected case, lower row - corrected case, cross-sections taken at 210 $\mu$m for first frame (A1 and A2) and 290$\mu$m for second frame (B1 and B2)}
			\label{img:Correction to real angle}
		\end{center}
	\end{figure*}

	Comparison of the extracted angle data with and without applied correction is presented at Fig. \ref{img:Correction to real angle}. The plasma border was limited by density level $n_e=10^{18}$ cm$^{-3}$ to remove areas with large error at the sides of the field distribution. The abelization procedure was carried out using Fourier method.

	\begin{figure*}
		\begin{center}
			\includegraphics[width=0.45\linewidth]{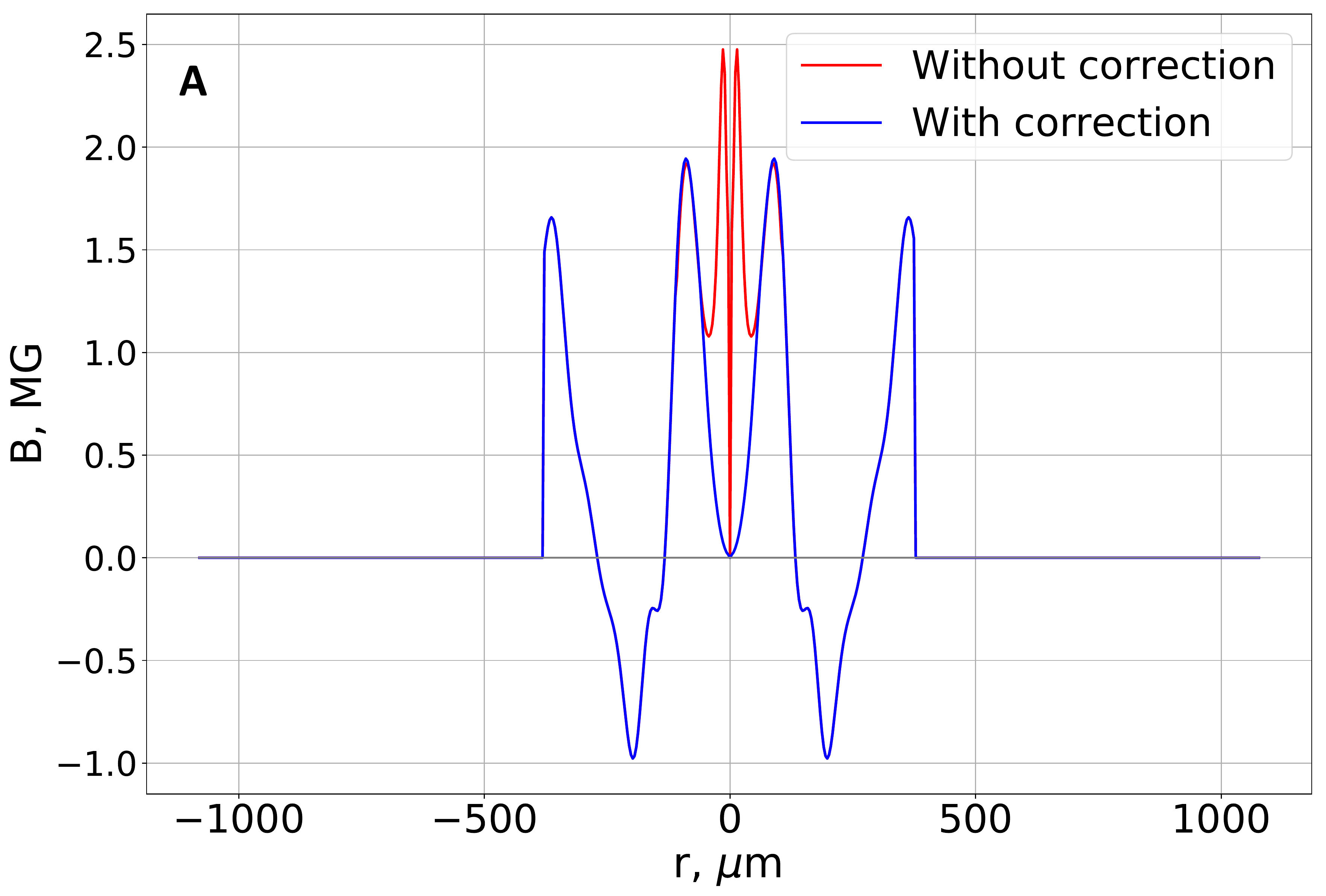}
			\includegraphics[width=0.45\linewidth]{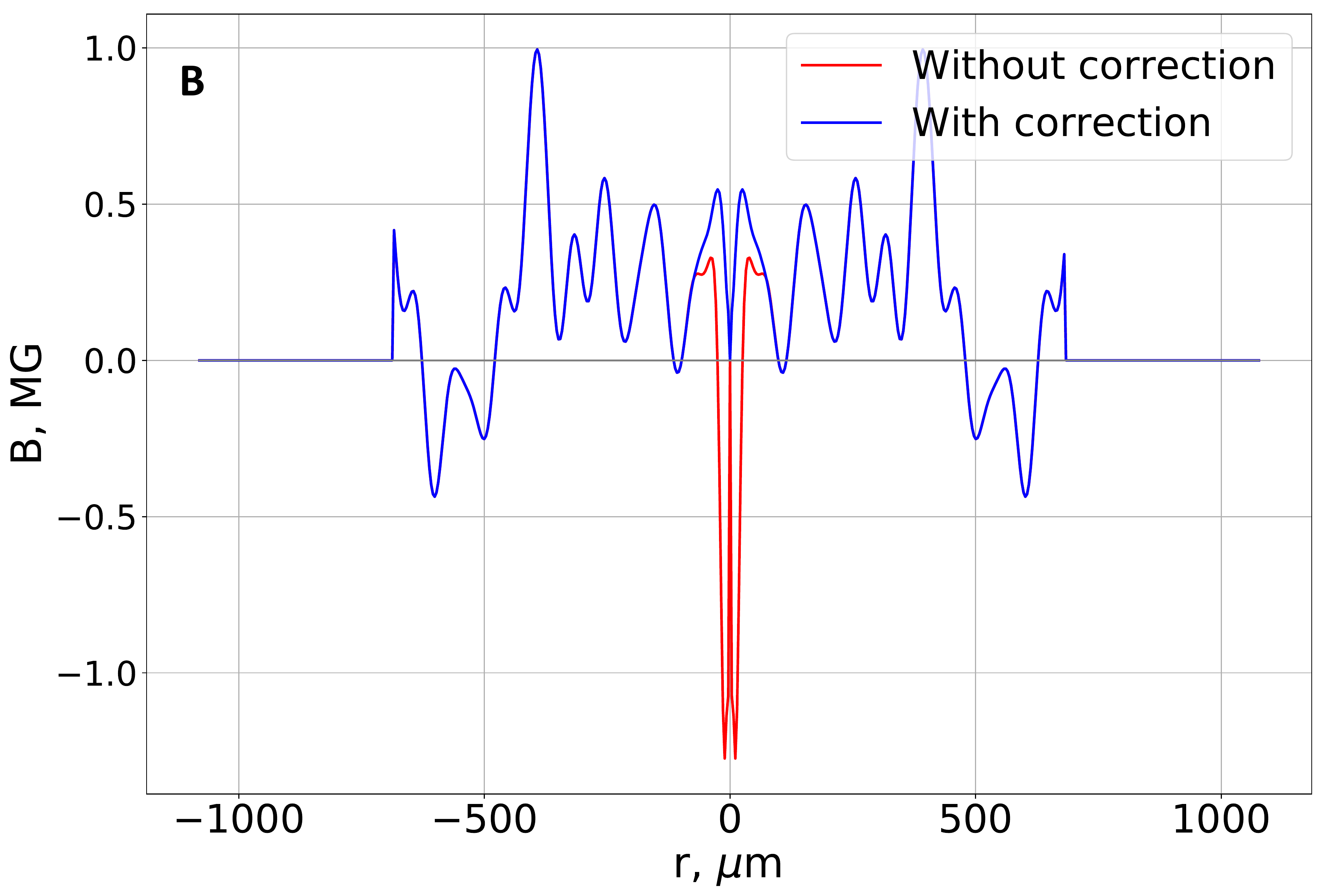}
			%\hfil
			%\includegraphics[width=0.3\linewidth]{52992_3_ND_corr_and_nocorr_520.pdf}
			\caption{Cross-sections of the calculated magnetic field taken at 210 $\mu$m for the first frame (A), 290$\mu$m for the second frame (B) of shot \#52992.}
			\label{img:real mag field slice}
		\end{center}
	\end{figure*}
	The fields calculated without using the angle modification technique, presented in this paper, have high value near the symmetry axis. Applying the angle correction method removes such high-value narrow positive and negative field peaks, leaving everything else the same. It can also be seen, that despite plasma border restrictions the field value increases at the sides, which is most likely connected to a presence of the measured angle background - according to our theoretical calculations using the simulated data, the error in that region increases non-linearly with the increase of the background value. The field data in a cross-section for visual comparison is presented at Fig. \ref{img:real mag field slice}.
	
\begin{figure}
		\begin{center}
			\begin{subfigure}[b]{0.99\linewidth}
				\includegraphics[width=0.49\linewidth]{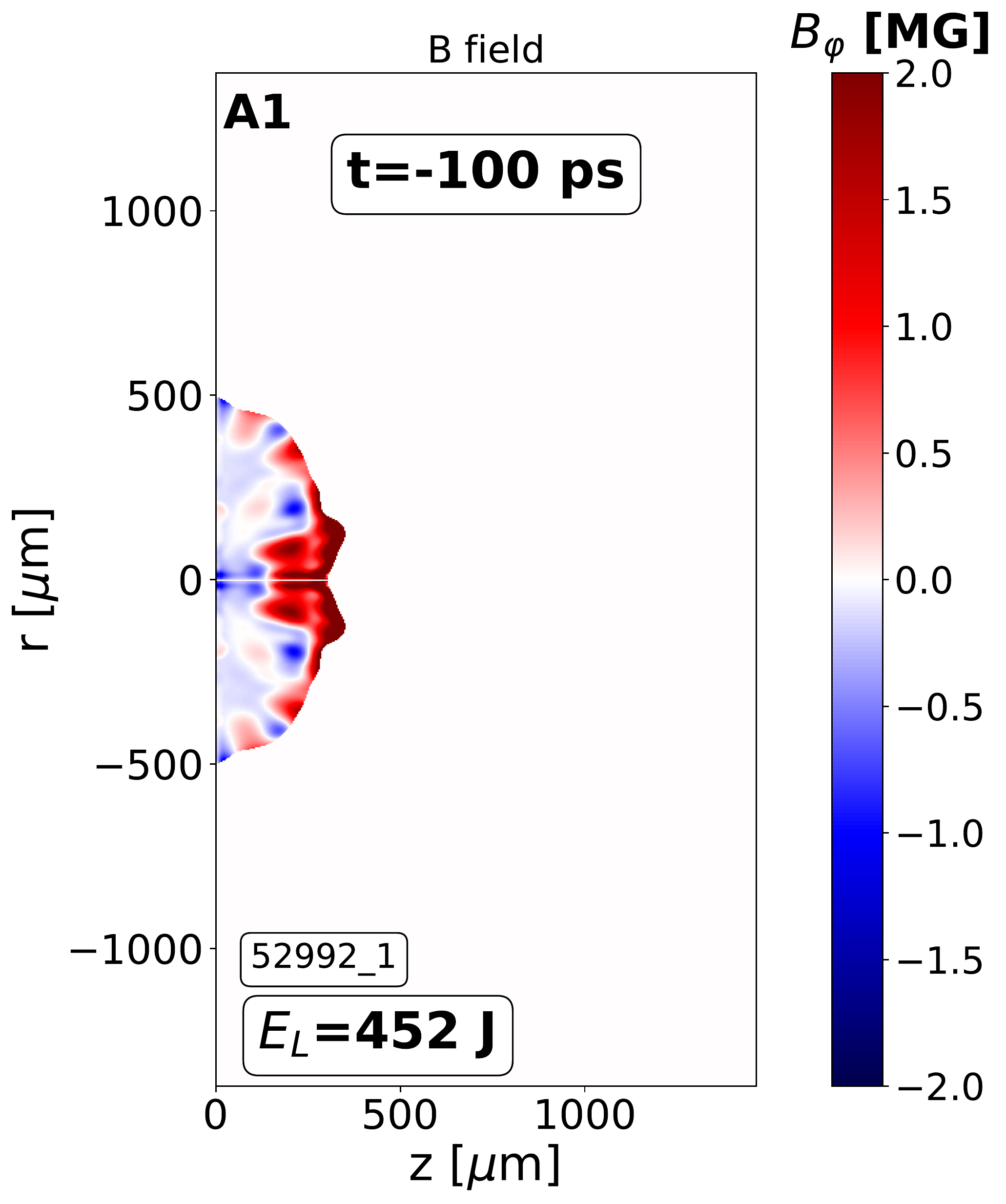}
				\hfil
				\includegraphics[width=0.49\linewidth]{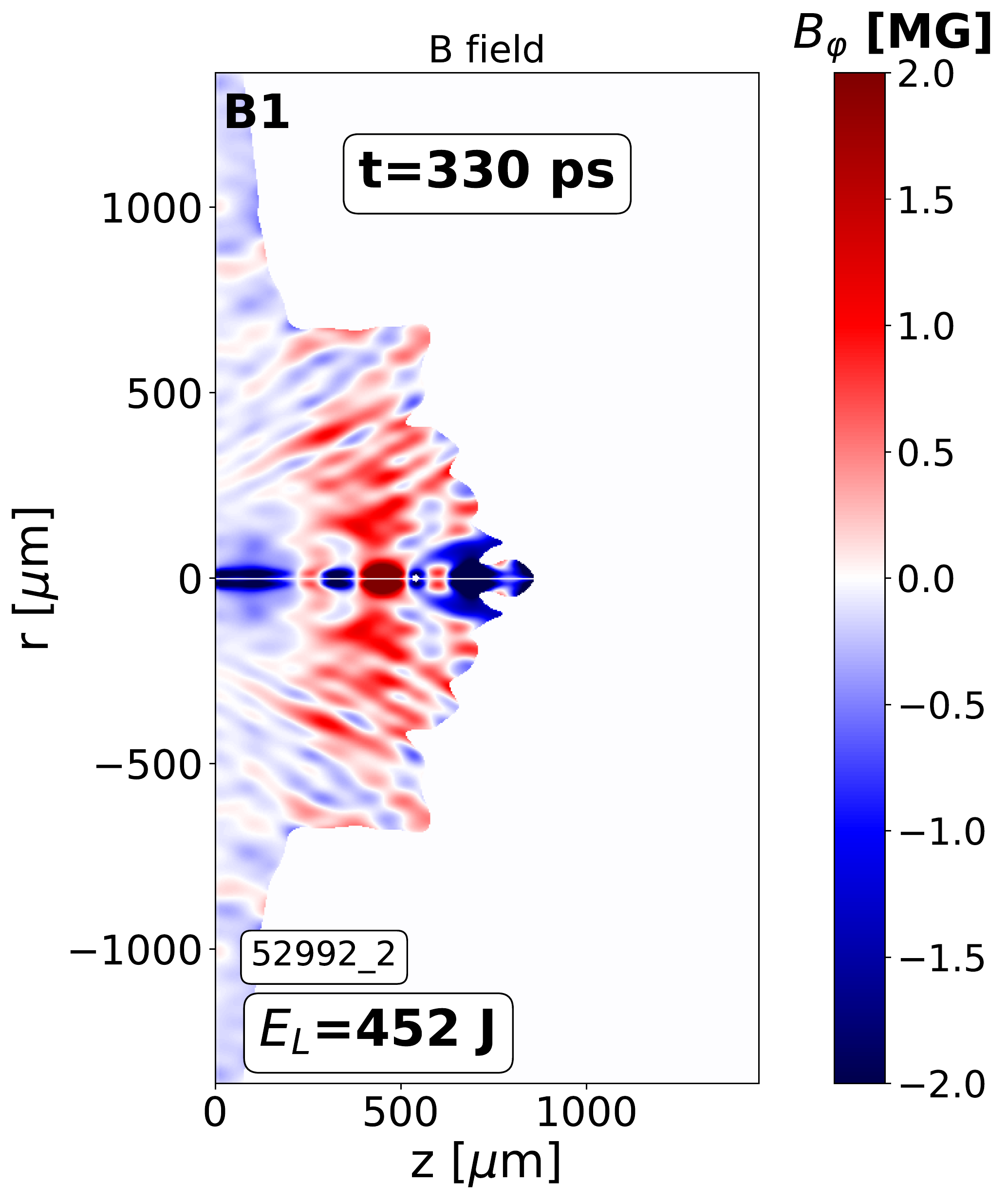}
				%\hfil
				%\includegraphics[width=0.3\linewidth]{New_Dens_field_nocorr_52992_3.pdf}
			\end{subfigure}
			\vspace{1ex}
			\begin{subfigure}[b]{0.99\linewidth}
				\includegraphics[width=0.49\linewidth]{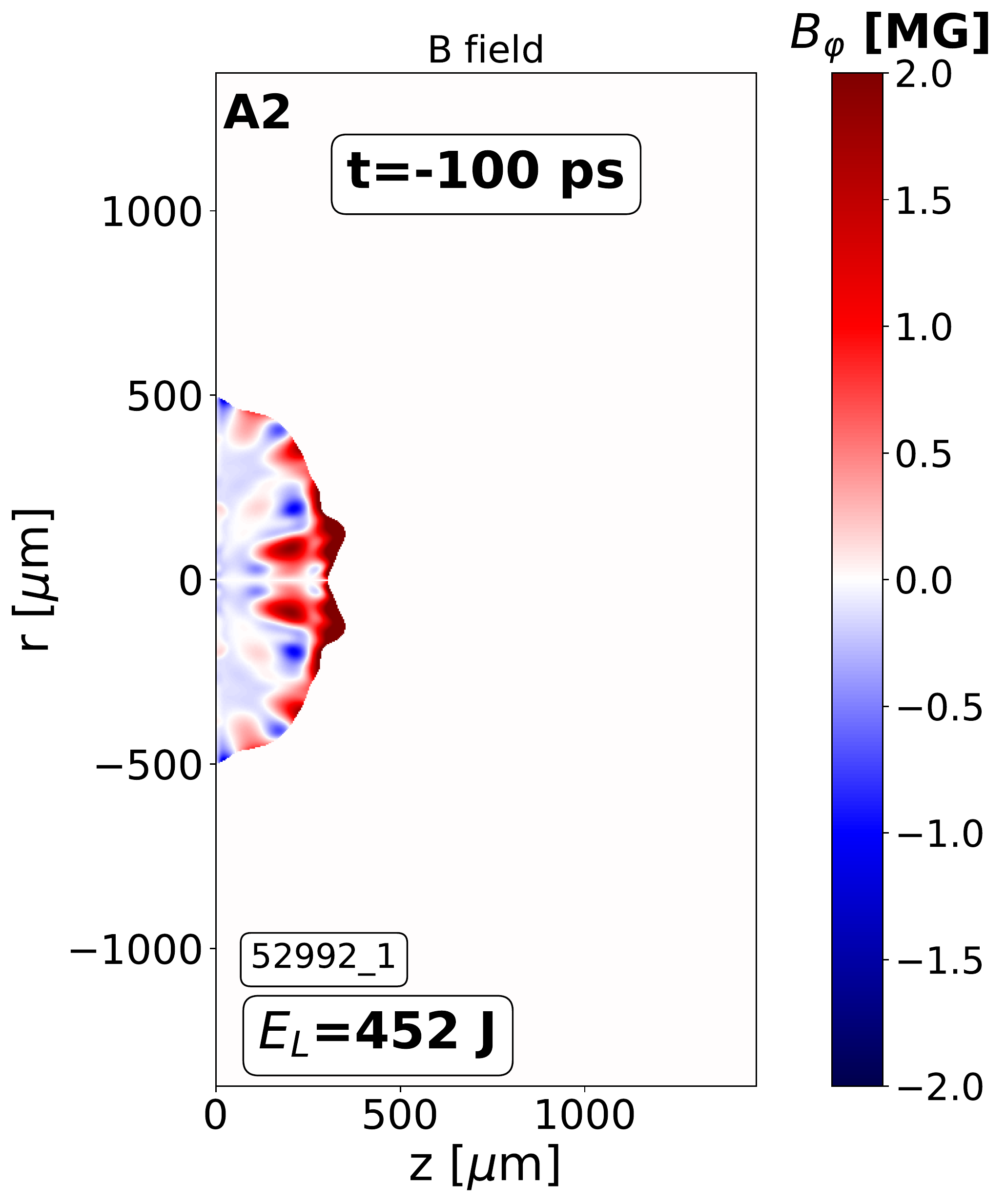}
				\hfil
				\includegraphics[width=0.49\linewidth]{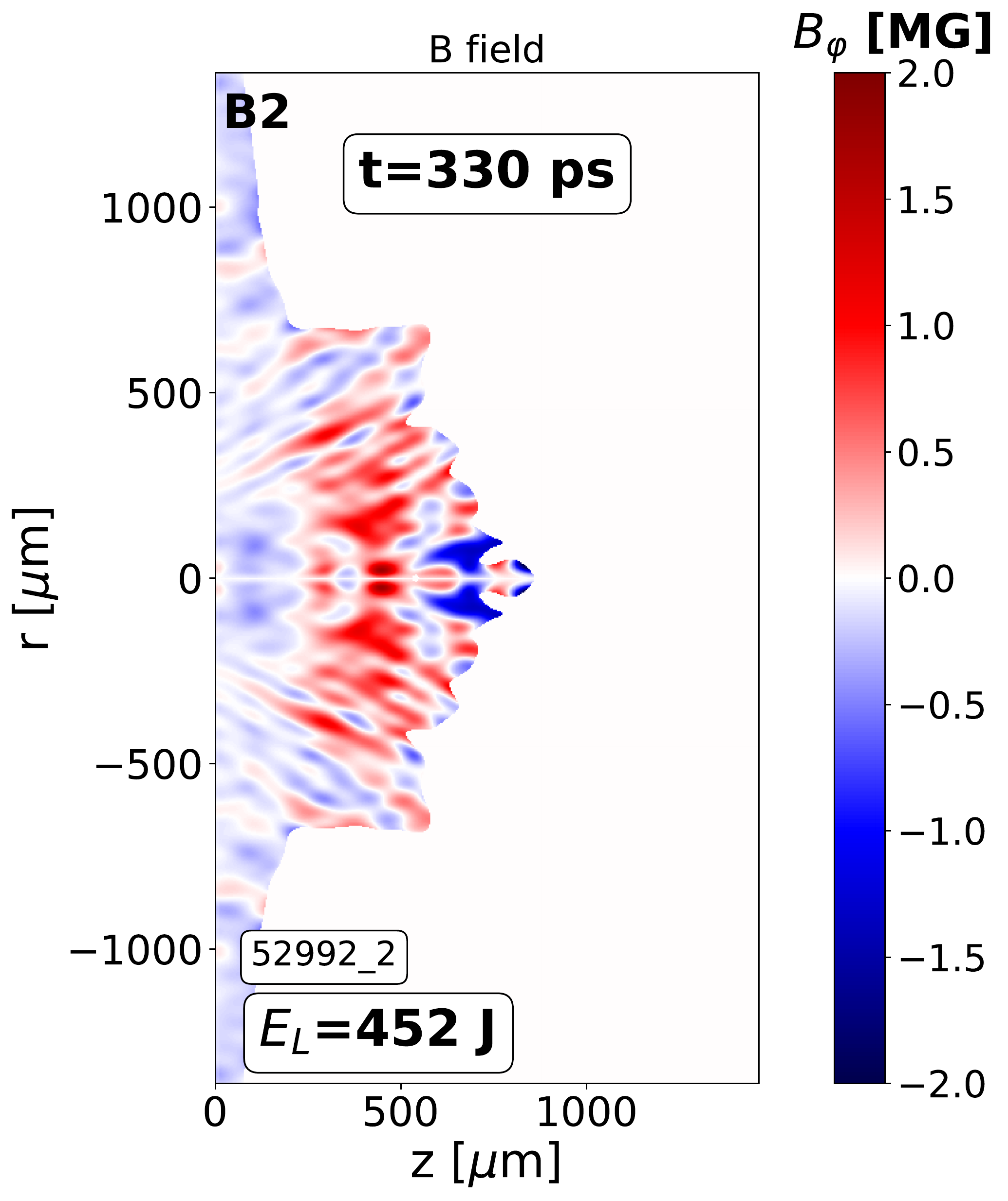}
				%\hfil
				%\includegraphics[width=0.3\linewidth]{New_Dens_field_corr_52992_3.pdf}
			\end{subfigure}
			\caption{Calculated magnetic field distributions without correction (upper row) and with applied correction (lower row) for the first frame (A1 and A2) and the second frame (B1 and B2) of shot \#52992.}
			\label{img:real mag field}
		\end{center}
	\end{figure}

	The complete field distributions are presented in Fig. \ref{img:real mag field}. From this figures one can make following conclusion: for the non-corrected case there is a high-value field contained in thin cylinder along the chosen symmetry axis. The correction leads to disappearance of such field behaviour near the axis, but does not influence other regions. The more accurate information on the structure of SMF
	also will lead to the more accurate information on the distribution of current density in the ablative plasma, which is extremely important in the description of laser-plasma interaction, namely the hot electron generation.
	
	\bibliography{articlebib} 

\begin{thebibliography}{33}
\expandafter\ifx\csname natexlab\endcsname\relax\def\natexlab#1{#1}\fi
\providecommand{\url}[1]{\texttt{#1}}
\providecommand{\href}[2]{#2}
\providecommand{\path}[1]{#1}
\providecommand{\DOIprefix}{doi:}
\providecommand{\ArXivprefix}{arXiv:}
\providecommand{\URLprefix}{URL: }
\providecommand{\Pubmedprefix}{pmid:}
\providecommand{\doi}[1]{\href{http://dx.doi.org/#1}{\path{#1}}}
\providecommand{\Pubmed}[1]{\href{pmid:#1}{\path{#1}}}
\providecommand{\bibinfo}[2]{#2}
\ifx\xfnm\relax \def\xfnm[#1]{\unskip,\space#1}\fi
%Type = Article
\bibitem[{Dostal et~al.(2017)Dostal, Dudzak, Pisarczyk, Pfeifer, Huynh,
  Chodukowski, Kalinowska, Krousky, Skala, Hrebicek, Medrik, Golasowski, Juha,
  and Ullschmied}]{Dostal-17}
\bibinfo{author}{J.~Dostal}, \bibinfo{author}{R.~Dudzak},
  \bibinfo{author}{T.~Pisarczyk}, \bibinfo{author}{M.~Pfeifer},
  \bibinfo{author}{J.~Huynh}, \bibinfo{author}{T.~Chodukowski},
  \bibinfo{author}{Z.~Kalinowska}, \bibinfo{author}{E.~Krousky},
  \bibinfo{author}{J.~Skala}, \bibinfo{author}{J.~Hrebicek},
  \bibinfo{author}{T.~Medrik}, \bibinfo{author}{J.~Golasowski},
  \bibinfo{author}{L.~Juha}, \bibinfo{author}{J.~Ullschmied},
\newblock \bibinfo{title}{Synchronizing single-shot high-energy iodine
  photodissociation laser pals and high-repetition-rate femtosecond ti:sapphire
  laser system},
\newblock \bibinfo{journal}{Review of Scientific Instruments}
  \bibinfo{volume}{88} (\bibinfo{year}{2017}) \bibinfo{pages}{045109}.
%Type = Article
\bibitem[{Ivanov et~al.(2017)Ivanov, Tsymbalov, Shulyapov, Krestovskikh,
  Brantov, Bychenkov, Volkov, and Savel'ev}]{Plasma_Interferometry_2017}
\bibinfo{author}{K.~A. Ivanov}, \bibinfo{author}{I.~N. Tsymbalov},
  \bibinfo{author}{S.~A. Shulyapov}, \bibinfo{author}{D.~A. Krestovskikh},
  \bibinfo{author}{A.~V. Brantov}, \bibinfo{author}{V.~Y. Bychenkov},
  \bibinfo{author}{R.~V. Volkov}, \bibinfo{author}{A.~B. Savel'ev},
\newblock \bibinfo{title}{Prepulse controlled electron acceleration from solids
  by a femtosecond laser pulse in the slightly relativistic regime},
\newblock \bibinfo{journal}{Physics of Plasmas} \bibinfo{volume}{24}
  (\bibinfo{year}{2017}) \bibinfo{pages}{063109}.
%Type = Article
\bibitem[{Hough(2005)}]{Hough}
\bibinfo{author}{J.~Hough},
\newblock \bibinfo{title}{Polarimetry techniques at optical and infrared
  wavelengths},
\newblock \bibinfo{journal}{Astronomical Polarimetry: Current Status and Future
  Directions} \bibinfo{volume}{343} (\bibinfo{year}{2005}).
%Type = Article
\bibitem[{Santos et~al.(2015)Santos, Bailly-Grandvaux, Giuffrida,
  Forestier-Colleoni, Fujioka, Zhang, Korneev, Bouillaud, Dorard, Batani,
  Chevrot, Cross, Crowston, Dubois, Gazave, Gregori, D'Humi{\`{e}}res, Hulin,
  Ishihara, Kojima, Loyez, Marqu{\`{e}}s, Morace, Nicola{\"{i}}, Peyrusse,
  Poy{\'{e}}, Raffestin, Ribolzi, Roth, Schaumann, Serres, Tikhonchuk, Vacar,
  and Woolsey}]{Santos-NJP2015}
\bibinfo{author}{J.~J. Santos}, \bibinfo{author}{M.~Bailly-Grandvaux},
  \bibinfo{author}{L.~Giuffrida}, \bibinfo{author}{P.~Forestier-Colleoni},
  \bibinfo{author}{S.~Fujioka}, \bibinfo{author}{Z.~Zhang},
  \bibinfo{author}{P.~Korneev}, \bibinfo{author}{R.~Bouillaud},
  \bibinfo{author}{S.~Dorard}, \bibinfo{author}{D.~Batani},
  \bibinfo{author}{M.~Chevrot}, \bibinfo{author}{J.~E. Cross},
  \bibinfo{author}{R.~Crowston}, \bibinfo{author}{J.-L. J.-L. Dubois},
  \bibinfo{author}{J.~Gazave}, \bibinfo{author}{G.~Gregori},
  \bibinfo{author}{E.~D'Humi{\`{e}}res}, \bibinfo{author}{S.~Hulin},
  \bibinfo{author}{K.~Ishihara}, \bibinfo{author}{S.~Kojima},
  \bibinfo{author}{E.~Loyez}, \bibinfo{author}{J.-R. J.-R. Marqu{\`{e}}s},
  \bibinfo{author}{A.~Morace}, \bibinfo{author}{P.~Nicola{\"{i}}},
  \bibinfo{author}{O.~Peyrusse}, \bibinfo{author}{A.~Poy{\'{e}}},
  \bibinfo{author}{D.~Raffestin}, \bibinfo{author}{J.~Ribolzi},
  \bibinfo{author}{M.~Roth}, \bibinfo{author}{G.~Schaumann},
  \bibinfo{author}{F.~Serres}, \bibinfo{author}{V.~T. Tikhonchuk},
  \bibinfo{author}{P.~Vacar}, \bibinfo{author}{N.~Woolsey},
\newblock \bibinfo{title}{{Laser-driven platform for generation and
  characterization of strong quasi-static magnetic fields}},
\newblock \bibinfo{journal}{New Journal of Physics} \bibinfo{volume}{17}
  (\bibinfo{year}{2015}) \bibinfo{pages}{083051}.
%Type = Article
\bibitem[{Pisarczyk et~al.(2019)Pisarczyk, Santos, Dudzak, Zaras-Szyd{\l}owska,
  Ehret, Rusiniak, Dostal, Chodukowski, Renner, Gus'kov, Korneev, Burian,
  Vlachos, Kochetkov, Makaruk, Rosinski, Kalal, Krupka, Pfeifer, Klir,
  Cikhardt, Krasa, Singh, Borodziuk, Krus, Juha, Hrebicek, Golasowski, and
  Skala}]{Pisarczyk2019}
\bibinfo{author}{T.~Pisarczyk}, \bibinfo{author}{J.~Santos},
  \bibinfo{author}{R.~Dudzak}, \bibinfo{author}{A.~Zaras-Szyd{\l}owska},
  \bibinfo{author}{M.~Ehret}, \bibinfo{author}{Z.~Rusiniak},
  \bibinfo{author}{J.~Dostal}, \bibinfo{author}{T.~Chodukowski},
  \bibinfo{author}{O.~Renner}, \bibinfo{author}{S.~Gus'kov},
  \bibinfo{author}{P.~Korneev}, \bibinfo{author}{T.~Burian},
  \bibinfo{author}{C.~Vlachos}, \bibinfo{author}{I.~Kochetkov},
  \bibinfo{author}{D.~Makaruk}, \bibinfo{author}{M.~Rosinski},
  \bibinfo{author}{M.~Kalal}, \bibinfo{author}{M.~Krupka},
  \bibinfo{author}{M.~Pfeifer}, \bibinfo{author}{D.~Klir},
  \bibinfo{author}{J.~Cikhardt}, \bibinfo{author}{J.~Krasa},
  \bibinfo{author}{S.~Singh}, \bibinfo{author}{S.~Borodziuk},
  \bibinfo{author}{M.~Krus}, \bibinfo{author}{L.~Juha},
  \bibinfo{author}{J.~Hrebicek}, \bibinfo{author}{J.~Golasowski},
  \bibinfo{author}{J.~Skala},
\newblock \bibinfo{title}{Elaboration of 3-frame complex interferometry for
  optimization studies of capacitor-coil optical magnetic field generators},
\newblock \bibinfo{journal}{Journal of Instrumentation} \bibinfo{volume}{14}
  (\bibinfo{year}{2019}) \bibinfo{pages}{C11024--C11024}.
%Type = Article
\bibitem[{Pisarczyk et~al.(1990)Pisarczyk, Rupasov, Sarkisov, and
  Shikanov}]{Pisarczyk90}
\bibinfo{author}{T.~Pisarczyk}, \bibinfo{author}{A.~Rupasov},
  \bibinfo{author}{G.~Sarkisov}, \bibinfo{author}{S.~Shikanov},
\newblock \bibinfo{title}{Faraday-rotation method for magnetic-field
  diagnostics in a laser plasma},
\newblock \bibinfo{journal}{Journal of Russian Laser Research - J RUSS LASER
  RES} \bibinfo{volume}{11} (\bibinfo{year}{1990}) \bibinfo{pages}{1--32}.
%Type = Article
\bibitem[{Pisarczyk et~al.(2015)Pisarczyk, Gus'kov, Dudzak, Chodukowski,
  Dostal, Demchenko, Korneev, Kalinowska, Kalal, Renner, Šmíd, Borodziuk,
  Krouský, Ullschmied, Hrebicek, Medrik, Golasowski, Pfeifer, Skala, and
  Pisarczyk}]{Pisarczyk2015}
\bibinfo{author}{T.~Pisarczyk}, \bibinfo{author}{S.~Gus'kov},
  \bibinfo{author}{R.~Dudzak}, \bibinfo{author}{T.~Chodukowski},
  \bibinfo{author}{J.~Dostal}, \bibinfo{author}{N.~Demchenko},
  \bibinfo{author}{P.~Korneev}, \bibinfo{author}{Z.~Kalinowska},
  \bibinfo{author}{M.~Kalal}, \bibinfo{author}{O.~Renner},
  \bibinfo{author}{M.~Šmíd}, \bibinfo{author}{S.~Borodziuk},
  \bibinfo{author}{E.~Krouský}, \bibinfo{author}{J.~Ullschmied},
  \bibinfo{author}{J.~Hrebicek}, \bibinfo{author}{T.~Medrik},
  \bibinfo{author}{J.~Golasowski}, \bibinfo{author}{M.~Pfeifer},
  \bibinfo{author}{J.~Skala}, \bibinfo{author}{P.~Pisarczyk},
\newblock \bibinfo{title}{Space-time resolved measurements of spontaneous
  magnetic fields in laser-produced plasma},
\newblock \bibinfo{journal}{Physics of Plasmas} \bibinfo{volume}{22}
  (\bibinfo{year}{2015}) \bibinfo{pages}{102706}.
%Type = Article
\bibitem[{{Korobkin} and {Serov}(1966)}]{Korobkin1966}
\bibinfo{author}{V.~V. {Korobkin}}, \bibinfo{author}{R.~V. {Serov}},
\newblock \bibinfo{title}{Investigation of the magnetic field of a spark
  produced by focusing laser radiation},
\newblock \bibinfo{journal}{Soviet Journal of Experimental and Theoretical
  Physics Letters} \bibinfo{volume}{4} (\bibinfo{year}{1966})
  \bibinfo{pages}{70}.
%Type = Article
\bibitem[{STAMPER et~al.(1971)STAMPER, Papadopoulos, SUDAN, DEAN, MCLEAN, and
  DAWSON}]{STAMPER1971}
\bibinfo{author}{J.~STAMPER}, \bibinfo{author}{K.~Papadopoulos},
  \bibinfo{author}{R.~SUDAN}, \bibinfo{author}{S.~DEAN},
  \bibinfo{author}{E.~MCLEAN}, \bibinfo{author}{J.~DAWSON},
\newblock \bibinfo{title}{Spontaneous magnetic fields in laser-produced
  plasmas},
\newblock \bibinfo{journal}{Physical Review Letters - PHYS REV LETT}
  \bibinfo{volume}{26} (\bibinfo{year}{1971}) \bibinfo{pages}{1012--1015}.
%Type = Article
\bibitem[{Basov et~al.(1987)Basov, Volovski, Gamalii, Denus, Pisarchik,
  Rupasov, Sarkisov, Sklizkov, and Tikhonchuk}]{Basov}
\bibinfo{author}{N.~Basov}, \bibinfo{author}{E.~Volovski},
  \bibinfo{author}{E.~Gamalii}, \bibinfo{author}{S.~Denus},
  \bibinfo{author}{T.~Pisarchik}, \bibinfo{author}{A.~Rupasov},
  \bibinfo{author}{G.~Sarkisov}, \bibinfo{author}{G.~Sklizkov},
  \bibinfo{author}{A.~Tikhonchuk, VT annd~Shikanov},
\newblock \bibinfo{title}{Detection of spontaneous magnetic fields in a laser
  plasma in the delfin-1 device},
\newblock \bibinfo{journal}{JETP LETTERS} \bibinfo{volume}{45}
  (\bibinfo{year}{1987}) \bibinfo{pages}{213--217}.
%Type = Article
\bibitem[{Pisarczyk et~al.(2020)Pisarczyk, Kalal, Gus'kov, Batani, Renner,
  Santos, Dudzak, Zaras-Szydlowska, Chodukowski, Rusiniak, Dostal, Krasa,
  Krupka, Kochetkov, Singh, Cikhardt, Burian, Krus, Pfeifer, Cristoforetti,
  Gizzi, Baffigi, Antonelli, Demchenko, Rosinski, Terwinska, Borodziuk, Kubes,
  Ehret, Juha, Skala, and Korneev}]{TadeuszPPCF}
\bibinfo{author}{T.~Pisarczyk}, \bibinfo{author}{M.~Kalal},
  \bibinfo{author}{S.~Gus'kov}, \bibinfo{author}{D.~Batani},
  \bibinfo{author}{O.~Renner}, \bibinfo{author}{J.~J. Santos},
  \bibinfo{author}{R.~Dudzak}, \bibinfo{author}{A.~Zaras-Szydlowska},
  \bibinfo{author}{T.~Chodukowski}, \bibinfo{author}{Z.~Rusiniak},
  \bibinfo{author}{J.~Dostal}, \bibinfo{author}{J.~Krasa},
  \bibinfo{author}{M.~Krupka}, \bibinfo{author}{I.~Kochetkov},
  \bibinfo{author}{S.~Singh}, \bibinfo{author}{J.~Cikhardt},
  \bibinfo{author}{T.~Burian}, \bibinfo{author}{M.~Krus},
  \bibinfo{author}{M.~Pfeifer}, \bibinfo{author}{G.~Cristoforetti},
  \bibinfo{author}{L.~A. Gizzi}, \bibinfo{author}{F.~Baffigi},
  \bibinfo{author}{L.~Antonelli}, \bibinfo{author}{N.~Demchenko},
  \bibinfo{author}{M.~Rosinski}, \bibinfo{author}{D.~Terwinska},
  \bibinfo{author}{S.~Borodziuk}, \bibinfo{author}{P.~Kubes},
  \bibinfo{author}{M.~Ehret}, \bibinfo{author}{L.~Juha},
  \bibinfo{author}{J.~Skala}, \bibinfo{author}{P.~Korneev},
\newblock \bibinfo{title}{Hot electron retention in laser plasma created under
  terawatt sub-nanosecond irradiation of cu targets},
\newblock \bibinfo{journal}{Plasma Physics and Controlled Fusion}
  \bibinfo{volume}{62} (\bibinfo{year}{2020}).
%Type = Article
\bibitem[{Pisarczyk et~al.(2017)Pisarczyk, Gus'kov, Chodukowski, Dudzak,
  Korneev, Demchenko, Kalinowska, Dostal, Zaras-Szydlowska, Borodziuk, Juha,
  Cikhardt, Krasa, Klir, Cikhardtova, Kubes, Krousky, Krus, Ullschmied,
  Jungwirth, Hrebicek, Medrik, Golasowski, Pfeifer, Renner, Singh, Kar, Ahmed,
  Skala, and Pisarczyk}]{kinetic}
\bibinfo{author}{T.~Pisarczyk}, \bibinfo{author}{S.~Y. Gus'kov},
  \bibinfo{author}{T.~Chodukowski}, \bibinfo{author}{R.~Dudzak},
  \bibinfo{author}{P.~Korneev}, \bibinfo{author}{N.~N. Demchenko},
  \bibinfo{author}{Z.~Kalinowska}, \bibinfo{author}{J.~Dostal},
  \bibinfo{author}{A.~Zaras-Szydlowska}, \bibinfo{author}{S.~Borodziuk},
  \bibinfo{author}{L.~Juha}, \bibinfo{author}{J.~Cikhardt},
  \bibinfo{author}{J.~Krasa}, \bibinfo{author}{D.~Klir},
  \bibinfo{author}{B.~Cikhardtova}, \bibinfo{author}{P.~Kubes},
  \bibinfo{author}{E.~Krousky}, \bibinfo{author}{M.~Krus},
  \bibinfo{author}{J.~Ullschmied}, \bibinfo{author}{K.~Jungwirth},
  \bibinfo{author}{J.~Hrebicek}, \bibinfo{author}{T.~Medrik},
  \bibinfo{author}{J.~Golasowski}, \bibinfo{author}{M.~Pfeifer},
  \bibinfo{author}{O.~Renner}, \bibinfo{author}{S.~Singh},
  \bibinfo{author}{S.~Kar}, \bibinfo{author}{H.~Ahmed},
  \bibinfo{author}{J.~Skala}, \bibinfo{author}{P.~Pisarczyk},
\newblock \bibinfo{title}{Kinetic magnetization by fast electrons in
  laser-produced plasmas at sub-relativistic intensities},
\newblock \bibinfo{journal}{Physics of Plasmas} \bibinfo{volume}{24}
  (\bibinfo{year}{2017}) \bibinfo{pages}{102711}.
%Type = Article
\bibitem[{Gong et~al.(2019)Gong, Mackenroth, Yan, and Arefiev}]{Gong}
\bibinfo{author}{Z.~Gong}, \bibinfo{author}{F.~Mackenroth},
  \bibinfo{author}{X.~Yan}, \bibinfo{author}{A.~Arefiev},
\newblock \bibinfo{title}{Radiation reaction as an energy enhancement mechanism
  for laser-irradiated electrons in a strong plasma magnetic field},
\newblock \bibinfo{journal}{Scientific Reports} \bibinfo{volume}{9}
  (\bibinfo{year}{2019}).
%Type = Article
\bibitem[{Remington et~al.(2006)Remington, Drake, and
  Ryutov}]{RevModPhys.78.755}
\bibinfo{author}{B.~A. Remington}, \bibinfo{author}{R.~P. Drake},
  \bibinfo{author}{D.~D. Ryutov},
\newblock \bibinfo{title}{Experimental astrophysics with high power lasers and
  $z$ pinches},
\newblock \bibinfo{journal}{Rev. Mod. Phys.} \bibinfo{volume}{78}
  (\bibinfo{year}{2006}) \bibinfo{pages}{755--807}.
%Type = Article
\bibitem[{Strozzi et~al.(2015)Strozzi, Perkins, Marinak, Larson, Koning, and
  Logan}]{strozzi_2015}
\bibinfo{author}{D.~J. Strozzi}, \bibinfo{author}{L.~J. Perkins},
  \bibinfo{author}{M.~M. Marinak}, \bibinfo{author}{D.~J. Larson},
  \bibinfo{author}{J.~M. Koning}, \bibinfo{author}{B.~G. Logan},
\newblock \bibinfo{title}{Imposed magnetic field and hot electron propagation
  in inertial fusion hohlraums},
\newblock \bibinfo{journal}{Journal of Plasma Physics} \bibinfo{volume}{81}
  (\bibinfo{year}{2015}) \bibinfo{pages}{475810603}.
%Type = Article
\bibitem[{Hohenberger et~al.(2014)Hohenberger, Theobald, Hu, Anderson, Betti,
  Boehly, Casner, Fratanduono, Lafon, Meyerhofer, Nora, Ribeyre, Sangster,
  Schurtz, Seka, Stoeckl, and Yaakobi}]{Hohenberger}
\bibinfo{author}{M.~Hohenberger}, \bibinfo{author}{W.~Theobald},
  \bibinfo{author}{S.~X. Hu}, \bibinfo{author}{K.~S. Anderson},
  \bibinfo{author}{R.~Betti}, \bibinfo{author}{T.~R. Boehly},
  \bibinfo{author}{A.~Casner}, \bibinfo{author}{D.~E. Fratanduono},
  \bibinfo{author}{M.~Lafon}, \bibinfo{author}{D.~D. Meyerhofer},
  \bibinfo{author}{R.~Nora}, \bibinfo{author}{X.~Ribeyre},
  \bibinfo{author}{T.~C. Sangster}, \bibinfo{author}{G.~Schurtz},
  \bibinfo{author}{W.~Seka}, \bibinfo{author}{C.~Stoeckl},
  \bibinfo{author}{B.~Yaakobi},
\newblock \bibinfo{title}{Shock-ignition relevant experiments with planar
  targets on omega},
\newblock \bibinfo{journal}{Physics of Plasmas} \bibinfo{volume}{21}
  (\bibinfo{year}{2014}) \bibinfo{pages}{022702}.
%Type = Article
\bibitem[{Kalal(2003)}]{Kalal2003}
\bibinfo{author}{M.~Kalal},
\newblock \bibinfo{title}{Complex interferometry},
\newblock \bibinfo{journal}{Modern Topics in Physics}  (\bibinfo{year}{2003})
  \bibinfo{pages}{267--272}.
%Type = Article
\bibitem[{Kalal et~al.(1987)Kalal, Nugent, and Luther-Davies}]{Kalal:87}
\bibinfo{author}{M.~Kalal}, \bibinfo{author}{K.~A. Nugent},
  \bibinfo{author}{B.~Luther-Davies},
\newblock \bibinfo{title}{Phase-amplitude imaging: its application to fully
  automated analysis of magnetic field measurements in laser-produced plasmas},
\newblock \bibinfo{journal}{Appl. Opt.} \bibinfo{volume}{26}
  (\bibinfo{year}{1987}) \bibinfo{pages}{1674--1679}.
%Type = Article
\bibitem[{Zaraś-Szydłowska et~al.(2020)Zaraś-Szydłowska, Pisarczyk,
  Chodukowski, Rusiniak, Dudzak, Dostal, Kalal, Kochetkov, Krupka, Borodziuk,
  and Pisarczyk}]{AgaArt}
\bibinfo{author}{A.~Zaraś-Szydłowska}, \bibinfo{author}{T.~Pisarczyk},
  \bibinfo{author}{T.~Chodukowski}, \bibinfo{author}{Z.~Rusiniak},
  \bibinfo{author}{R.~Dudzak}, \bibinfo{author}{J.~Dostal},
  \bibinfo{author}{M.~Kalal}, \bibinfo{author}{I.~Kochetkov},
  \bibinfo{author}{M.~Krupka}, \bibinfo{author}{S.~Borodziuk},
  \bibinfo{author}{P.~Pisarczyk},
\newblock \bibinfo{title}{Implementation of amplitude–phase analysis of
  complex interferograms for measurement of spontaneous magnetic fields in
  laser generated plasma},
\newblock \bibinfo{journal}{AIP Advances} \bibinfo{volume}{10}
  (\bibinfo{year}{2020}) \bibinfo{pages}{115201}.
%Type = Article
\bibitem[{Zaras-Szydlowska et~al.(2017)Zaras-Szydlowska, Pisarczyk,
  Chodukowski, Kalinowska, Dudzak, Dostal, and Borodziuk}]{AgaBelfast}
\bibinfo{author}{A.~Zaras-Szydlowska}, \bibinfo{author}{T.~Pisarczyk},
  \bibinfo{author}{T.~Chodukowski}, \bibinfo{author}{Z.~Kalinowska},
  \bibinfo{author}{R.~Dudzak}, \bibinfo{author}{J.~Dostal},
  \bibinfo{author}{S.~Borodziuk},
\newblock \bibinfo{title}{Complex interferometry application for spontaneous
  magnetic field determination in laser-produced plasma},
\newblock \bibinfo{journal}{44th EPS Conference on Plasma Physics, P5-214,
  Belfast, Northern Ireland}  (\bibinfo{year}{2017}).
%Type = Article
\bibitem[{Kalal et~al.(1988)Kalal, Luther-Davies, and Nugent}]{Kalal1998}
\bibinfo{author}{M.~Kalal}, \bibinfo{author}{B.~Luther-Davies},
  \bibinfo{author}{K.~Nugent},
\newblock \bibinfo{title}{Phase‐amplitude imaging: The fully automated
  analysis of megagauss magnetic field measurements in laser‐produced
  plasmas},
\newblock \bibinfo{journal}{Journal of Applied Physics} \bibinfo{volume}{64}
  (\bibinfo{year}{1988}) \bibinfo{pages}{3845 -- 3850}.
%Type = Inbook
\bibitem[{Bracewell(2000)}]{Bracewell}
\bibinfo{author}{R.~N. Bracewell}, \bibinfo{title}{The Fourier transform and
  its applications}, \bibinfo{edition}{3rd ed.} ed.,
  \bibinfo{publisher}{McGraw-Hill}, \bibinfo{year}{2000}, p.
  \bibinfo{pages}{353}.
%Type = Article
\bibitem[{Kasperczuk et~al.(1978)Kasperczuk, Paduch, Pokora, and
  Wereszczynski}]{Kasper1978}
\bibinfo{author}{A.~Kasperczuk}, \bibinfo{author}{M.~Paduch},
  \bibinfo{author}{L.~Pokora}, \bibinfo{author}{Z.~Wereszczynski},
\newblock \bibinfo{title}{Numerical solution to abel integral equation in
  optical diagnostics of axisymmetrical objects},
\newblock \bibinfo{journal}{Journal of Technical Physics} \bibinfo{volume}{19}
  (\bibinfo{year}{1978}) \bibinfo{pages}{137--150}.
%Type = Article
\bibitem[{Hansen and Law(1985)}]{Hansen:85}
\bibinfo{author}{E.~W. Hansen}, \bibinfo{author}{P.~Law},
\newblock \bibinfo{title}{Recursive methods for computing the abel transform
  and its inverse},
\newblock \bibinfo{journal}{J. Opt. Soc. Am. A} \bibinfo{volume}{2}
  (\bibinfo{year}{1985}) \bibinfo{pages}{510--520}.
%Type = Article
\bibitem[{Kasperczuk and Pisarczyk(2001)}]{TadKasp2001}
\bibinfo{author}{A.~Kasperczuk}, \bibinfo{author}{T.~Pisarczyk},
\newblock \bibinfo{title}{Application of automated interferometric system for
  investigation of the behaviour of a laser-produced plasma in strong external
  magnetic fields},
\newblock \bibinfo{journal}{Optica Applicata} \bibinfo{volume}{31}
  (\bibinfo{year}{2001}) \bibinfo{pages}{571--597}.
%Type = Book
\bibitem[{Ghiglia and Pritt(1998)}]{Ghiglia1998TwoDimensionalPU}
\bibinfo{author}{D.~C. Ghiglia}, \bibinfo{author}{M.~D. Pritt},
  \bibinfo{title}{Two-Dimensional Phase Unwrapping: Theory, Algorithms, and
  Software}, \bibinfo{publisher}{Wiley}, \bibinfo{year}{1998}. \URLprefix
  \url{https://catalog.hathitrust.org/Record/003984153}.
%Type = Article
\bibitem[{Sindhu et~al.(2017)Sindhu, Kumar, and Shankar}]{Phase2017}
\bibinfo{author}{K.~Sindhu}, \bibinfo{author}{V.~Kumar},
  \bibinfo{author}{A.~Shankar},
\newblock \bibinfo{title}{A review on phase unwrapping techniques in
  interferometry},
\newblock \bibinfo{journal}{International Journal of Advanced Research in
  Electronics and Communication Engineering} \bibinfo{volume}{6}
  (\bibinfo{year}{2017}) \bibinfo{pages}{798--802}.
%Type = Article
\bibitem[{Takeda et~al.(1982)Takeda, Ina, and Kobayashi}]{Takeda82}
\bibinfo{author}{M.~Takeda}, \bibinfo{author}{H.~Ina},
  \bibinfo{author}{S.~Kobayashi},
\newblock \bibinfo{title}{Fourier-transform method of fringe-pattern analysis
  for computer-based topography and interferometry},
\newblock \bibinfo{journal}{J. Opt. Soc. Am.} \bibinfo{volume}{72}
  (\bibinfo{year}{1982}) \bibinfo{pages}{156--160}.
%Type = Article
\bibitem[{Zhong and Weng(2005)}]{Zhong:05}
\bibinfo{author}{J.~Zhong}, \bibinfo{author}{J.~Weng},
\newblock \bibinfo{title}{Phase retrieval of optical fringe patterns from the
  ridge of a wavelet transform},
\newblock \bibinfo{journal}{Opt. Lett.} \bibinfo{volume}{30}
  (\bibinfo{year}{2005}) \bibinfo{pages}{2560--2562}.
%Type = Article
\bibitem[{Hipp et~al.(2004)Hipp, Woisetschläger, Reiterer, and Neger}]{IDEA}
\bibinfo{author}{M.~Hipp}, \bibinfo{author}{J.~Woisetschläger},
  \bibinfo{author}{P.~Reiterer}, \bibinfo{author}{T.~Neger},
\newblock \bibinfo{title}{Digital evaluation of interferograms},
\newblock \bibinfo{journal}{Measurement: Journal of the International
  Measurement Confederation} \bibinfo{volume}{36} (\bibinfo{year}{2004})
  \bibinfo{pages}{53--66}.
%Type = Misc
\bibitem[{Flacco and Vinci(2013)}]{Neutrino}
\bibinfo{author}{A.~Flacco}, \bibinfo{author}{T.~Vinci},
  \bibinfo{title}{Neutrino},
  \bibinfo{howpublished}{\url{http://web.luli.polytechnique.fr/Neutrino/}},
  \bibinfo{year}{2013}. \bibinfo{note}{Accessed: 2020-04-12}.
%Type = Article
\bibitem[{{Arevalillo-Herráez} et~al.(2016){Arevalillo-Herráez}, {Villatoro},
  and {Gdeisat}}]{Quality}
\bibinfo{author}{M.~{Arevalillo-Herráez}}, \bibinfo{author}{F.~R.
  {Villatoro}}, \bibinfo{author}{M.~A. {Gdeisat}},
\newblock \bibinfo{title}{A robust and simple measure for quality-guided 2d
  phase unwrapping algorithms},
\newblock \bibinfo{journal}{IEEE Transactions on Image Processing}
  \bibinfo{volume}{25} (\bibinfo{year}{2016}) \bibinfo{pages}{2601--2609}.
%Type = Article
\bibitem[{Kalal(2016)}]{Kalal_2016}
\bibinfo{author}{M.~Kalal},
\newblock \bibinfo{title}{Complex interferometry potential in case of
  sufficiently stable diagnostic system},
\newblock \bibinfo{journal}{Journal of Instrumentation} \bibinfo{volume}{11}
  (\bibinfo{year}{2016}) \bibinfo{pages}{C06002--C06002}.

\end{thebibliography}
\end{document}